%% file: main_journal_arxiv.tex
\documentclass[journal, twoside, 10pt, a4paper]{IEEEtran}

\usepackage{xstring}
\usepackage[utf8]{inputenc} 
\usepackage[T1]{fontenc}
\usepackage{url}              %
\usepackage{cite}             %

\usepackage[cmex10]{amsmath}  %
\usepackage{mleftright}       %
\mleftright                   %

\usepackage{graphicx}         %
\usepackage{booktabs}         %

\usepackage{algorithm}    %
\usepackage[noend]{algpseudocode}
\usepackage{mathtools}

\usepackage{preamble}
\makeatletter%
\newcommand{\onetwo}[2]{
\if@twocolumn%
#2	
\else%
#1
\fi
}
\makeatother

\begin{document}
\IEEEoverridecommandlockouts %

\title{Stronger Polarization for the Deletion Channel}

 \author{%
   \IEEEauthorblockN{Dar Arava, Ido Tal
   }
   
   \IEEEauthorblockA{                   Department of Electrical and Computer Engineering,\\
Technion, Haifa 32000, Israel.\\
                     \{aravadar@campus, idotal@ee\}.technion.ac.il}
\thanks{This work was supported in part by the United-States Israel Binational Science Foundation (BSF) under Grant No.~2018218, by the National Science Foundation (NSF) under Grant CIF-2212437, and by the German Research Foundation (DFG) via the German--Israeli Project Cooperation (DIP). The work was presented in part at the International Symposium on Information Theory (ISIT) 2023. }
                   }

\maketitle

\begin{abstract}
In this paper we show a polar coding scheme for the deletion channel with a probability of error that decays roughly like $2^{-\sqrt{\Lambda}}$, where $\Lambda$ is the length of the codeword. That is, the same decay rate as that of seminal polar codes for memoryless channels.  This is stronger than prior art in which the square root is replaced by a cube root. Our coding scheme is similar yet distinct from prior art. The main differences are: 1) Guard-bands are placed in almost all polarization levels; 2) Trellis decoding is applied to the whole received word, and not to segments of it. As before, the scheme is capacity-achieving. The price we pay for this improvement is a higher decoding complexity, which is nonetheless still polynomial, $O(\Lambda^4)$.
\end{abstract}

\section{Introduction}
\label{sec:introduction}

Deletion errors, along with insertion errors, arise in communication channels with symbol-timing mismatch \cite{MBT:10p}. These synchronization errors are also common in polymer-based storage solutions \cite{HMG:19p}. 

The simplest theoretical model for these errors is the deletion channel with a constant deletion probability. The channel output is a sub-string of the symbols in the input. Deletions occur according to an  i.i.d. process that deletes each input symbol with probability
$\delta$.

\label{sec:polar codes}
Polar codes \cite{Arikan:09p} for a deletion channel with a fixed deletion probability were first presented in \cite{TPFV:22p}. See also \cite{TTVM:17c,TFV:17c,TFVL:17c,TFV:21p}, which use polar codes for weaker settings.
In \cite{TPFV:22p}, the authors show that for a fixed regular hidden-Markov input process and a fixed parameter $\nu \in (0,\frac{1}{3})$, their coding
scheme approaches the mutual information rate between the input process and the channel output.
The encoding and decoding complexities are $O(\Lambda \log \Lambda)$ and $O(\Lambda^{1+3\nu})$, respectively, where $\Lambda$ is the codeword length. Furthermore, for any $0 < \nu' < \nu$ and large enough $\Lambda$, the probability of a decoding block error is at most $2^{-\Lambda^{\nu'}}$. The authors also show that there exists a  sequence of regular hidden-Markov input processes for which the mutual information rate approaches the deletion channel capacity. This result also follows as a special case of the work of Li and
Tan \cite{LiTan:21p}, which proved the above for finite-order Markov processes.

We extend \cite{TPFV:22p}, and show that with simple alternations to the encoding and decoding schemes, the error probability decreases as $2^{-\Lambda^{\beta'}}$ where $\beta' \in (0,\frac{1}{2})$. Recall that this is stronger than the result in \cite{TPFV:22p}, in which the error exponent is $\nu' \in (0,\frac{1}{3})$.

The structure of our paper is as follows. Section~\ref{sec:mainTheorem} presents the main results of our work, still without all the definitions in place. Section~\ref{sec:notation} completes these definitions and sets up key notation. It also introduces key concepts we inherit from \cite{TPFV:22p}, and compares and contrasts our results to those in \cite{TPFV:22p}. The first three subsections of Section~\ref{sec:codingScheme} detail both our encoding and decoding schemes. Thus, the practitioner who only wishes to implement our scheme may decide to only read up to this point. The section concludes with a complexity analysis and simulation results. Section~\ref{sec:firstKeyLemma} contains our first key lemma, Lemma~\ref{lemmaStep}. This lemma bounds the evolution of the Bhattacharyya parameter when applying `$-$' and `$+$' polarization operations. It contains an additive term which is not present in simpler channels. Section~\ref{section:strongerwithpenalty} contains our second key lemma, Lemma~\ref{lemmaZproccess}, which shows that despite this additive penalty, strong polarization still occurs. Section~\ref{sec:deletion-proof} uses these two key lemmas in order to prove our main result.
\section{Main theorem}
\label{sec:mainTheorem}
Our main result builds upon the guard-band function $g$ introduced in \cite{TPFV:22p}. We will define $g$ shortly. For now, we note that $g(\bfx,n_0,\xi)$ recursively transforms $\bfx$, a word of length $2^n$, into a slightly longer word, where the length is controlled by the parameter $\xi$, and $n-n_0$ is the recursion depth. We say that $g(\bfx,n_0,\xi)$ is the result of adding guard-bands to $\bfx$.

Throughout the paper, we assume a deletion channel with a \emph{fixed} deletion probability. We also assume a fixed regular hidden Markov input distribution (see \cite[Subsection II-D]{TPFV:22p} for the formal definition). Denote by $\mathcal{I}$ the information rate between an input distributed according to this distribution and the corresponding output of the deletion channel.  Denote by $Z$ and $K$ the Bhattacharyya parameter and the total-variation, respectively (see, for example, \cite[Section III]{ShuvalTal:19c}).

Here is our ``stronger polarization'' theorem. It is stronger than the ``weaker polarization'' theorem proved in \cite{TPFV:22p}. In  \cite{TPFV:22p}, the parameters $0 < \beta' < \beta < \frac{1}{2}$ below are replaced with $0 < \nu' < \nu < \frac{1}{3}$, which implies a weaker decay to the error probability bound.

\begin{theorem}[Stronger polarization]
\label{theoremZstrong}
Let $\delta\in(0,1)$ be a fixed deletion probability of the deletion channel. Fix $\epsilon \in (0,1)$, $\xi \in (0,\frac{1}{6})$, and  $0 < \beta' < \beta < \frac{1}{2}$.
There exist $\nzeroth(\epsilon,\delta,\xi)$ and $\nthone(\epsilon,\beta,\beta',n_0)$ such that the following holds.
Take $n_0 \geq \nzeroth(\epsilon,\delta,\xi)$ and $n \geq \nthone(\epsilon,\beta,\beta',n_0)$.
Let $\bfX$ be of length $N=2^n$. The vector $\bfX$ is partitioned into blocks of length $2^{n_0}$, and each block is independently distributed according to the fixed regular hidden Markov input distribution.
Let $\bfU$ be the polar transform of $\bfX$. 
Denote by $\bfY$ the result of transmitting $g(\bfX, n_0, \xi)$ through the deletion channel.
The fraction of indices $i$ for which
\begin{IEEEeqnarray}{rCl}
	Z(U_i|U_1^{i-1},\bfY) &<& 2^{-{N^{\beta }}} < \frac{1}{2N} \cdot 2^{-\Lambda^{\beta'}} \label{BhatCond}\\
	K(U_i|U_1^{i-1}) &<& 2^{-{N^{\beta }}} < \frac{1}{2N} \cdot 2^{-\Lambda^{\beta'}} \label{TVCond}
\end{IEEEeqnarray}
is at least $\mathcal{I}-\epsilon$, where $\Lambda$ is the length of $g(\bfX, n_0,\xi)$. Furthermore,
\[
	\frac{N}{\Lambda} > 1 - \epsilon \; .
\]
\end{theorem}

By using the Honda-Yamamoto scheme \cite{HondaYamamoto:13p,KoradaUrbanke:10p}, we get the following corollary.
\begin{coro}
The above implies a coding scheme with rate $\mathcal{I}-2\epsilon$ and probability of error at most $2^{-\Lambda^{\beta'}}$, where $\Lambda$ is the length of the transmitted codeword.
\end{coro}

\section{Preliminaries} \label{sec:notation}
In this section we set up some notation and summarize key concepts from \cite{TPFV:22p}.

\subsection{Three related channels}
We now introduce three related channels: the deletion channel, the trimming channel, and their composition, the trimmed deletion channel.

\textbf{Deletion Channel (DC)} The deletion channel is the channel we are to code over. As its name implies, it takes a binary vector and deletes each bit with probability $\delta$. Thus, the output of the channel is typically shorter than its input. We will often denote a random vector that is an input to such a channel by $\bfG$ and denote the corresponding output by $\bfY$. 

The following two channels were introduced in \cite{TPFV:22p}, and are concepts we will need for our results as well.

\textbf{Trimming Channel (TC)} The trimming channel takes a binary vector and removes from it all leading and trailing zeros. Note that the trimming channel is deterministic. We will often denote the input to this channel by either $\bfY$ or $\bfZ$. We denote the trimming operation by appending a `$*$' as a superscript. Thus, the outputs corresponding to $\bfY$ and $\bfZ$ will be $\bfY^*$ and $\bfZ^*$, respectively.

\textbf{Trimmed Deletion Channel (TDC)} The trimmed deletion channel is the composition of the above two channels. Thus, if the input to the channel is $\bfG$, then we first pass $\bfG$ through the deletion channel and obtain $\bfY$, and then pass $\bfY$ through the trimming channel, which yields $\bfY^*$.

\input{./tikz_schemes/scheme_TDC}

\subsection{Weak polarization for the TDC}

We next state a weak polarization theorem for the trimmed deletion channel. The theorem follows easily by combining the weak polarization theorem \cite[Theorem~20]{TPFV:22p} with \cite[Lemma 1]{ShuvalTal:19.2p}, which bounds the Bhattacharyya parameter $Z$ and total variation $K$ by monotonic functions of the conditional entropy. As we will see, we will use this theorem to reach the stronger polarization rate stated in our main theorem.

\begin{theorem}[Weak polarization for the trimmed deletion channel] \label{theoremWeakPolarization}
Fix $\symstart \in (0, 1)$ and let $N_0 = 2^{n_0}$. For a given
fixed regular hidden Markov input distribution, let $\bfX$ 
be a random vector of length $N_0$ distributed according to the input distribution. Let $\bfY^*$ be the
result of passing $\bfX$ through a trimmed deletion channel with deletion probability $\delta$. Denote $\bfU$ as the polar transform of $\bfX$.
For all $\epsold>0$ there exists an $\nzerothtwo(\epsold,\symstart,\delta)$ s.t. if $n_0\geq\nzerothtwo$ then the fraction of indices $i$ for which
\[
	Z(U_i|U_1^{i-1},\bfY^*)<\symstart \quad \mbox{and} \quad 
K(U_i|U_1^{i-1})<\symstart 
\]
is larger than $\mathcal{I}-\epsold$.
\end{theorem}
Note that in the above, and throughout the paper, the information rate $\mathcal{I}$ is
\[\mathcal{I}\triangleq \lim_{N_0\to\infty}\frac{I(\bfX;\bfY)}{N_0} \; , \]
where $\bfX$ is of length $|\bfX|=N_0$ and is drawn according to the fixed regular hidden Markov input distribution and $\bfY$ is the corresponding output of the deletion channel. In \cite[Section V]{TPFV:22p} it was shown that $\mathcal{I}$ is indeed well defined.

We end this section by noting that Theorem~\ref{theoremWeakPolarization} does not employ guard-bands. However, in order to prove stronger polarization, we will employ guard-bands (as was also done in \cite{TPFV:22p}).

\subsection{Blocks and guard-bands} \label{subsectionPrelim}
Recall that in the main theorem, $\bfX$ was partitioned into independent blocks of length $N_0=2^{n_0}$. There are $N_1 = \frac{N}{N_0}$ such blocks, and we denote them by $\bfX(1),\bfX(2),\ldots,\bfX(N_1)$. That is, $\bfX$ is the concatenation of the above $N_1$ blocks,
 \[\bfX= \bfX(1)\odot\bfX(2)\odot\cdots\odot\bfX(N_1).\]
 We denote the first and second halves of $\bfX$ by $\bfX_\RN{1}$ and $\bfX_\RN{2}$. Denoting the length of a vector by $|\cdot|$, we have $|\bfX_\RN{1}| = |\bfX_\RN{2}| = \frac{N}{2}$ and
  \begin{align*} 
 \bfX&= \bfX_\RN{1}\odot\bfX_\RN{2}.
  \end{align*}
  Note that $\bfX_\RN{1}$ and $\bfX_\RN{2}$ are independent, a convention that will also hold in other places in which we use the ``$\RN{1}$'' and ``$\RN{2}$'' subscripts.

Recall that the function $g$ mentioned previously transforms a vector $\bfX$ of length $2^n$ into a slightly longer vector with ``guard-bands''. We now define $g$ recursively, and note that it adds the guard-bands between blocks. For a vector $\bfX$ of length $\leq 2^{n_0}$, $g(\bfX,n_0,\xi)$ is simply the identity function. For a vector $\bfX$ of length greater than $2^{n_0}$,
  \begin{equation} \label{GBrule}
	  g(\bfX,n_0,\xi) \triangleq  g(\bfX) \triangleq\underbrace{g(\bfX_\RN{1})}_{\triangleq \bfG_\RN{1}}\odot \underbrace{\stackrel{{\ell_n}}{\overlinesegment{\strut 000\ldots00}}}_{\triangleq \bfG_\Delta} \odot \underbrace{g(\bfX_\RN{2})}_{\triangleq \bfG_\RN{2}}\;.
  \end{equation}
That is, we add
\begin{equation}
	\label{ln}
    \ell_n= \left\lfloor2^{(1-\xi)(n-1)}\right\rfloor 
\end{equation}
`$0$' symbols between the first and second halves of $\bfX$, and apply $g$ recursively on each half. Note that $\xi > 0$ is a ``small'' constant that we will constrain later.
To summarize: $\bfX$ is a concatenation of $2^{n-n_0}$ independent blocks, each of length $N_0 = 2^{n_0}$. The function $g(\bfX,n_0,\xi)$ adds a guard-band of `$0$' symbols between each two blocks, and the length of these guard-bands varies. 
Here is an illustration, for the case in which $n=n_0+2$:

\centerline{
\input{./tikz_schemes/scheme_gX_small}
}

We are not concerned by the added GB bits, since for a large enough $n_0$,  the effect they have on the rate is negligible. The following lemma shows this. 
\begin{lemma}[GBs have a negligible effect on the rate] 
    \label{lemmaRateWithGBs}
    For all $\epsilon>0$ and $\xi>0$ there exists an $\nzerothtwo(\epsilon,\xi)$ such that if $n_0\geq\nzerothtwo$ then
    \[\frac{|\bfX|}{|g(\bfX,n_0,\xi)|}\triangleq\frac{N}{\Lambda}>1-\epsilon\;.\]
\end{lemma}

\begin{IEEEproof}
    
We set $\nzerothtwo\triangleq-\frac{1}{\xi}\log_2(\epsilon\cdot(1-2^{-\xi}))$ and bound $\Lambda$ as follows:
\begin{IEEEeqnarray}{lCl}\IEEEnonumber\label{Glessthan1.5N} 
\Lambda  \triangleq  |g(\bfX)|&%
  \eqann[<]{a}&  |\bfX|\cdot\left(1+\frac{2^{-(\xi n_0+1)}}{1-2^{-\xi}}\right) 
  \\\IEEEyessubnumber & <&  |\bfX|\cdot\left(1+\frac{2^{-\xi n_0}}{1-2^{-\xi}}\right) 
  \label{Glessthan1.5Na} 
  \\ \IEEEyessubnumber
 & \eqann[\leq]{b} & N\cdot\left(1+\epsilon\right) \; , \IEEEyessubnumber \label{Glessthan1.5Nb} 
\end{IEEEeqnarray}
where \eqannref{a} is by \cite[Lemma 22]{TPFV:22p} and \eqannref{b} follows by recalling that $n_0\geq\nzerothtwo$.
Finally:
\begin{IEEEeqnarray}{rl}\IEEEnonumber*
\frac{N}{\Lambda}\stackrel{\eqref{Glessthan1.5N}}{>} \frac{1}{1+\epsilon}=1-\frac{\epsilon}{1+\epsilon}>1-\epsilon \; .
\end{IEEEeqnarray}
\end{IEEEproof}

We remind the reader that $\bfG = \bfG_\RN{1} \odot \bfG_\Delta \odot \bfG_\RN{2}$ is passed through the DC. We denote the output of this channel by $\bfY$, and denote the parts corresponding to  $\bfG_\RN{1}$, $\bfG_\Delta$, and  $\bfG_\RN{2}$ by $\bfY_\RN{1}$, $\bfY_\Delta$, and  $\bfY_\RN{2}$, respectively. We further denote the application of the TC on $\bfY$ by $\bfZ \triangleq \bfY^*$, and denote the parts corresponding to $\bfY_\RN{1}$, $\bfY_\Delta$, and  $\bfY_\RN{2}$ by $\bfZ_\RN{1}$, $\bfZ_\Delta$, and  $\bfZ_\RN{2}$, respectively. See Fig. \ref{fig:XGYZ}, which is essentially \cite[Figure~5]{TPFV:22p}. Note that, in general, $\bfZ_\RN{1}$ is formed by trimming off only the left side of $\bfY_\RN{1}$. Hence, typically, $\bfZ_\RN{1} \neq (\bfY_\RN{1})^*$ and $\bfZ_\RN{2} \neq (\bfY_\RN{2})^*$. Also, note that in the typical case, $\bfZ_\Delta = \bfY_\Delta$.

\begin{figure} 
\begin{center}
  \input{./tikz_schemes/scheme_XGYZ}
\caption{The random variables $\bfX$, $\bfG$, $\bfY$, and $\bfZ$.}
\label{fig:XGYZ} %
\end{center}
\end{figure}

\subsection{An overview of this work compared to \cite{TPFV:22p}}
Throughout this paper we inherit concepts from \cite{TPFV:22p}, both in our coding scheme and in our analysis. All the concepts we use from \cite{TPFV:22p}  are defined in this work as well, to ensure this paper is self-contained.   To further highlight what is new, this section outlines the main ideas in \cite{TPFV:22p} and the differences in this work. Every topic mentioned in this section will be explained in more detail later on.

The encoder in \cite{TPFV:22p} adds GBs to separate the polar coded word into blocks. We do the same, while slightly changing constrains on parameters that control how many GBs are added and what their lengths are ($\xi$ and $n_0$). We also slightly change the rule on how to select the information indices in the polar coding scheme. However, in essence, we use the same encoder. 

The decoding steps in \cite{TPFV:22p} are as follows. First, the decoder uses the GBs to process the received word into trimmed blocks. In a good scenario, this processing yields the TDC outputs of the blocks of $\bfX$. Next, the decoder builds a trellis for each trimmed block. The trellises incorporate the joint probability of the deletion channel input and output. 
Lastly, the trellises are processed according to the successive cancellation polar decoder. 

Our decoder does not perform the first step of partitioning and trimming the output. Instead, we build one big trellis using the whole received word. As the trellises in  \cite{TPFV:22p}, this trellis incorporates the joint probability of the deletion channel input and output. As in \cite{TPFV:22p}, we then process the trellis according to  the successive cancellation algorithm. Unlike in  \cite{TPFV:22p}, our trellis includes sections that correspond to GBs. Thus, we expand the definition from  \cite{TPFV:22p} on how to build the trellis and how to process it such that it deals with the GB sections as well. 

The processing of $\bfY$ into trimmed blocks in \cite{TPFV:22p} is subject to errors. For example, if the middle index of the received word does not fall within the middle GB, the decoder in  \cite{TPFV:22p} will likely fail. Thus, our coding scheme which does not contain such a preliminary step has a higher probability of decoding successfully. This comes at a cost in decoding complexity, $\mathcal{O}(\Lambda^4)$ in our case compared to  at most $\mathcal{O}(\Lambda^2)$ in \cite{TPFV:22p}.
 
Regarding the analysis of the error probability, \cite{TPFV:22p} shows that weak polarization occurs for the TDC after $n_0$ polar transforms (without adding GBs). Then, they show strong polarization for a ``block-TDC'' output. That is, essentially, for input $g(\bfX)$ the ``block-TDC''  output is the TDC outputs of each block in $\bfX$. We remind that, in a good scenario,  the result of the partition and trimming operation in the first step of the decoder of \cite{TPFV:22p} is exactly the  ``block-TDC'' output. Lastly, they bound the probability of failing in partitioning and trimming $\bfY$  into the ``block-TDC'' output (i.e. the probability of not being in the good scenario). This error term is what ultimately constrained their error exponent to at most $\frac{1}{3}$.

We remind that our decoder does not perform partition and trimming to the received vector $\bfY$. However, when analyzing the performance of our coding scheme we will use the TDC. Since the TDC is a degraded version of the deletion channel, we may use the bounds on its error probability to bound the error probability of the DC as well.   We will show a recursive relation on how the TDC polarizes. This recursive relation will be pivotal in our proof. We will see that the recursive relation is reached while bounding the probability of successfully partitioning the TDC output of $g(\bfX)$ to the TDC outputs of $g(\bfX_{\RN{1}})$ and $g(\bfX_{\RN{2}})$. This is not to be confused with the requirement in \cite{TPFV:22p} of successfully partitioning the received vector to the TDC outputs of the blocks $\bfX(1),\ldots,\bfX(N_1)$, which is less probable. When bounding the probability of the former, we will use similar arguments as in the bound of the latter in \cite{TPFV:22p}. To conclude ---  in \cite{TPFV:22p} it was natural to understand why the authors analyze a channel that trims the output (since the decoder applies trimming), whereas in our setting this might seem like a detour to the reader. However, we do not know how to reach our destination (an analysis of the error probability) without taking this detour.

\section{Coding scheme for the Deletion Channel}
\label{sec:codingScheme}

\subsection{Encoder}\label{subsec:EncoderDeletion}
The encoding steps are as follows. The vector $\bfu = u_1^N$ is produced successively,
starting from $u_1$ and ending in $u_N$. If the current index $i$ satisfies \eqref{BhatCond} and \eqref{TVCond}, then $u_i$
is set to an information bit.
Otherwise, $u_i$ is a ``frozen bit''. Its value is set according to the distribution $\Prb(U_i = u_i|U_1^{i-1} = u_1^{i-1})$,
where $u_1^{i-1}$ are the realizations that were
set in the previous indices. Note that this distribution is derived from the regular hidden Markov input distribution we set for the independent blocks. That is, if all bits were frozen, we would get that the blocks $\bfx(1),\bfx(2),\ldots,\bfx(N_1)$ of $\bfx = \mathcal{A}(\bfu)$ are independently drawn from the regular hidden Markov process. For the calculation of $\Prb(U_i = u_i|U_1^{i-1} = u_1^{i-1})$ we may
use the Markov generalization 
of Honda-Yamamoto successive cancellation encoding for
 input distributions with memory \cite{Wang+:15c, Wang+:14a}.  This is done by recursively tracking the distribution of the hidden Markov input and does not require the deletion channel trellis. 
We assume the random draws are common to both
the encoder and the decoder. This is typically implemented
using a pseudo-random number generator, common to both
sides: if the pseudo-random number $0 \leq r_i \leq 1$ drawn for this stage is such that $\Prb(U_i = 0 |U_1^{i-1}=u_1^{i-1})\leq r_i$, we set $u_i = 0$. Otherwise, we set $u_i = 1$. We note that in practice, one may wish to simplify the above by picking a frozen $u_i$ so that it maximizes $\Prb(U_i = u_i |U_1^{i-1}=u_1^{i-1})$\cite{Mondelli+:18p, ChouBloch:15c}.
We next transform $\bfu$ to $\bfx = \mathcal{A}(\bfu)$, using the standard Arıkan transform presented in the seminal paper \cite{Arikan:09p}.
For the final step, we add guard-bands of $0$'s to $\bfx$, creating $\bfg=g(\bfx,n_0,\xi)$. We transmit $\bfg$ through the deletion channel. The transform adding guard-bands is as in \eqref{GBrule}.

\begin{algorithm}
\caption{Encoder}\label{encoder}
\begin{algorithmic}[1]
\Function{Encode}{information bits, $n_0$, $\xi$}
\For{$i=0$\textbf{ to} $N-1$}
\If{ $i$ satisfies \eqref{BhatCond} and \eqref{TVCond}}
\State $u_i\gets\text{next information bit}$
\Else  \Comment{$u_i$ is dynamically frozen}
\State $u_i\gets\text{draw from input distribution, given $u_1^{i-1}$ }$
\EndIf
 \EndFor
\State  $\bfx\gets\mathcal{A}(\bfu)$ \Comment{\Arikan transform}
\State  $\bfg\gets g(\bfx,n_0,\xi)$ \Comment{add GBs by \eqref{GBrule}}
\State $\textbf{return } \bfg$
\EndFunction
\end{algorithmic}
\end{algorithm}
Our encoder is the same as that in \cite{TPFV:22p}, with the following differences. In our encoder, the selection of information indices is according to \eqref{BhatCond}  and \eqref{TVCond}. In \cite{TPFV:22p}, these conditions are  replaced with (essentially, see Equations (101) and (102) in \cite{TPFV:22p}):
\onetwo{
\[
 Z(U_i|U_1^{i-1},(\bfY(1))^*,...,(\bfY(N_1))^*)<2^{-N^{\nu}} \; \text{ and }\;K(U_i|U_1^{i-1})<2^{-N^{\nu}}  \;,
\]
}{
\begin{multline*}
 Z(U_i|U_1^{i-1},(\bfY(1))^*,...,(\bfY(N_1))^*)<2^{-N^{\nu}} \\\text{ and }\;K(U_i|U_1^{i-1})<2^{-N^{\nu}}  \;,
\end{multline*}
}
where  $\nu\in\left(0,\frac{1}{3}\right)$. That is, the deletion channel output $\bfY$ 
in \eqref{BhatCond} is replaced with the TDC outputs of the blocks of $\bfX$, and the upper bounds in \eqref{BhatCond}  and \eqref{TVCond} are replaced with a weaker bound. We also differ in the selection of $n_0$, i.e.\ the step from which we start adding guard-bands. Still, the encoding complexity remains $O(\Lambda\log \Lambda)$ for a codeword length of  $|\bfg|=\Lambda$ \cite[Sub-claim 9]{TPFV:22p}. We note that the encoding complexity also increases polynomially with the number of states. This is due to the computation of $\Prb(U_i = u_i |U_1^{i-1} = u_1^{i-1})$ when setting the frozen bit values \cite[Theorem 2]{Wang+:15c}. In total, the encoding complexity is $O(|\mathcal{S}|^3\Lambda\log \Lambda)$, where $|\mathcal{S}|$ is the number of states in the hidden Markov process. Thus, for a fixed input distribution we get $O(\Lambda\log \Lambda)$.

\subsection{High level decoder description}\label{subsec:DeletionDecoder1}
Our decoder is a generalization of the one described in \cite[Subsection IV]{TPFV:22p}. That is, a base trellis is constructed, and then `$-$' and `$+$' operations are applied to it. One major difference is that in our case, the base trellis corresponds to all of the received word (Panel~(a) in Figure~\ref{fig:trellis_transform_with_GBs}). This is in contrast to \cite{TPFV:22p}, in which $N/N_0$ base trellises are constructed --- one for each block (Figure~\ref{fig:prior_art_decoder}). Since we operate on a larger trellis, our complexity is $O(\Lambda^4)$, as opposed to at most $O(\Lambda^2)$ in \cite{TPFV:22p}. As explained in Theorem~\ref{theoremZstrong}, this added complexity is compensated for by a reduced probability of error. That is, we reach the same asymptotic bound as seminal polar codes \cite{ArikanTelatar:09c}.

As in \cite{TPFV:22p}, we will use a trellis $\mathcal{T}$ to represent the joint probability of the deletion channel input and output. 
We perform   `$-$' and `$+$' operations on  $\mathcal{T}$, which merge two-edge paths in $\mathcal{T}$ and result in trellises with a reduced number of columns:  $\mathcal{T}^{[0]}$ and $\mathcal{T}^{[1]}$, respectively.  
The mechanics of our decoder will differ from those of  \cite[Section IV]{TPFV:22p} in one main point: our trellis contains sections corresponding to guard-bands, and such sections will be treated differently than what was described in \cite[Section IV]{TPFV:22p} when applying `$-$' and `$+$' operations.

The decoder recursively performs `$-$' and `$+$' transforms on $\mathcal{T}$  as follows. First, we perform $n$ `$-$' transforms, creating $\mathcal{T}^{[000 \ldots 00]}$. For simplicity, assume for now that the hidden Markov input distribution contains only one state (that is, the input distribution is memoryless). In the resulting trellis, we consider the pair of single-edge paths from the  upper-left vertex to the lower-right vertex, which represent the two possible values for $\hat{u}_1$. The decision on $\hat{u}_1$ (if it is not frozen) is by the most probable value, i.e.\ the edge with the largest probability.
Using $\hat{u}_1$, we next create:
\[
	\mathcal{T}^{\stackrel{n}{\overlinesegment{ [000 \ldots01]}}}=\left(\mathcal{T}^{\stackrel{n-1}{\overlinesegment{ [000 \ldots0]}}}\right)^{\strut [1]} \; .
\]
 We use $\mathcal{T}^{[000 \ldots01]}$ to decide on $\hat{u}_2$. We repeat this procedure such that with trellis $\mathcal{T}^{[b_1 b_2\ldots b_{n}]}$ we decide on the value of $\hat{u}_{ i(b_1,\ldots,b_n)}$ (if it is not frozen), where $i(b_1,\ldots,b_n)$ is the index with binary representation $b_1,\ldots,b_n$, see (\ref{eq:iBinaryRepresentation}). If $i$ is a frozen index, we mimic the encoder operation to get $\hat{u}_{ i}$ based on the previous $\hat{u}_1^{i-1}$.  See Figure \ref{fig:RecursiveTrellisTransforms} for an illustration of the decoding process.
The next three subsections fill in the details missing in the above description.
\begin{figure} 
\input{./tikz_schemes/scheme_trellis_polarization}
\caption{Recursive trellis transforms.} \label{fig:RecursiveTrellisTransforms}
\end{figure}

\onetwo{
\begin{figure*}
	\begin{center}
\input{./tikz_schemes/scheme_trellis_evolution_spaced}
\caption{Trellis evolution in the decoder. Panel~(a) is the initial trellis $\mathcal{T}$. White rectangles are the block trellises which correspond to the blocks of $\bfX$. In block trellises, both edges with label `$1$' and edges with label `$0$' exist. The blue rectangles mark the GB trellises, which correspond to the GBs (have only `$0$' labels). The number of sections in each block trellis or GB trellis in $\mathcal{T}$ is the length of the  corresponding block or GB. Panel~(b) shows the trellis after collapsing each of the GB trellises into a single section and after the first $n_0$ polar transforms. Panel~(c) shows how the next polar transform is applied, and is divided into two steps: merging a block trellis with a GB trellis and then merging the resulting trellis with a block trellis (either a `$-$' or a `$+$' merge).
} %
\label{fig:trellis_transform_with_GBs} 
\end{center}
\end{figure*}
}
{
\begin{figure}
	\begin{center}
\input{./tikz_schemes/scheme_trellis_evolution}
\caption{Trellis evolution in the decoder. Panel~(a) is the initial trellis $\mathcal{T}$. White rectangles are the block trellises which correspond to the blocks of $\bfX$. In block trellises, both edges with label `$1$' and edges with label `$0$' exist. The blue rectangles mark the GB trellises, which correspond to the GBs (have only `$0$' labels). The number of sections in each block trellis or GB trellis in $\mathcal{T}$ is the length of the  corresponding block or GB. Panel~(b) shows the trellis after collapsing each of the GB trellises into a single section and after the first $n_0$ polar transforms. Panel~(c) shows how the next polar transform is applied, and is divided into two steps: merging a block trellis with a GB trellis and then merging the resulting trellis with a block trellis (either a `$-$' or a `$+$' merge).
} %
\label{fig:trellis_transform_with_GBs} 
\end{center}
\end{figure}
}

\begin{figure}
	\begin{center}
\input{./tikz_schemes/smallTrellises-noBackground}
\caption{An illustration of the prior-art decoder \cite{TPFV:22p}. The initial step partitions the channel output $\bfy$ into trimmed outputs of each block. In this example  $n=n_0+2$, hence there are four blocks. The next step builds a block trellis for each block. These trellises are then processed according to the successive cancellation decoder, by recursively performing `$-$' and `$+$' transforms to them. } %
\label{fig:prior_art_decoder} 
\end{center}
\end{figure}

\subsection{Trellis operations}\label{subsec:DeletionDecoder2}We first describe how to build the initial trellis $\mathcal{T}$. We will then describe the transforms performed on the trellis. For an extended description on how to build $\mathcal{T}$ and how to perform `$-$' and `$+$' operations on it, see \cite[Sections III and IV]{TPFV:22p}. We will present the main idea and provide a visual example, while emphasizing the differences in our case.

The trellis $\mathcal{T}$ is built according to the received binary vector $\bfy$. It also factors in the deletion probability $\delta$, the stationary probabilities of the input process states, and the transition probabilities between them.

\subsubsection{Vertices}
  The trellis consists of a grid of vertices with $|\bfy|+1$ rows and  $|\bfg|+1$ columns (where $\bfg\triangleq g(\bfx)$ is the transmitted codeword we are to decode).  We will denote by $i$ a vertex row index, by $j$ a vertex column index and by $\indj$ (a dot-less $j$) the corresponding index of the bit $g_{j}$ in the vector $\bfx$. That is, if $g_j$ is a bit that originates from the blocks of $\bfx$ (i.e. is not a GB bit) then $g_j=x_{\indj}$.  For some of the columns there will also be $|\mathcal{S}|$ ``layers'' of vertices, which correspond to the possible states of the input distribution.
 Specifically, for a column $j-1$, where $j\in\{1,\ldots,|\bfg|\}$, if $g_j$ originates from $\bfx$ then we must consider the possible input states of $S_{\indj-1}$ in the trellis. For this column we denote by $v_{i,j-1,s}$ the vertex in row $i$ column $j-1$ and layer (state) $s\in\mathcal{S}$. The vertex $v_{i,j-1,s}$ corresponds to the event in which the output corresponding to $g_1^{j-1}$ is $y_1^i$ and the state of the input process before transmitting $x_\indj$ was $S_{\indj-1}=s$.   Otherwise, if $g_j$ is a GB bit, we denote by $v_{i,j-1}$ the vertex in row $i$ and column $j-1$ (no layers in this column).  The vertex $v_{i,j-1}$ corresponds to the event in which the output corresponding to $g_1^{j-1}$ is $y_1^i$. The last column in $\mathcal{T}$ will also not have multiple layers, $v_{i,|\bfg|}$.  As we will shortly see, the only vertex of interest in this column is the vertex $v_{|\bfy|,|\bfg|}$, which corresponds to the event in which the output corresponding to $\bfg$ is $\bfy$, which is always the case. The total set of vertices is:
 
\begin{multline*}
\left\{v_{i,j-1,s}:\; \mathsmall{
\begin{array}{l}
i\in\{0,\ldots,|\bfy|\},\\
 j\in\{1,\ldots,|\bfg|\} \text{ is an index in a block}\\
 s\in\mathcal{S}
 \end{array}}\right\}
 \\
 \bigcup\left\{v_{i,j-1}:\;
 \mathsmall{
\begin{array}{l}i\in\{0,\ldots,|\bfy|\},\\
 j \in\{1,\ldots,|\bfg|\} \text{ is an index in a GB} \end{array}}
 \right\}
 \\
 \bigcup\left\{v_{i,|\bfg|}:\;\mathsmall{i\in\{0,\ldots,|\bfy|\}} \right\}\;.    
\end{multline*}

We set weights for the vertices in the first and last columns of the trellis as in \cite{TPFV:22p}. That is, for   $v_{0,0,s}$, where $s \in\mathcal{S}$, we set $\mathrm{weight}(v_{0,0,s}) =
\pi(s)$, the
stationary probability of state $s$ in the input process. All other vertices in the first column have $\mathrm{weight}(v_{i,0,s}) = 0$. By this we effectively force all paths to
start at a vertex $v_{0,0,s}$, and incorporate
the probability of starting the path at the state $s$. 
For the last column we set $\mathrm{weight}(v_{|\bfy|,|\bfg|} ) = 1$ and $\mathrm{weight}(v_{i,|\bfg|} ) = 0$ for $i\in\{0,1,\ldots,|\bfy|-1\}$. By this
we effectively force all paths to end at the vertex $v_{|\bfy|,|\bfg|}$.
That is, at the end of a path, $|\bfg|$ symbols have been
transmitted, and of these, $|\bfy|$ have been received.

\subsubsection{Edges}
We briefly explain how the edges are set in the trellis $\mathcal{T}$.  Each edge $e$ connects a vertex in some column $j-1$ to a vertex in column $j$, and corresponds to the input bit $g_{j}$ ($j\in\{1,...,|\bfg|\}$). Hence, the edge label is $g_j$. We refer to the edges from column $j-1$ to column $j$ as the edges in trellis section $j$. 
Notice that if $e$ is an outgoing edge of some vertex $v_{i,j-1,s}$ and is an incoming edge of some other vertex $v_{i',j,s'}$, it corresponds to the transition of states in the input process from $S_{\indj-1}=s$  to $S_{\indj}=s'$.
 \begin{itemize}
     \item 
A \emph{diagonal} edge  connects a vertex in row $i-1$ to a vertex in row $i$, and corresponds to output bit $y_{i}$ ($i\in\{1,\ldots,|\bfy|\}$).
This edge corresponds to the event where the bit $y_i$ is the result of $g_j$ passing through the deletion channel and not being deleted.
As in \cite{TPFV:22p}, the edge label is the value $y_i$.
\item
A \emph{horizontal} edge connects a vertex in row $i-1$ to a vertex in row $i-1$. 
This edge corresponds to the event where
$g_j$ was deleted. If $g_j$ originates from $\bfx$, there are two parallel horizontal edges for the two possible cases, one with the label `$0$' for $g_j=0$ and one with the label `$1$' for $g_j=1$. Conversely,  if $g_j$ is a GB bit, there is a single horizontal edge with label `$0$'.
 \end{itemize}
Notice that if $g_j$ is a GB bit, all the edges in section $j$ (from  column $j-1$ to column $j$) are labeled `$0$' (no `$1$' edges), since $g_j=0$.

The edge weights in the initial trellis $\mathcal{T}$ are the conditional probabilities of the events they correspond to. For example, the edge \[v_{i-1,j-1,s}\stackrel{e}{\to} v_{i,j,s'}\] connects the vertex in row $i-1$, column $j-1$, and layer $s$  to the vertex in row $i$, column $j$, and layer $s'$.
This edge corresponds to the ``no deletion'' event:  $g_j = x_\indj$ is the input to the channel, there was no deletion hence $y_i = g_j = x_\indj$, and the input process states before and after this happened were $S_{\indj-1}=s$ and $S_{\indj}=s'$.
The edge weight is the probability of this event, conditioned on $S_{\indj-1}=s$:
\[\mathrm{weight}(e) = (1-\delta)\cdot\Prb(X_\indj=y_i,S_{\indj}=s'|S_{\indj-1}=s)\;.\]  
Note that the term $\Prb(X_\indj=y_i,S_{\indj}=s'|S_{\indj-1}=s)$ is determined by the input process.
See Table~\ref{table:trellisedgeweights}  for the weights and labels of all the edges in $\mathcal{T}$.

\begin{table*}[t]
\begin{threeparttable}
\centering %
\caption{Edge weights and labels in trellis $\mathcal{T}$.} \label{table:trellisedgeweights}
\begin{tabular}{!{\vrule width 1pt}c!{\vrule width 1pt}l|l!{\vrule width 1pt}}
\Cline{2-3}{1pt}
    \multicolumn{1}{c!{\vrule width 1pt}}{} &\thead{$\begin{array}{c} \text{horizontal edge} \\ \text{(deletion)} \\  i'=i \end{array}$}  & \thead{$\begin{array}{c} \text{diagonal edge} \\ \text{(non-deletion)} \\i'=i+1\end{array} $} \\
\noalign{\hrule height 1pt}
$\begin{array}{c}v_{i,j-1,s}\stackrel{e}{\to} v_{i',j,s'} \\ (g_j \text{ is a bit from }\bfx)\end{array}$   
&  $\begin{array}{l}
x\triangleq\mathrm{label}(e)\in\{0,1\} \text{ (two parallel edges)}, \\
\mathrm{weight}(e)=\delta\cdot\Prb(X_\indj=x,S_{\indj}=s'|S_{\indj-1}=s)%
\end{array}$  
&  $\begin{array}{l}\mathrm{label}(e)=y_{i'},\\\mathrm{weight}(e)=(1-\delta)\cdot\Prb(X_\indj=y_{i'},S_{\indj}=s'|S_{\indj-1}=s) \end{array}$  
\\
\hline
$\begin{array}{c}v_{i,j-1}\stackrel{e}{\to} v_{i',j} \\ (g_j \text{ is a bit from a GB})\end{array}$ 
& $\begin{array}{l}\mathrm{label}(e)=0,\\\mathrm{weight}(e)=\delta\end{array} $
& if $y_{i'}=0$: $\begin{array}{l}\mathrm{label}(e)=y_{i'}=0,\\\mathrm{weight}(e)=1-\delta\end{array} \qquad \qquad\text{else: no edge} $
\\
\hline
 $\begin{array}{c}v_{i,j-1}\stackrel{e}{\to} v_{i',j,s'} \\ (g_j \text{ is a last bit in a GB})\end{array} $  
 &$\begin{array}{l}\mathrm{label}(e)=0,\\\mathrm{weight}(e)=\delta\cdot\pi(s') \end{array} $
 & if $y_{i'}=0$: $\begin{array}{l}\mathrm{label}(e)=y_{i'}=0,\\\mathrm{weight}(e)=(1-\delta)\cdot\pi(s') \end{array} \; \text{else: no edge} $
 \\
 \hline
 $\begin{array}{c}v_{i,j-1,s}\stackrel{e}{\to} v_{i',j}\\ (g_j \text{ is a last bit in a block of }\bfx)\end{array}$ 
 & $\begin{array}{l}x\triangleq\mathrm{label}(e)\in\{0,1\}\text{ (two parallel edges)},\\\mathrm{weight}(e)=\delta\cdot\Prb(X_\indj=x|S_{\indj-1}=s)\end{array} $ 
 & $\begin{array}{l}\mathrm{label}(e)=y_{i'},\\\mathrm{weight}(e)=(1-\delta)\cdot\Prb(X_\indj=y_{i'}|S_{\indj-1}=s)\end{array} $ \\
\noalign{\hrule height 1pt}
\end{tabular}

\begin{tablenotes}
            \item \small The first row regards edges in the sections of the trellis that correspond to the blocks of $\bfx$. The weights and labels of the edges in these sections are as defined in \cite{TPFV:22p}. The second row regards edges in GB sections of the trellis.
		    The third and forth rows regard the edges in the ``stitches'' between blocks and GBs. The third row describes edges in the last section of each GB. We denoted by $\pi(s)$ the probability that the regular hidden-Markov process starts at state $s\in\mathcal{S}$. We incorporated $\pi(s)$ in the edge weights this way since each block is independently drawn (a new initial state for each block). The fourth row describes the edges in the last section of each block. Here we disregard the state $S_{\indj}$, since it is of no consequence. That is, the GB bits are always `$0$'; they are \emph{not} drawn from the hidden Markov input distribution. Thus, the weights in the fourth row are equal to the marginalization of the corresponding weights in the first row over all $s'$. 
    \end{tablenotes}
    \end{threeparttable}
\end{table*}

\subsubsection{Block and GB trellises}
Up to this point, in this section we have defined how to build the initial trellis $\mathcal{T}$. We now set up the definitions of a `block' trellis and a `guard-band' (GB) trellis which we will use soon to define the polar transforms performed on $\mathcal{T}$. 
The initial trellis $\mathcal{T}$ that we have described is a concatenation of `block' trellises and `guard-band' trellises. A block trellis consists of consecutive sections with edges starting in vertices of the form $v_{i,j-1,s}$. Consequently, section $j$ is part of a `block' trellis iff $j$ is an index originating from a block in $\bfx$.

Informally, a GB trellis is a block trellis without the edges labeled `$1$'. Formally, it consists of consecutive sections with edges starting in vertices of the form $v_{i,j-1}$. Consequently, section $j$ is part of a `guard-band' trellis iff $j$ is an index originating from a guard-band added by the function $g$.

The first $N_0$ sections in $\mathcal{T}$ are the block trellis $\mathcal{T}^{\mathrm{B-1}}$. The following $\ell_{n_0+1}$ sections are the GB trellis $\mathcal{T}^{\mathrm{GB-1}}$. The next $N_0$ sections are  $\mathcal{T}^{\mathrm{B-2}}$, followed by $\ell_{n_0+2}$ sections of $\mathcal{T}^{\mathrm{GB-2}}$ and so on (where the lengths of the guard-bands are according to the recursive function $g$ defined in \eqref{GBrule}).
That is,
\[\mathcal{T}\triangleq\mathcal{T}^{\mathrm{B-1}}
\odot\mathcal{T}^{\mathrm{GB-1}}
\odot\mathcal{T}^{\mathrm{B-2}}
\odot\mathcal{T}^{\mathrm{GB-2}}
...\odot\mathcal{T}^{\mathrm{GB-}(N_1-1)}
\odot\mathcal{T}^{\mathrm{B-}N_1}\;.\]
See an example of $\mathcal{T}$ in Panel~(a) in Figure~\ref{fig:trellis_transform_with_GBs} where $n=n_0+2$. See an example of a block trellis in  Figure~\ref{fig:Blocktrellis}, and  an example of a GB trellis in  Figure~\ref{fig:GBtrellis}. In the latter two figures, the code length is $|\bfg|=11$ bits. Specifically,
$N_0=2^{n_0}=4$ and $n=n_0+1=3$. That is, only one GB is added to $\bfx$ (see also Panel~(a) in Figure~\ref{fig:with_GB_sub-trellis}) and it is of length $\ell_{3}=3$ bits. In this example (that is, in Figures~\ref{fig:Blocktrellis}--\ref{fig:with_GB_sub-trellis}) three bits were deleted in the channel, and the received vector is $\bfy=01100010$  ($|\bfy|=8$). Lastly we note that in this example the input distribution was memoryless, $|\mathcal{S}|=1$, hence only a single layer is depicted in Figure~\ref{fig:Blocktrellis}.

\subsubsection{Trellis polar transforms} \label{subsec:TrellisPolarTransforms} We now briefly describe the polar transforms performed on $\mathcal{T}$. Our trellis consists of both block trellises and GB trellises, whereas in  \cite{TPFV:22p} there were only block trellises in a ``trimmed'' version (contrast Figure~\ref{fig:trellis_transform_with_GBs} with Figure~\ref{fig:prior_art_decoder}). 
Hence, we need to  define how to process these GB trellises along the polarization transforms.

A key operation which we will define shortly is the ``merge'' operation, which collapses two sections into one. It will serve two purposes: collapsing a GB trellis into a single section (getting from the left part of Figure~\ref{fig:GBtrellis} to the right part) and merging this section into the block section to its left (getting from Panel~(a) in Figure~\ref{fig:with_GB_sub-trellis} to Panel~(b)).

A preliminary step of our decoding algorithm is to collapse each GB trellis into a single section. For each guard-band trellis this is performed as follows, see Figure~\ref{fig:GBtrellis}. Recall that merging two sections together results in one section. We repeatedly merge the rightmost section in the GB trellis with the GB section to its left, until only one section remains. That is, first merge the rightmost section with its neighbor to the left, then merge the section created with the next neighbor to the left and so on, up to the leftmost section. %

We now define the merge operation. A merge operation is always performed on a pair of consecutive sections, where the right section is from a GB trellis (has only `$0$' edges). It combines all the possible two-edge paths connecting two vertices,
\begin{equation}\label{eq:e1e2edges}
    \alpha\stackrel{e_1}{\to}\beta\stackrel{e_2}{\to}\gamma\;,
\end{equation}
where all $e_1$ edges have the same label and all $e_2$ edges are labeled `$0$', into one single edge
\[\alpha\stackrel{\tilde{e}}{\to}\gamma\;.\]
We remind that edges always connect a vertex in some column to a vertex in the following column, hence $\alpha$ is a vertex in the leftmost column in the sections merged,  $\beta$ is a vertex in the middle column, and $\gamma$ is in the rightmost column. The label and weight of the resulting edge are
\onetwo{
	\[
    \mathrm{label}(\tilde{e}) = \mathrm{label}(e_1)
    \quad \mathrm{and} \quad \mathrm{weight}(\tilde{e}) = \sum_{e_1,e_2}\mathrm{weight}(e_1)\cdot\mathrm{weight}(e_2)\;.
\]
}{
\begin{multline*}
    \mathrm{label}(\tilde{e}) = \mathrm{label}(e_1)
    \\ 
    \text{and  } \mathrm{weight}(\tilde{e}) = \sum_{e_1,e_2}\mathrm{weight}(e_1)\cdot\mathrm{weight}(e_2)\;.
    \end{multline*}
}

We note that since the right section merged is always from a GB trellis with only `$0$' labels, setting the label of $\tilde{e}$ to $\mathrm{label}(e_1)$ is the same as setting it to: $\mathrm{label}(e_1)$ xor $\mathrm{label}(e_2)$. Thus, the new edges created by the merge operation are created just as in the minus transform defined in \cite[Definition 5]{TPFV:22p}.

The first $n_0$ `$-$' and `$+$' transforms on $\mathcal{T}$ are as defined in  \cite[Definitions 5,6]{TPFV:22p}, and are performed on the block trellises. We refer to this as the ``without GB'' phase. After this phase, each block trellis has collapsed into one section,  see Figure~\ref{fig:Blocktrellis}. 
In the following $n_1\triangleq n-n_0$ transforms, which we refer to as ``with GB'' transforms, we perform the following operation on each `sub-trellis'.  Each sub-trellis is denoted by $s\mathcal{T}$ and consists of three consecutive sections. Specifically, we define
\begin{equation}
\label{eq:sub-trellis}
    s\mathcal{T}  = \mathcal{T}^{\mathrm{B-}j}\odot\mathcal{T}^{\mathrm{GB-}j}\odot\mathcal{T}^{\mathrm{B-}(j+1)},\quad j\in\{1,...,2^{n-\lambda}\}\text{ is odd,}
\end{equation}
and $\lambda \geq n_0$ is the polarization depth we are currently at (and wish to enlarge by $1$). We remind that after the first $n_0$ transforms and after merging the GB paths, $\mathcal{T}^{\mathrm{B-}(j+1)}, \mathcal{T}^{\mathrm{GB-}j},$ and $\mathcal{T}^{\mathrm{B-}(j+1)}$ each consist of one single section. That is, each $s\mathcal{T}$ includes a GB section between two non-guard-band sections, see Panel~(a) in Figure~\ref{fig:with_GB_sub-trellis}. Thus, we first merge the left section, $\mathcal{T}^{\mathrm{B-}j}$, and the GB section, $\mathcal{T}^{\mathrm{GB-}j}$. This results in two-sections with no GB sections (Panel~(b) in Figure~\ref{fig:with_GB_sub-trellis}). We may now perform the `$-$' or `$+$' transform as in the ``without GB'' phase. 
See Figure~\ref{fig:trellis_transform_with_GBs} for an illustration.

In the resulting trellis, we think of the operation described above on each $s\mathcal{T}$ as producing a block trellis (even though $s\mathcal{T}$ consisted of a middle section that is the result of merging a GB trellis). Since we have not changed the even index GB trellises, we have produced a trellis with a structure as in
\eqref{eq:sub-trellis}, with $\lambda + 1$ in place of $\lambda$. Hence, we can continue the recursion.
\begin{figure}[ht]  
\centering 
        \input{tikz_schemes/tikz_blockTrellis}
\caption{Left: The first block trellis $\mathcal{T}^{\mathrm{B-1}}$. Right: The same block trellis after $n_0=2$ minus polar transforms. Red edges are labeled `$1$' and blue edges are labeled `$0$'. To improve legibility, a gray edge represents two parallel edges, red and blue (the two edges do not have the same weight necessarily). In this example $n_0=2$, thus there are four sections in each block trellis. } \label{fig:Blocktrellis} 
\end{figure}  

\begin{figure}[ht]  
\centering 
        \input{tikz_schemes/tikz_GBtrellis}
 \caption{Left: The GB trellis $\mathcal{T}^{\mathrm{GB-1}}$. Right: The same GB trellis after merging all sections. Blue edges are labeled `$0$' (in GB trellises all edges are labeled `$0$'). In this example $n=3$, $n_0=2$ and $\ell_3 = 3$, hence $\mathcal{T}^{\mathrm{GB-1}}$ is the only GB trellis in $\mathcal{T}$ and has 3 sections.} \label{fig:GBtrellis}          
\end{figure}

\begin{figure*}[ht]          
\input{tikz_schemes/scheme_trellis_T00}
 \caption{
 Panel~(a) shows the trellis $\mathcal{T}^{[00]}$. That is, $\mathcal{T}$ after $n_0=2$ minus transforms and after merging all the GB paths in the GB trellis, $\mathcal{T}^{\mathrm{GB-1}}$.
 The two block trellises $\mathcal{T}^{\mathrm{B-1}}$ and $\mathcal{T}^{\mathrm{B-2}}$ are identical since they are identical in the initial trellis $\mathcal{T}$, and we performed only minus transforms. This is not the case for other transforms (for example in the other trellises $\mathcal{T}^{[01]},\mathcal{T}^{[10]}$ and $\mathcal{T}^{[11]}$ the block trellises are not generally identical). 
 The next polar transform, performed on $\mathcal{T}^{[00]}$, will be a ``with GB'' transform. That is, first merge the paths of the left block trellis $\mathcal{T}^{\mathrm{B-1}}$ with the GB trellis $\mathcal{T}^{\mathrm{GB-1}}$. The result of this step is shown in Panel~(b). Then perform a regular polar transform (``without GB'' transform) on the result. After this step we get $\mathcal{T}^{[000]}$ which is shown in Panel~(c).  In this example $n=3$, $n_0=2$ and $\ell_3 = 3$, hence we get this result for the received vector $\bfy$.  We note that  $n_1\triangleq n-n_0=1$ in this example, thus there is only one sub-trellis $s\mathcal{T}$ in $\mathcal{T}^{[00]}$, which is  $\mathcal{T}^{[00]}$ itself. As the polarization steps progress, the trellis becomes quite busy and hard to follow. To clarify the tangled scheme, we highlighted some edges as examples. The bold red edge in Panel~(b) was created from the bold edges in Panel~(a). As shown, only one path exists between row $i=3$ and $i=8$ in the block trellis and GB trellis we merge. This edge takes the label of the block trellis edge in the path (which is labeled `$1$'). The dashed gray arrow in Panel~(b), which represents two parallel edges with different labels and likely different weights, was created from the dashed edges in Panel~(a). As shown in Panel~(a), there exists a path from row $i=4$ to $i=8$ which gives the label `$0$' (the gray arrow from $i=4$ to $i=7$ also represents an edge with label `$0$'). There also exist two paths that give the label `$1$' (one passes through $i=7$ along the same gray arrow and one goes straight from $i=4$ to $i=8$). The dotted arrow in Panel~(c) was created from the dotted paths in Panel~(b). This arrow represents two parallel edges: blue and red. The decision on $\hat{u}_1$, if it is not frozen, is according to the higher weighted edge of the two (a higher weight indicates a higher likelihood).   }\label{fig:with_GB_sub-trellis}
\end{figure*}

\subsection{Decoding complexity}\label{subsec:DeletionDecoderComplexity}%
In this section we show that the decoding complexity is $O(|\mathcal{S}|^3 \Lambda^4)$. This will follow by first recalling that in \cite[Subsection IV-D]{TPFV:22p}, the complexity of decoding a trellis with $\Lambda$ sections and no GBs is $O(|\mathcal{S}|^3 \Lambda^4)$, and then showing that the processing of the GBs does not change the order of the decoding complexity. In \cite{TPFV:22p} this complexity can be further reduced to roughly $O(|\mathcal{S}|^3\Lambda^2)$. On the upside, we do not require a successful partitioning   of $\bfy$ to the TDC outputs of each block $\bfx(1),\bfx(2),\ldots,\bfx(N_1)$, whereas in \cite{TPFV:22p} it is crucial for correct decoding. Thus, and as proven herein, we reach a better probability of correct decoding.

Regarding the details, we will first show that the complexity of collapsing each of the GB trellises into one section is $\mathcal{O}(\Lambda^3)$.  Then, we will analyze the complexity of performing the polar transforms on the trellis (both ``with GBs'' and ``without GBs'') and show that it is $\mathcal{O}(|\mathcal{S}|^3 \Lambda^4)$.

During the process of collapsing the largest GB trellis in $\mathcal{T}$, which consists of $\ell_n$ sections, we also get the results of collapsing each of the smaller GB trellises. In detail, recall that in a GB trellis with $\ell$ sections, the first $\ell -1$   sections (from the left) are identical and created according to the second row in Table~\ref{table:trellisedgeweights}. The right-most section in each GB trellis is created according to the third row in Table~\ref{table:trellisedgeweights}. Observe that in the second and third rows of Table~\ref{table:trellisedgeweights} the section index $j$ does not play a part. Hence, any two suffixes of equal length corresponding to different GB trellises are the same. When collapsing the largest GB trellis, we merge the right-most section with the one to its left, then with the next section to its left and so on, until $\ell_n$ sections were merged. Hence, we can pause the merging iterations after $\ell_{n_0+1},\ell_{n_0+2},\ldots,\ell_{n-1}$ sections were merged and extract the rightmost section as the result of collapsing GB trellises which consist of $\ell_{n_0+1},\ell_{n_0+2},\ldots,\ell_{n-1}$ sections, respectively.

We bound the number of calculations when collapsing a GB trellis with  $\ell_n$ sections by
\[\sum_{m=1}^{\ell_n-1}(\Lambda+1)\cdot2\cdot(m+1)\in\mathcal{O}(\Lambda^3)\;,\]
where $m$ is the iteration of the merge operations. In each merge we sum over all possible two-edge paths. The paths may start at $|\bfy|+1$ possible starting vertices (in each row of the trellis) which is bounded by $\Lambda+1$.   The number of outgoing edges of the starting vertex is at most $2$ (diagonal and horizontal edges). The number of outgoing edges in the middle vertex of a path is at most $m+1$ (since these are edges of a section created after performing $m-1$ merges).

We now bound the complexity of the remaining decoding steps, after the GB trellises are collapsed into single sections. Let $\lambda$ denote the polarization depth as before, i.e.\ the number of `$+$' or `$-$' transforms performed on $\mathcal{T}$ (see Figure~\ref{fig:RecursiveTrellisTransforms}).
We will now show that for a fixed regular hidden Markov input distribution with $|\mathcal{S}|$ states, the complexity of our decoder is bounded by
\begin{multline*}
	\overbrace{    \sum_{\lambda=0}^{n_0-1} \underbrace{2^{\lambda+1}}_{\text{(a)}} \cdot\underbrace{4(\Lambda+1)|\mathcal{S}|^3(2^{\lambda}+1)^2}_{\text{(b1)}}\cdot \underbrace{2^{n-\lambda-1}}_{\text{(c1)}} }^{\text{``without GB'' phase}}
 + \\
	   \overbrace{ \sum_{\lambda=n_0}^{n-1} \underbrace{2^{\lambda+1}}_{\text{(a)}} \cdot\underbrace{8(\Lambda+1)|\mathcal{S}|^3(C\cdot2^{\lambda+1}+1)^2}_{\text{(b2)}}\cdot \underbrace{2^{n-\lambda-1}}_{\text{(c2)}}}^{\text{``with GB'' phase}} \\
	    \in O(|\mathcal{S}|^3\Lambda^4)\;.
\end{multline*}
\begin{itemize}
  \item[] 
 \eqannref{a} is the number of times we perform a polar transform on trellises of depth $\lambda$, see an illustration in Figure~\ref{fig:RecursiveTrellisTransforms}. 
    \item[] \eqannref{b1} bounds the number of calculations performed on each pair of sections in a trellis of depth $\lambda<n_0$.
    \item[] \eqannref{c1} is the number of pairs of consecutive sections (first odd then even indexed) in a trellis of depth $\lambda<n_0$.
       \item[] \eqannref{b2} bounds the number of calculations on each sub-trellis of a given trellis of depth $\lambda\geq n_0$. The constant $C$ will soon be defined.
   \item[] \eqannref{c2} is the number of sub-trellises in a trellis of depth $\lambda$, see \eqref{eq:sub-trellis}. 
\end{itemize}

Before proving \eqannref{b1} and \eqannref{b2},  we remind the reader that we sum over all possible two-edge paths in pairs of consecutive sections. %
The proof will follow from bounding the number of two-edge paths in each pair of sections
in a trellis of polarization depth $\lambda$. 

We first show \eqannref{b1}. In \eqannref{b1} we have  $\lambda<n_0$, i.e. we only work on block trellis sections after $\lambda$ ``without GB'' transforms. Each section in a block trellis of depth $\lambda\leq n_0$ was created from $2^{\lambda}$ sections in the base trellis $\mathcal{T}$. Thus, in depth $\lambda$, we have $2^\lambda+1$ possible `types' of edges, where the edge type refers to how many rows the edge crosses. For example, in the initial trellis $\lambda=0$ there are two types: horizontal edges and diagonal edges that cross one row. In a block trellis after $\lambda=2$ ``without GB'' transforms we have five types: horizontal edges and diagonal edges which cross one, two, three or four rows (see Figure~\ref{fig:Blocktrellis}).

Each edge can be labeled either `$0$' or `$1$', and edges of a given type can enter at most $|\mathcal{S}|$ vertices. Thus, we get at most $2\cdot(2^{\lambda}+1)\cdot |\mathcal{S}|$  outgoing edges for each vertex. Hence, there are at most $(2\cdot(2^{\lambda}+1) |\mathcal{S}|)^2$ two-edge paths starting from each vertex.

There are $|\bfy|+1$ starting vertex rows, which is upper bounded by $\Lambda+1$, and each row is at most $|\mathcal{S}|$ layers deep.
In total, the number of calculations performed when collapsing two sections into one for $\lambda<n_0$ is bounded by \eqannref{b1}, that is 
\[4(\Lambda+1)|\mathcal{S}|^3(2^{\lambda}+1)^2\;.\]

For \eqannref{b2}, we first bound the number of edge types in each sub-trellis. As before, the number of possible edge types in a section of depth $\lambda$ is equal to the number of sections in $\mathcal{T}$ that produced it plus 1. For depth $\lambda$:
\begin{itemize}
    \item The block sections were each created from $\Lambda_{\lambda}$ sections in the base trellis $\mathcal{T}$, where $\Lambda_{\lambda}$ is the length of $g(\bfx)$ and $\bfx$ is of length $2^\lambda$. Thus, we have 
 $\Lambda_\lambda+1$ types in each block section.
    \item The middle GB section in each sub-trellis $s\mathcal{T}$ was created from $\ell_{\lambda+1}$ sections in the base trellis  (which we collapsed into a single section as described previously), see \eqref{GBrule} and \eqref{eq:sub-trellis}. Thus, we have at most 
	    $\ell_{\lambda+1}+1$ edge types in the GB section of each sub-trellis. For example, see Panel~(a) in Figure~\ref{fig:with_GB_sub-trellis} which shows a sub-trellis of depth $\lambda=2$. In this figure the middle GB trellis of $s\mathcal{T}$ was created from $\ell_3=3$ GB sections which collapsed into one, and there are four edge types (horizontal edges and diagonal edges that cross 1-3 rows).   Since $\ell_{\lambda+1}\stackrel{\eqref{ln}}{<}2^{\lambda}\leq\Lambda_\lambda$, we can bound the possible types of edges in the GB section by $\Lambda_{\lambda}+1$ (which is the bound of types in the block sections).
\end{itemize} 

  We have shown that the number of edge types in each sub-trellis of depth $\lambda$ is at most $\Lambda_\lambda+1$. By the same logic used for \eqannref{{b1}},  in each sub-trellis there are now at most \[(2\cdot(\Lambda_{\lambda}+1) |\mathcal{S}|)^2\] two-edge paths from each starting vertex.
  Thus, using the same arguments as the \eqannref{b1} case, the number of steps for merging the GB trellises into the block trellises (``step 1'' in Panel~(c), Figure~ \ref{fig:trellis_transform_with_GBs}, and  transitioning from Panel~(a) to Panel~(b) in Figure~\ref{fig:with_GB_sub-trellis}) is at most
\[4(\Lambda+1)|\mathcal{S}|^3(\Lambda_\lambda+1)^2\;.\]

We bound the number of edge types in a sub-trellis after the merge by $\Lambda_{\lambda+1}+1$, since the resulting block section was created from $\Lambda_\lambda+\ell_{\lambda+1}<\Lambda_\lambda+\ell_{\lambda+1}+\Lambda_\lambda=\Lambda_{\lambda+1}$ sections of $\mathcal{T}$. Now recall that we must also apply a `$-$' or `$+$' transform to the resulting trellis (``step 2'' in Panel~(c), Figure~ \ref{fig:trellis_transform_with_GBs}, and  transitioning from Panel~(b) to Panel~(c) in Figure~\ref{fig:with_GB_sub-trellis}). By the same logic as before, this requires at most
\[4(\Lambda+1)|\mathcal{S}|^3(\Lambda_{\lambda+1}+1)^2\]
steps. Thus, since $\Lambda_\lambda \leq \Lambda_{\lambda+1}$, we may bound the total number of steps by
\[8(\Lambda+1)|\mathcal{S}|^3(\Lambda_{\lambda+1}+1)^2 \; .\]

Lastly, we loosely bound $\Lambda_\lambda$ for all $n_0\geq1$ and $\xi>0$ by $\Lambda_\lambda\stackrel{\eqref{Glessthan1.5Na}}{<}C\cdot2^\lambda$, where $C\triangleq\frac{1}{1-2^{-\xi}}$ is a constant, which gives \eqannref{b2}.

\subsection{Simulation results}

\begin{figure*}[tb]
	\begin{center}
 \includegraphics[draft=false,scale=0.5]{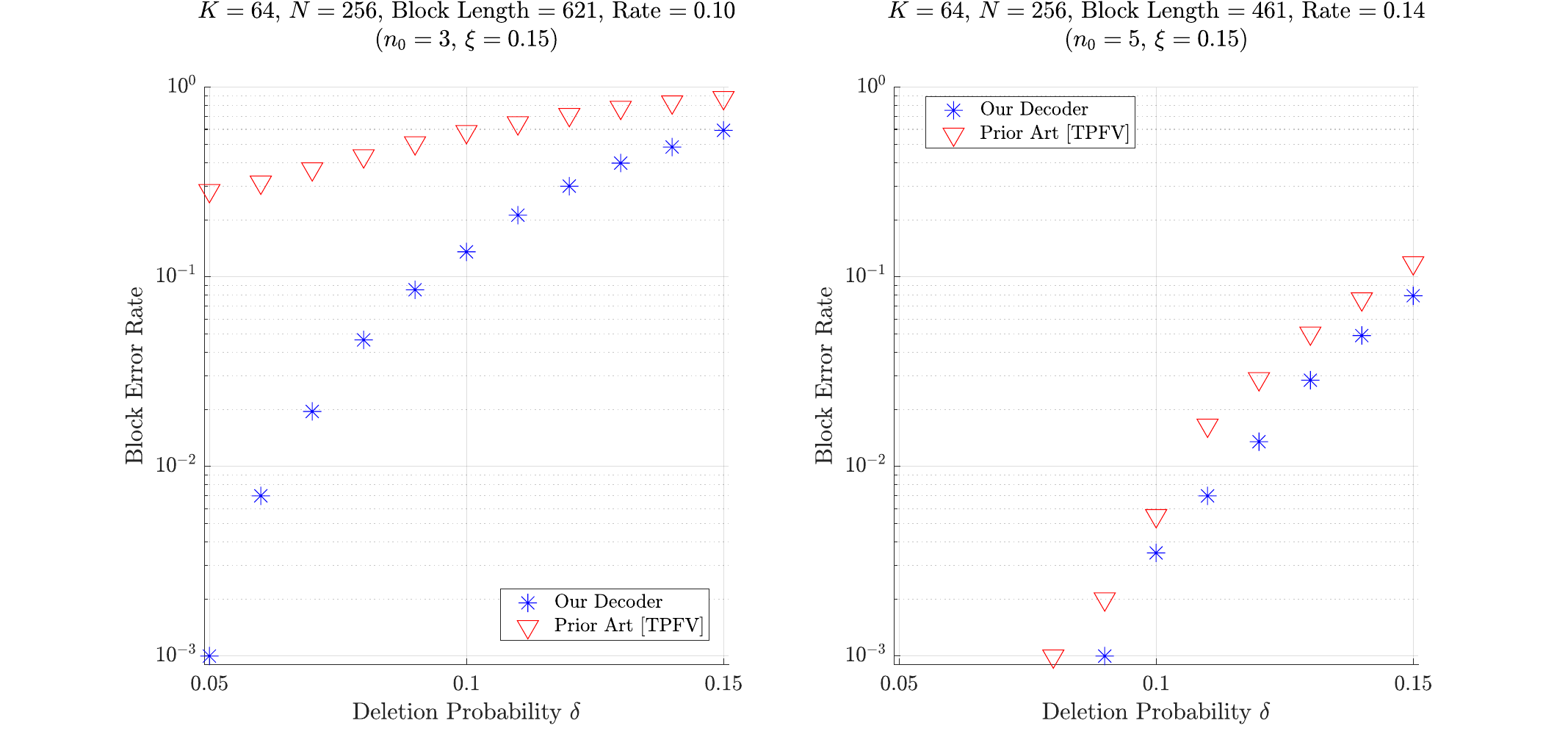}

\caption{The block error rate of the decoder in this work compared to the prior-art decoder \cite{TPFV:22p}. The input distribution was fixed to a memoryless uniform distribution, $\mathrm{Bernoulli}(\frac{1}{2})$. The block-error rate was averaged over 2000 simulations.
 In the comparison above between our decoder and the one of \cite{TPFV:22p}, the same encoder was used and the channel deletions were the same.  That is, the decoders in this figure were fed with the same channel output $\bfy$ and the only difference was the decoding algorithm applied to it. The information indices were selected as the $K=64$ indices with the lowest error probability in a genie decoder. The genie decoder error probability for each index was estimated via Monte Carlo using 2000 test blocks with deletion probability $\delta = 0.1$.}%
\label{fig:simulationresults} 
\end{center}
\end{figure*}

Figure~\ref{fig:simulationresults}  shows simulation results comparing the performance of two decoders: the one of prior art \cite{TPFV:22p} and the one described and analyzed in this paper.  Our decoder outperforms the previous one in all the scenarios tested. This is expected, since we decode based on the trellis of the entire output as opposed to processing the output into blocks, and then using trellises for each block.

It became apparent after establishing numerical results that parameters in short code lengths need to be optimized. For the asymptotic case, when $N\to\infty$, we will detail in the proof of Theorem~\ref{theoremZstrong} how to select $n_0$ and $\xi$, the parameters controlling the number of GBs and their lengths, and how to select a large enough $n$ to reach the desired code rate and block error probability. However, when working with shorter lengths, we have no simple analytical way of choosing optimal values for these parameters. %

\begin{figure*}[tb]
 \begin{center}
 \includegraphics[draft=false, scale=0.5]{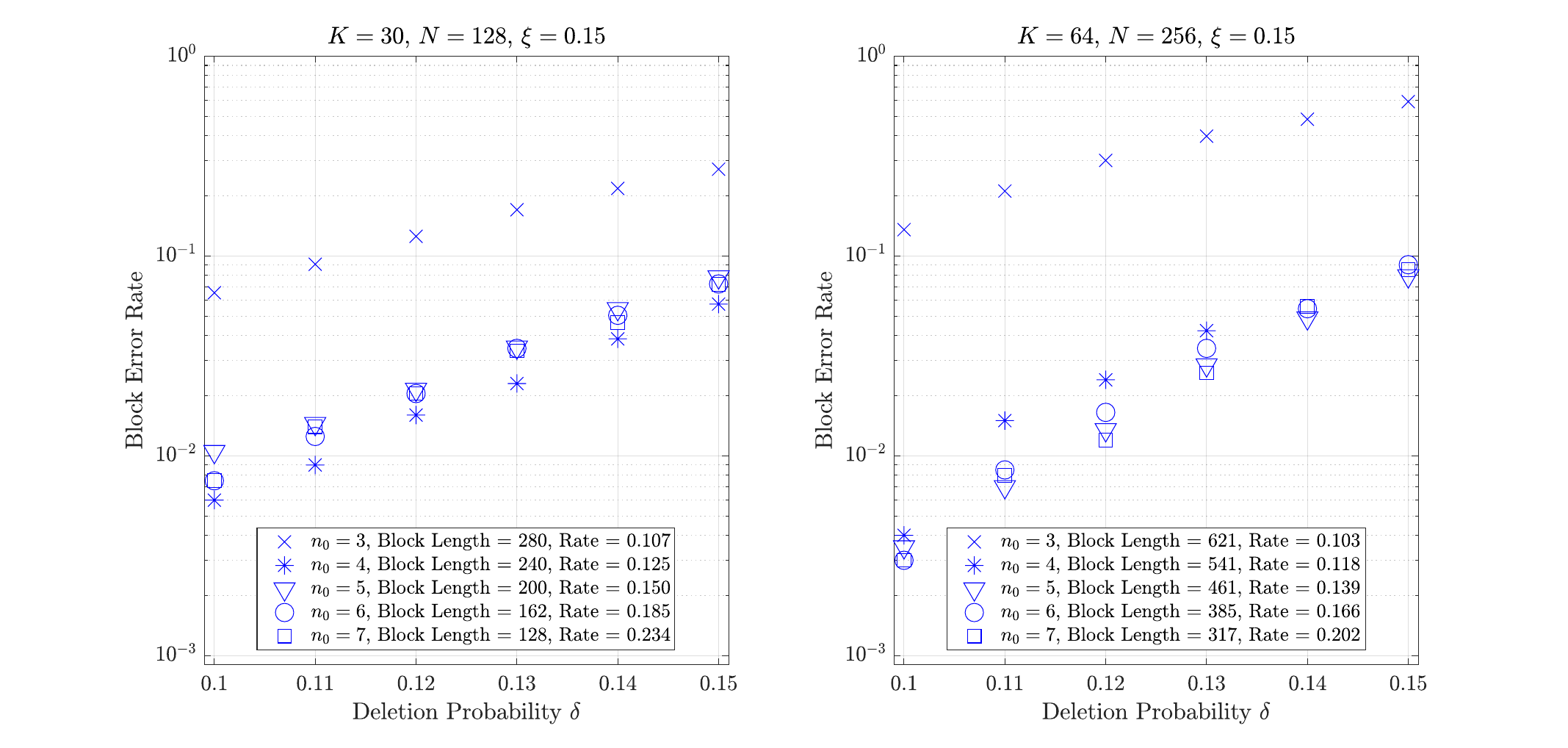} 
 \end{center}

\caption{The block error rate of the decoder in this work over various selections of $n$ and $n_0$. As in Figure~\ref{fig:simulationresults}, the input distribution was fixed to a memoryless uniform distribution, $\mathrm{Bernoulli}(\frac{1}{2})$. The block-error rate was averaged over 2000 simulations.
  The information indices were selected as the $K$ indices with the lowest error probability in a genie decoder. The genie decoder error probability for each index was estimated via Monte Carlo using 2000 test blocks with deletion probability $\delta = 0.1$.}%
\label{fig:simulationresults_n0effect} 
\end{figure*}

Consider the choice of $n_0$ for some finite $n$. This paper shows that in an asymptotic setting ($n \to \infty$), the parameter $n_0$ should be ``large enough'', for two reasons. First, if $n_0$ is small the rate is negatively affected: Lemma~\ref{lemmaRateWithGBs} promises that adding guard-bands has a negligible effect on the rate, if $n_0$ is large enough. Secondly, a small probability of the complement of the `$\GBM$' event is key to proving Lemma~\ref{lemmaStep} below, and this probability goes to $0$ as $n_0$ increases (Lemma~\ref{lemma_PrnegGBM}). On the other hand, for a fixed $n$, we cannot take $n_0$ too large, since Lemma~\ref{lemmaZproccess} below  only guarantees a strong polarization phase when $n$ is sufficiently far from $n_0$. That is, the inequality on $Z_n$ in \eqref{ZstepWithTailnew} is only guaranteed to hold for $n \geq n_0$.
The above trade-off is demonstrated in  Figure~\ref{fig:simulationresults_n0effect}. For example, in the left panel ($N=128$) transitioning from $n_0=3$ to $n_0=4$ demonstrates improvement, whereas continuing to $n_0=5$ shows a decline. 

\section{First key lemma --- the Bhattacharyya evolution for the trimmed-deletion channel}
\label{sec:firstKeyLemma}
The proof of our main theorem hinges on two key lemmas. The first lemma is specific to our setting, the trimmed-deletion channel, and is stated and proved in this section. The second lemma is more general, and will be stated and proved in Section \ref{section:strongerwithpenalty}.
These two lemmas will be used in our proof of Theorem~\ref{theoremZstrong}, in Section \ref{sec:deletion-proof}.

In the seminal paper \cite[Proposition 5]{Arikan:09p}, it was shown that in a memoryless setting, a `$+$' transform squares the Bhattacharyya parameter, while a `$-$' transform at most doubles it. This was the key property used to prove strong polarization in \cite{ArikanTelatar:09c}. Our first key lemma states a similar --- yet weaker --- claim for our setting. The second key lemma shows that such a claim is still strong enough to prove strong polarization with the same exponent as in the memoryless case, $\beta\in\left(0,\frac{1}{2}\right)$.

We first set up some additional notation. We denote the \Arikan\ transform of the vector $\bfX$ by $\bfU = \mathcal{A}(\bfX)$. Recall that the two halves of $\bfX$ are $\bfX_{\RN{1}}$ and  $\bfX_{\RN{2}}$. Their \Arikan\ transforms are denoted $\bfV\triangleq\mathcal{A}(\bfX_\RN{1})$ and $\bfV'\triangleq\mathcal{A}(\bfX_\RN{2})$, and we have 
\[
	U_{2j-1} = V_j + V'_j \; , \quad U_{2j} = V'_j \; , %
\]
where addition is modulo $2$. As in the seminal paper, the binary vector corresponding to $i-1$ is denoted $b_1,\ldots,b_{n}$. That is, for $1\leq i \leq N = 2^n$,
\begin{equation}
	\label{eq:iBinaryRepresentation}
    i  = i(b_1,\ldots,b_n) =  1+\sum_{k=1}^{n} b_k 2^{n-k} . 
\end{equation}
   The following lemma is cardinal to proving the stronger polarization stated in Theorem~\ref{theoremZstrong}. The proof will be given in the following subsections. Recall that $\delta$ is the deletion rate, and  that the guard-band length is given in \eqref{ln}, and is a function of $\xi$.
\begin{lemma}[Recursive bounds for the  Bhattacharyya parameter of the TDC]
	\label{lemmaStep} Fix a regular and non-degenerate hidden-Markov input distribution. Let $\bfX=\mathcal{A}(\bfU)$ be of length $N=2^n$, comprised of i.i.d.\ blocks of length $N_0 = 2^{n_0}$, each distributed according to the input distribution. Let $\xi \in (0,\frac{1}{6})$ be a fixed parameter and let $\bfY^* = \bfZ_\RN{1} \odot \bfZ_\Delta \odot \bfZ_\RN{2}$ be the result of transmitting $g(\bfX,n_0,\xi)$ through the TDC.
Then, there exists $\nzerothtwo(\xi,\delta)$  s.t.\ for $n_0\geq\nzerothtwo$ and all $n>n_0$ the following holds. Let $1 \leq i \leq N$ and $j = \lfloor (i+1)/2 \rfloor$. Then,
\begin{IEEEeqnarray}{l}
\IEEEyesnumber\label{ZstepBothOrig}\IEEEyessubnumber*
\label{ZstepSmallerOrigA}
 Z(U_i| U_1^{i-1}, \bfY^*)  \leq  Z(U_i|U_1^{i-1},\bfZ^*_\RN{1},\bfZ^*_\RN{2}) + 2^{-N^{\frac{2}{3}} }
    \\  \label{ZstepSmallerOrigB}
\quad \leq \begin{cases}
    \mathsmall{2\cdot Z(V_j|V_1^{j-1},\bfZ^*_\RN{1}) + 2^{-N^{\frac{2}{3}} }}
    &\mathsmall{\mbox{if }b_n=0 \; \mintrans}
    \\
	\mathsmall{ Z(V_j|V_1^{j-1},\bfZ^*_\RN{1})^2 + 2^{-N^{\frac{2}{3}} }} &\mathsmall{\mbox{if }b_n=1 \; \plustrans \; .}
\end{cases}  
\end{IEEEeqnarray}
\end{lemma}

We draw the reader's attention to several important points. First, note that in \eqref{ZstepSmallerOrigA}, there is  an additive penalty of $2^{-N^{\frac{2}{3}} }$, associated with conditioning on $\bfZ^*_\RN{1},\bfZ^*_\RN{2}$ as opposed to conditioning on $\bfY^*$. That is, there is a price to be paid for conditioning on the pair of TDC outputs corresponding to $g(\bfX_\RN{1})$ and $g(\bfX_\RN{2})$, as opposed to conditioning on the TDC output corresponding to $g(\bfX_\RN{1} \odot \bfX_\RN{2})$. Informally, this is because in the former we have been given the correct partitioning of the output into two halves (that are then further processed by the TC). The inequality \eqref{ZstepSmallerOrigB} shows us why such a penalty is worth paying: since $\bfZ^*_\RN{1}$ and $\bfZ^*_\RN{2}$ are independent and identically distributed, we may now use the standard arguments in \cite{Arikan:09p} to reach a recursive relation. This relation is indeed weaker than the one of the seminal paper for memoryless channels  \cite[
Proposition 5]{Arikan:09p} in which the additive penalty of $2^{-N^{\frac{2}{3}}}$ does not exist. 
To conclude, the lemma allows us to track the evolution of the Bhattacharyya parameter after each polarization step.

The proof of Lemma~\ref{lemmaStep} will be broken into three conceptual parts. In the first part, we define the ``Guard-Band in Middle'' event, termed $\GBM$. That is, the event that the middle symbol of the TDC output originated from the middle guard-band. In the second part, we show that under $\GBM$, $\bfY^*$ can be processed to yield $\bfZ^*_\RN{1}$ and $\bfZ^*_\RN{2}$. This essentially gives us the first term on the RHS of \eqref{ZstepSmallerOrigA}. In the third part, we show that the $\GBM$ event is very likely. That is, the additive penalty of $2^{-N^{\frac{2}{3}}}$ in \eqref{ZstepSmallerOrigA} comes from bounding the probability that $\GBM$ does not occur.%

We note that in \cite{AravaTal:23c}  we have shown a weaker bound for the Bhattacharyya parameter of this setting. A multiplicative penalty of $\frac{3N}{2}$ existed in the upper bound in \eqref{ZstepBothOrig}.  That is,
\[ Z(U_i| U_1^{i-1}, \bfY^*)  \leq  \frac{3N}{2} \cdot Z(U_i|U_1^{i-1},\bfZ^*_\RN{1},\bfZ^*_\RN{2}) + 2^{-N^{\frac{2}{3}} }.\]
We will now be able to avoid this harsh multiplicative penalty by employing the effect of channel degradation \cite[page 207]{RichardsonUrbanke:08b}, \cite[Lemma 3]{TalVardy:13p}, which will be demonstrated in the upcoming proof.

\subsection{The $\GBM$ event} \label{subsec:GBM}
In this subsection we define the ``Guard-band in Middle'' ($\GBM$) event, related notation, and consequences. Recall from Section~\ref{sec:notation} and Figure~\ref{fig:XGYZ} that $\bfY^* = \bfZ = \bfZ_\RN{1} \odot \bfZ_\Delta \odot \bfZ_\RN{2}$ is the result of passing $\bfG_\RN{1} \odot \bfG_\Delta \odot \bfG_\RN{2}$ through the TDC. The $\GBM$ event occurs if $\bfZ$ is not empty and its middle index
\begin{equation} \label{eq:midIndex}
	i_{\text{mid}} \triangleq \left\lfloor\frac{|\bfZ|+1}{2}\right\rfloor \;,
\end{equation}
falls within $\bfZ_\Delta$. That is, $\GBM$ occurs iff $Z_{i_{\text{mid}}}$ originates from the middle guard-band $\bfG_\Delta$, see Figure~\ref{fig:GBM}. The complementary event is denoted $\GBMc$, see Figure~\ref{fig:GBMc}.

We denote the left and right halves of $\bfZ = \bfY^*$ as  $\bfZL=(Z_1,\ldots,Z_{i_{\text{mid}}})$ and $\bfZR=(Z_{i_{\text{mid}}+1},\ldots,Z_{|\bfZ|})$, see Figure~\ref{fig:GBM}. 
The main utility of the $\GBM$ event is this (again, see Figure~\ref{fig:GBM}): since $\bfZ_\Delta$ contains only `$0$' symbols, under $\GBM$
\begin{IEEEeqnarray}{lCl}
\IEEEyesnumber\label{ZLRequalZ12}
(\bfZL)^*&=&(\bfZ_{\RN{1}})^* \IEEEyessubnumber \; , \\
(\bfZR)^*&=&(\bfZ_{\RN{2}})^* \IEEEyessubnumber \; .
\end{IEEEeqnarray}
That is, under $\GBM$, the simple operation of trimming the two halves of $\bfY^*$ is assured to give us $\bfZ_{\RN{1}}^* \triangleq (\bfZ_{\RN{1}})^*$ and  $\bfZ_{\RN{2}}^* \triangleq (\bfZ_{\RN{2}})^*$. This simple observation will be used in the next subsection in order to state a recursive relation.
\begin{figure} 
        \begin{center}
        \begin{tikzpicture}[xscale=1.2,yscale=1]
\node at (-3.15,1) {$\bfZ =\bfY^*$};
        \draw [very thick,draw=black] (-2.0,1.25) rectangle (-0.60,0.75) node[pos=.5] {$\bfZ_{\RN{1}}$};
        \draw [very thick,draw=black,fill=blue!70!black] (-0.60,1.25) rectangle (0.30,0.75) node[pos=.5,white] {$\bfZ_{\Delta}$};
        \draw [very thick,draw=black] (0.30,1.25) rectangle (2.0,0.75) node[pos=.5] {$\bfZ_{\RN{2}}$};

    \draw [{-latex'},draw=black] (0,1.6) --(0,1.25) node[above=8pt] {$i_\mathrm{mid}$};
    
    \draw [decorate,
        decoration = {calligraphic brace,mirror, 
            amplitude=6pt},thick] (-2,0.75)--(0,0.75)  node[pos=0.5,below  =5pt]{$\bfZL$} ;
    \draw [decorate,
        decoration = {calligraphic brace, mirror,
            amplitude=6pt},thick] (0,0.75)--(2,0.75)  node[pos=0.5,below  =5pt]{$\bfZR$} ;
    \draw[-{latex'[scale=1pt]},green!50!black, thick,opacity=1] (-1,0.1)--(-1.25,-0.75) node[left,scale=0.9pt,pos=.4] {TC};    
    \draw[-{latex'[scale=1pt]},green!50!black,opacity=1, thick] (1,0.1)--(1.25,-0.75) node[left,scale=0.9pt,pos=.4] {TC}; 

        \draw [very thick,draw=black] (0.5,-0.8) rectangle (2,-1.3) node[pos=.5] {$\bfZR^*$};
        \node at (2.4,-1.05){$=\bfZ^*_{\RN{2}}$} ;
        \draw [very thick,draw=black] (-2.0,-0.8) rectangle (-0.9,-1.3) node[pos=.5] {$\bfZL^*$};  \node at (-0.5,-1.05){$=\bfZ^*_{\RN{1}}$}   ;
    \draw[dashed] (-0.9,-0.8) -- (0,0.75); 
    \draw[dashed] (0.5,-0.8) -- (0,0.75);
        \draw[dashed] (2,-0.8) -- (2,0.75);
                \draw[dashed] (-2,-0.8) -- (-2,0.75);
        \end{tikzpicture}
        \caption{An illustration of the $\GBM$ event.} %
        \label{fig:GBM}
\end{center}
\end{figure}
\begin{figure} 
	\begin{center}
\input{tikz_schemes/GBMc-noBackground}
        \caption{An illustration of the complementary event, $\neg\GBM$.} %
        \label{fig:GBMc}
\end{center}
\end{figure}

\subsection{Bounding the Bhattacharyya parameter using $\GBM$} \label{subsec:boundingZusingGBM}
In this subsection, we derive an upper bound on the Bhattacharyya parameter $Z(U_i|U_1^{i-1},\bfY^*)$. In the previous subsection we defined $\bfZL^*$ and $\bfZR^*$ as the left and right halves of  $\bfZ\triangleq\bfY^*$ after trimming their  trailing and leading zeros, respectively.  That is, $\bfZL^*, \bfZR^*$ are the result of some processing of $\bfY^*$. This processing is illustrated in Fig \ref{fig:GBM} and Fig \ref{fig:GBMc}. We now consider a channel which is a degradation of the TDC channel. The input of the channel is the same as the case of the TDC, but the outputs of the channel are $\bfZL^*, \bfZR^*$. Clearly this is a degraded channel of the TDC, since $\bfY^*$, the output of the TDC is further processed into $\bfZL^*, \bfZR^*$.
The Bhattacharyya parameter of the degraded channel is larger or equal to the Bhattacharyya of the TDC \cite[page 207]{RichardsonUrbanke:08b}\cite[Lemma 3]{TalVardy:13p}, hence:
 \begin{IEEEeqnarray}{lCl} \label{bhattchanneldeg}
	 Z(U_i|U_1^{i-1},\bfY^*) &\leq& Z(U_i|U_1^{i-1},\bfZL^*, \bfZR^*) \; .
\end{IEEEeqnarray}

In order to save space we use the following shorthand in the upcoming probability expressions: $u_1^{i-1}$ is short for $U_1^{i-1} = u_1^{i-1}$, $\bfzL^*$ and $\bfzR^*$ are short for $\bfZL^*=\bfzL^*$ and $\bfZR^*=\bfzR^*$, respectively, and $0$ and $1$ are short for $U_i = 0$ and $U_i=1$, respectively. Furthermore, in our shorthand notation, a sum without subscripts is always over $u_1^{i-1},\bfzL^*,\bfzR^*$. To illustrate, we use both the long and short notation in the following expression for our quantity of interest. %
 \begin{IEEEeqnarray*}{lCl}
\IEEEeqnarraymulticol{3}{l}{	Z(U_i|U_1^{i-1},\bfZL^*, \bfZR^*) }\\
 &\triangleq& \sum_{u_1^{i-1},\bfzL^*,\bfzR^*}\sqrt{
\begin{matrix*}[l]\Prb(U_i=0,U_1^{i-1}=u_1^{i-1},\bfZL^*{=}\bfzL^*,\bfZR^*{=}\bfzR^*)
\\{} \times \Prb(U_i=1,U_1^{i-1}=u_1^{i-1},\bfZL^*{=}\bfzL^*,\bfZR^*{=}\bfzR^*)\end{matrix*}}
\\ &\triangleq& \sum \sqrt{\Prb(0,u_1^{i-1},\bfzL^*,\bfzR^*)\cdot\Prb(1,u_1^{i-1},\bfzL^*,\bfzR^*)}
\end{IEEEeqnarray*}

By the law of total probability over $\{\GBM,\GBMc\}$, the RHS above equals (for $n \geq n_0 + 1$, assuring a guard-band was added):
 \begin{IEEEeqnarray}{cl}
	 \label{ZgbmbothSums}
	 \IEEEnosubnumber\IEEEnonumber
     &=
     \sum\sqrt{\mathsmall{
     \begin{matrix*}[l]
	     \left(\Prb(0,u_1^{i-1},\bfzL^*,\bfzR^*,\GBM){+}\Prb(0,u_1^{i-1},\bfzL^*,\bfzR^*,\GBMc)\right) 
        \\
	{} \times \left(\Prb(1,u_1^{i-1},\bfzL^*,\bfzR^*,\GBM){+}\Prb(1,u_1^{i-1},\bfzL^*,\bfzR^*,\GBMc)\right)
     \end{matrix*}
     }}
     \\\IEEEnonumber %
     & \leq \sum \sqrt{     \mathsmall{ 
     \Prb(0,u_1^{i-1},\bfzL^*,\bfzR^*,\GBM) \cdot \Prb(1,u_1^{i-1},\bfzL^*,\bfzR^*,\GBM)
}} \quad + 
     \\\IEEEyesnumber \label{ZgbmTwoSums} 
     &  \sum \sqrt{\mathsmall{
	     \begin{matrix*}[l]
	    \mathsmall{\Prb(0,u_1^{i-1},\bfzL^*,\bfzR^*,\GBM) {\cdot} \Prb(1,u_1^{i-1},\bfzL^*,\bfzR^*,\GBMc)}
        \\
	\mathsmall{+ \Prb(0,u_1^{i-1},\bfzL^*,\bfzR^*,\GBMc) {\cdot}  \Prb(1,u_1^{i-1},\bfzL^*,\bfzR^*,\GBM)}
        \\
	\mathsmall{ + \Prb(0,u_1^{i-1},\bfzL^*,\bfzR^*,\GBMc) {\cdot} \Prb(1,u_1^{i-1},\bfzL^*,\bfzR^*,\GBMc)}
     \end{matrix*}
     }}
 \end{IEEEeqnarray}

We will now bound both sums in \eqref{ZgbmTwoSums}. For the first sum we have:
\begin{align*}
   &  \sum_{u_1^{i-1},\bfzL^*,\bfzR^*}
        \sqrt{
		\begin{matrix*}[l]
			\mathsmall{\Prb(0,u_1^{i-1},\bfZL^*{=}\bfzL^*,\bfZR^*{=}\bfzR^*,\GBM)}
            \\
	    \mathsmall{{} \times \Prb(1,u_1^{i-1},\bfZL^*{=}\bfzL^*,\bfZR^*{=}\bfzR^*,\GBM)}
        \end{matrix*}
        } 
\\
     & \stackrel{\eqref{ZLRequalZ12}}{=} \sum_{u_1^{i-1},\bfz_{\RN{1}}^*,\bfz_{\RN{2}}^*}
       \sqrt{
	       \begin{matrix*}[l]
	       \Prb(0,u_1^{i-1},\bfZ^*_{\RN{1}}{=}\bfz_{\RN{1}}^*,\bfZ^*_{\RN{2}}{=}\bfz_{\RN{2}}^*,\GBM) \\
	       {} \times \Prb(1,u_1^{i-1},\bfZ^*_{\RN{1}}{=}\bfz_{\RN{1}}^*,\bfZ^*_{\RN{2}}{=}\bfz_{\RN{2}}^*,\GBM)
        \end{matrix*}
        } 
\\
      & \leq \sum_{u_1^{i-1},\bfz_{\RN{1}}^*,\bfz_{\RN{2}}^*}
        \sqrt{
		\begin{matrix*}[l]
\Prb(0,u_1^{i-1},\bfZ^*_{\RN{1}}=\bfz_{\RN{1}}^*,\bfZ^*_{\RN{2}}=\bfz_{\RN{2}}^*) \\
{} \times \Prb(1,u_1^{i-1},\bfZ^*_{\RN{1}}=\bfz_{\RN{1}}^*,\bfZ^*_{\RN{2}}=\bfz_{\RN{2}}^*)
        \end{matrix*}
        } 
        \\
        & =  Z(U_i|U_1^{i-1},\bfZ^*_{\RN{1}},\bfZ^*_{\RN{2}})
\end{align*}

For the second sum in \eqref{ZgbmTwoSums} we have:
\begin{IEEEeqnarray*}{rCl}
	\IEEEeqnarraymulticol{3}{l}{
 \sum \sqrt{\mathsmall{
     \begin{matrix*}[l]
         \Prb(0,u_1^{i-1},\bfzL^*,\bfzR^*,\GBM) \cdot \Prb(1,u_1^{i-1},\bfzL^*,\bfzR^*,\GBMc)
        \\
	{} + \Prb(0,u_1^{i-1},\bfzL^*,\bfzR^*,\GBMc)\cdot \Prb(1,u_1^{i-1},\bfzL^*,\bfzR^*,\GBM)
        \\
	{} + \Prb(0,u_1^{i-1},\bfzL^*,\bfzR^*,\GBMc)\cdot \Prb(1,u_1^{i-1},\bfzL^*,\bfzR^*,\GBMc)
     \end{matrix*}
 }    }
}
\\
\quad & =& \sum \sqrt{\mathsmall{
     \begin{matrix*}[l]
         \Prb(0,u_1^{i-1},\bfzL^*,\bfzR^*,\GBM) \cdot \Prb(1,u_1^{i-1},\bfzL^*,\bfzR^*,\GBMc)
        \\
	{} + \Prb(0,u_1^{i-1},\bfzL^*,\bfzR^*,\GBMc)\cdot \Prb(1,u_1^{i-1},\bfzL^*,\bfzR^*)
     \end{matrix*}
     }}
\\
      & \leq & \sum \sqrt{
     \begin{matrix*}[l]
         \Prb(u_1^{i-1},\bfzL^*,\bfzR^*) \cdot \Prb(1,u_1^{i-1},\bfzL^*,\bfzR^*,\GBMc)
        \\
	{} + \Prb(0,u_1^{i-1},\bfzL^*,\bfzR^*,\GBMc) \cdot \Prb(u_1^{i-1},\bfzL^*,\bfzR^*)
     \end{matrix*}
     }
\\
      & =&  \sum \sqrt{
        \Prb(u_1^{i-1},\bfzL^*,\bfzR^*)\cdot \Prb(u_1^{i-1},\bfzL^*,\bfzR^*,\GBMc)
     }
\\
      & = &  \sum \Prb(u_1^{i-1},\bfzL^*,\bfzR^*)\sqrt{
        \Prb(\GBMc|u_1^{i-1},\bfzL^*,\bfzR^*)
     }
\\
      &\leq& \sqrt{\Prb(\GBMc)} \; ,
\end{IEEEeqnarray*}
where last inequality follows by the Jensen inequality, applied to the concave function $\sqrt{\ \cdot \ }$.

Combining the bounds for the two sums in (\ref{ZgbmbothSums}) yields
\begin{equation*}
    Z(U_i|U_1^{i-1},\bfY^*)\leq  Z(U_i|U_1^{i-1},\bfZ^*_{\RN{1}},\bfZ^*_{\RN{2}}) +\sqrt{\Prb(\GBMc)} \; .
\end{equation*}
To complete the proof of \eqref{ZstepSmallerOrigA}, it remains to show that the term  $\sqrt{\Prb(\GBMc)}$ is smaller than $2^{-N^{\frac{2}{3}}}$, for large enough $n$ and $n_0$. This will be shown in the upcoming lemma. Lastly, since $\bfZ^*_{\RN{1}}$ and $\bfZ^*_{\RN{2}}$ are i.i.d., the second inequality in our lemma, \eqref{ZstepSmallerOrigB}, is a direct consequence of \cite[Proposition 5]{Arikan:09p}.
Thus, the following lemma completes our proof.
\begin{lemma}[Upper bounding $\sqrt{\Prb(\GBMc)}$]
 \label{lemma_PrnegGBM}
Let $\bfX$ be of length $N=2^n$ and drawn as described in Lemma~\ref{lemmaStep}. Let $\bfY^*$ be the TDC output for input $g(\bfX,n_0,\xi)$. Let $\GBM$ be the event defined in Subsection~\ref{subsec:GBM}.  For a fixed deletion rate $\delta\in(0,1)$ and a guard-band length parameter $0<\xi<\frac{1}{6}$ used in \eqref{ln}, there exists an $\nzerothtwo(\xi,\delta)$, which is a function of the input distribution as well, such that:
 \begin{equation}     \label{eq:boundOnSqrtOfPGBMC}\sqrt{\Prb(\GBMc)}\leq2^{-N^{\frac{2}{3}}}
 \end{equation}
 for all  $n_0\geq\nzerothtwo(\xi,\delta)$ and $n>n_0$.
 \end{lemma}

Proving this proves Lemma~\ref{lemmaStep}: for it, we  take $\nzerothtwo$ to be the one in Lemma~\ref{lemma_PrnegGBM}. 

The proof of Lemma~\ref{lemma_PrnegGBM} follows from a strengthening of \cite[Lemma 23]{TPFV:22p}. That is, we show that there exists a threshold $\nzerothtwo$ and a constant $\theta > 0$ such that
\begin{equation}
\label{eq:strengthening}
	\Prb(\GBMc) < 2^{-\theta \cdot 2^{(1-2\xi)n}} \; ,
\end{equation}
for all $n_0 \geq \nzerothtwo$ and $n >n_0$. Thus, since we've required that the constant $\xi$ satisfy $\xi \in (0,\frac{1}{6})$, standard manipulations yield \eqref{eq:boundOnSqrtOfPGBMC}, for large enough $n$.

In \cite[Lemma 23]{TPFV:22p}, the RHS of \eqref{eq:strengthening} is weaker: $n$ is replaced by $n_0$. We give an outline of the differences between our proof of \eqref{eq:strengthening} and the proof of the weaker claim in  \cite[Lemma 23]{TPFV:22p}. For the full proof of Lemma~\ref{lemma_PrnegGBM} see Appendix~\ref{sec:highprobGBMfullproof}. The main difference lies in bounding the probability that too much of $\bfZ_{\RN{1}}$ is lost due to trimming.  The weaker result follows by showing that the probability of a certain prefix of the leftmost block being completely lost due to trimming and deletion is upper bounded by a term that decays exponentially with $N_0$, the length of the block. In our proof, we show that for \emph{any} prefix of $\bfG$, the number of block symbols is always greater than the number of guard-band symbols. Thus, the probability of such a prefix being lost due to deletion and trimming decays exponentially with its length. The stronger bound then follows by taking the prefix length to be proportional to $N$, as opposed to $N_0$.

\section{Second key lemma --- strong polarization despite a small additive penalty}\label{section:strongerwithpenalty}
In the previous section, we've stated Lemma~\ref{lemmaStep}, which gave upper bounds on the evolution of the Bhattacharyya parameter. Due to the added penalty in these bounds, we cannot use prior art in order to claim a polarization rate of roughly $2^{-\sqrt{N}}$. Indeed, in this section we state the second key lemma in the paper, Lemma~\ref{lemmaZproccess}, which implies such a rate for processes that evolve as in Lemma~\ref{lemmaStep}.
Lemma~\ref{lemmaZproccess} is stated quite generally, in the hope that it will be useful for other settings.
\begin{lemma}[Strong polarization despite an additive penalty] \label{lemmaZproccess} 
	Let $B_1,B_2,\ldots$ be i.i.d.\ uniformly distributed Bernoulli random variables.
Fix $\beta\in(0,\frac{1}{2})$, $\epsold >0$, $\symK\geq1$, and $\symg>\frac{1}{2}$. There exist a positive constant $\symstart=\symstart(\epsold,\symK)>0$ and a threshold $\nstartth=\nstartth(\epsold,\symK,\symg)$ such that the following holds. Fix $\nstartstart\geq\nstartth$ and let $Z_{\nstartstart},Z_{\nstartstart+1},Z_{\nstartstart+2},\ldots$ be a random process that satisfies:
\begin{equation}
    \label{ZstepWithTailnew}
Z_{n+1} \leq     \begin{cases}
    \symK \cdot Z_n +2^{-N^{\symg} }
    &\mbox{if }B_{n+1}=0 \;\mintrans 
    \\
    \symK \cdot Z_n^2 +2^{-N^{\symg} } &\mbox{if }B_{n+1}=1 \;\plustrans \; 
    \end{cases}
    \; , \quad n \geq \nstartstart
\end{equation}
and %
\begin{equation}
    \label{Zn0}
Z_{\nstartstart}\leq\symstart\;.
\end{equation}
There exists a threshold   $\nIIthnew = \nIIthnew(\epsold,\beta,\symK,\symg,\nstart) \geq \nstart$ starting from which strong polarization occurs with probability at least $1-\epsold$:
\begin{align}  
\Prb\left( Z_n<2^{-N^{\beta}}, \quad \forall n\geq \nIIthnew\right)\geq 1-\epsold  \; . \label{Zstrongnew}
\end{align}

\end{lemma}

 We notice that $Z_n$ provides a rather ``weak'' polarization step in \eqref{ZstepWithTailnew}. In memoryless channels,  $Z^{-}_{n+1}\leq 2\cdot Z_n$, and  $Z^{+}_{n+1}\leq Z_n^2$ \cite[z.2]{ArikanTelatar:09c}, which is a ``cleaner'' and stronger polarization step than our case. In the memoryless cases, strong polarization can be proven as in \cite{Tal:17.2p}. 
 In our case, we also have an additive term, $2^{-N^{\symg} }$. %
 Still, because the additive term decays to $0$ fast enough ($\gamma > 1/2$), strong polarization occurs, with the same exponent as the memoryless channels. 
 
We now prove Lemma \ref{lemmaZproccess}.
\begin{IEEEproof}
	Let $n_0 \geq \nstartth$ and let %
	$Z_{\nstartstart},Z_{\nstartstart+1},Z_{\nstartstart+2},\ldots$  satisfy \eqref{ZstepWithTailnew} and \eqref{Zn0}, for $\nstartth$ and $\symstart$ we have yet to commit to.
We first define an auxiliary process, starting at index $n_0$ with initial value $\symstart$:%
\begin{IEEEeqnarray}{lCl}
\IEEEyesnumber\label{Zstepnew}\IEEEyessubnumber* \label{Zstep1new}
\ZnC_{n+1}  &=& 2 \cdot\symK \cdot \begin{cases}
    \ZnC_n 
    &\mbox{if }B_{n+1}=0 \; \mintrans 
    \\
    \ZnC_n^2 &\mbox{if }B_{n+1}=1 \; \plustrans 
    \end{cases} \;\; , \quad n\geq \nstart
    \\\label{Zstep2new}
\ZnC_{\nstart}&=&\symstart
\end{IEEEeqnarray}

We now fix $\symga = (\gamma + 1/2)/2$ such that,
\begin{IEEEeqnarray}{rCl}\IEEEyesnumber\label{symga}
   \frac{1}{2}<&\symga&<\symg \; .
\end{IEEEeqnarray}
Let $\nIIthnew\geq\nstart$ be a parameter that will be committed to later on as well.
We define the following events for the processes $Z_n,\ZnC_n$:
\begin{IEEEeqnarray}{llCll}\IEEEyesnumber\IEEEyessubnumber* \label{events}
    & \eventSigmaanew &\triangleq& \{\ZnC_n\geq2^{-N^\symga}, \quad &\forall n\geq \nstart\} \label{events1} \; , \\
    & \eventSigmacnew &\triangleq& \{ \ZnC_n<2^{-N^{\beta}}, \quad &\forall n\geq \nIIthnew\} \label{events3}\; ,\\
    & \eventSigmadnew &\triangleq& \{Z_n\leq\ZnC_n, \quad &\forall n\geq \nstart\}\label{events4} \; .
\end{IEEEeqnarray}
The first two events bound the new process $\ZnC_n$, and the third event discusses a relation between  $\ZnC_n$ and the original process $Z_n$.  For the events above we list the following claims:
\begin{claim} \label{claimA}
    For all $\epsilon_a>0$ there exist $\symstart_a(\symK) >0$ and  $\nastartth(\symga,\epsilon_a,\symK,\symstart)$ s.t.\ if $0<\symstart\leq\symstart_a$ and $\nstart\geq \nastartth$ , then:
\begin{equation}\label{ZnCond3}
\Prb(\eventSigmaanew)>1-\epsilon_a \; .
\end{equation}
\end{claim}

\begin{claim} \label{claimB}
For all $\epsilon_b>0$ and $n_0 > 0$ there exist  $\symstart_b(\epsilon_b,\symK)>0$ and $\nbIIthnew(\beta,\epsilon_b,\symK,\symstart,\nstart)\geq\nstart$  s.t.\ if  $0<\symstart\leq\symstart_b$ and $ \nIIthnew\geq\nbIIthnew$ then:
\begin{equation}\label{ZnCond5}
\Prb(\eventSigmacnew)>1-\epsilon_b \; .
\end{equation}
\end{claim}
\begin{claim} \label{claimC}
There exists an $\ncstartth(\symg,\symga)$ s.t.\ if $ \nstart\geq \ncstartth$, then event $\eventSigmaanew$ implies $\eventSigmadnew$, i.e.:
\begin{equation}\label{ZnCond6}
\Prb(\eventSigmadnew|\eventSigmaanew)=1 \; .
\end{equation} \end{claim}

The proofs of Claims~\ref{claimA}--\ref{claimC} are given in the upcoming subsections.  The combination of the claims above proves our lemma. For this, we set:
\begin{equation} \label{epsilonabc} \epsilon_a=\epsilon_b=\frac{\epsilon'}{2}
\end{equation} 
and set: 
\begin{IEEEeqnarray}{lCl}\IEEEyesnumber\IEEEyessubnumber\label{symstartmin}
\symstart(\epsold,\symK) &\triangleq&\min\left\{\symstart_a,\symstart_b,\frac{1}{2}\right\} \; ,
\\\IEEEyessubnumber\label{nstartmax}
\nstartth(\epsold,\symK,\symg)&\triangleq&\max\{\nastartth,\ncstartth\} \; ,
\\\IEEEyessubnumber\label{nIImax}
\nIIthnew(\epsold,\beta,\symK,\symg,\nstart)&\triangleq&\max\{\nbIIthnew,\nstart\}\geq\nstart \; .
\end{IEEEeqnarray} 

Using the three claims above, we can bound the probabilities of events $\eventSigmacnew$ and $\eventSigmadnew$. For $\eventSigmacnew,$ we have:
\begin{equation}\label{eventCprob}
\begin{array}{ll}
    \Prb(\eventSigmacnew)&\eqann[>]{a} 1-\epsilon_b \;,
     \end{array}
\end{equation}
where in \eqannref{a} we applied \eqref{ZnCond5} from Claim \ref{claimB}, since the conditions are satisfied by our selection of $\symstart$ and $\nIIthnew$ in  \eqref{symstartmin} and \eqref{nIImax}, respectively. 
We next note that:
\begin{equation}\label{eventDprob}
\begin{array}{rcl}
    \Prb(\eventSigmadnew)&\geq &\Prb\left(\eventSigmadnew|\eventSigmaanew\right)
    \cdot \Prb\left(\eventSigmaanew\right)
    \\
    &\eqann[>]{a}&
 1\cdot(1-\epsilon_a) 
      = 1-\epsilon_a \;.
     \end{array}
\end{equation}
In \eqannref{a} we applied \eqref{ZnCond3} from Claim \ref{claimA}, and \eqref{ZnCond6} from Claim \ref{claimC}, since their conditions are satisfied by our selection of $\symstart$ and $\nstart$ in \eqref{symstartmin} and \eqref{nstartmax}, respectively.

By inspection, the intersection of $\eventSigmacnew$ and $\eventSigmadnew$ implies the event in \eqref{Zstrongnew}, by \eqref{nIImax}. Thus,
\begin{align*}
   & \Prb\left( Z_n<2^{-N^{\beta}}, \quad \forall n\geq \nIIthnew\right) \\
    & \geq \Prb\left(\left\{ \ZnC_n<2^{-N^{\beta}}, \quad \forall n\geq \nIIthnew\right\}\bigcap \left\{Z_n\leq\ZnC_n, \quad \forall n\geq \nstart\right\}\right) \\
    &= \Prb\left(\eventSigmacnew\cap\eventSigmadnew\right) \\
    &=1-\Prb\left(\neg\eventSigmacnew\cup\neg\eventSigmadnew\right)\;.
\end{align*}
 In the last equality we denoted events $\neg\eventSigmacnew,\neg\eventSigmadnew$ as the  complementary events of $\eventSigmacnew,\eventSigmadnew$ respectively. 
We now upper bound $\Prb\left(\neg\eventSigmacnew\cup\neg\eventSigmadnew\right)$:
\begin{IEEEeqnarray*}{rcl}
\Prb\left(\neg\eventSigmacnew\cup\neg\eventSigmadnew\right)&\leq&\Prb\left(\neg\eventSigmacnew\right)+\Prb\left(\neg\eventSigmadnew\right)
\\
&\stackrel{\eqref{eventCprob},\eqref{eventDprob}}{<}&
\epsilon_a+\epsilon_b
\\
&\stackrel{\eqref{epsilonabc}}{=}&\epsilon'\;.
\end{IEEEeqnarray*}
Thus:
 \[\Prb\left( Z_n<2^{-N^{\beta}}, \quad \forall n\geq \nIIthnew\right)\geq\Prb\left(\eventSigmacnew\cap\eventSigmadnew\right)\geq 1-\epsilon' \; . \] 
\end{IEEEproof}
\subsection{Proof of Claim \ref{claimA}}
\begin{IEEEproof} We first set: 
\begin{equation} \label{symstart_a}
    \symstart_a\triangleq\frac{1}{2\cdot\symK}\in\left(0,\frac{1}{2} \right] \; ,
\end{equation}
where the inclusion follows from requiring that $\symK \geq 1$ in Lemma~\ref{lemmaZproccess}.

We note that the process $\ZnC_n$ is positive for all $n\geq \nstart$, since it starts from a positive value of $\ZnC_{\nstart}=
\symstart
\in\left(0,\symstart_a\right]$ and in each step is either multiplied by $2\cdot\symK>0$ or squared and multiplied by $2\cdot\symK>0$ (both operations preserve positivity). 
By \eqref{Zstep1new} and since $\symK\geq1$:
\begin{equation} \label{ZstepLowerBound}
    \ZnC_n =
    2\cdot\symK \cdot \ZnC_{n-1}^{2^{B_{n}}}
    \geq \ZnC_{n-1}^{2^{B_{n}}}\;.
    \end{equation}
By recursively applying \eqref{ZstepLowerBound} we get:
\begin{IEEEeqnarray}{lCl}
	 \IEEEnosubnumber\IEEEnonumber
    \ZnC_n &\geq& \ZnC_{n-1}^{2^{B_{n}}}
    \\\IEEEnonumber
    &\geq&\left( \ZnC_{n-2}^{2^{B_{n-1}}}\right)^{2^{B_{n}}} 
   =\ZnC_{n-2}^{2^{B_{n-1}+{B_{n}}}}
   \\\IEEEnonumber
   &\geq &\ZnC_{n-3}^{2^{B_{n-2}+B_{n-1}+{B_{n}}}}
   \\ \IEEEyesnumber \label{recursive_lower_bound_with_Bi}
   &\geq& \ldots  \geq \ZnC_{\nstart}^{2^{\sum_{i=\nstart}^{n-1}B_{i+1}}}\;.
\end{IEEEeqnarray}

We also set $\rho\triangleq\symga-\frac{1}{2}\stackrel{\eqref{symga}}{>}0$, and define:
\begin{equation}
    \label{Delta_a}
    \Delta_a \triangleq \left\lceil\frac{1}{2\rho^2}\ln\left(\frac{1}{\epsilon_a\cdot(1-e^{-2\rho^2})}\right)\right\rceil>0,
\end{equation}
\begin{equation}
    \label{nstart_a_threshold}
    \nastartth \triangleq \left\lceil\frac{1}{\symga}\log_2(-\log_2\symstart)+\frac{1}{\symga}\Delta_a\right\rceil,
\end{equation}
which are well defined since $\symga\stackrel{\eqref{symga}}{>}\frac{1}{2}$, $\rho = \symga-\frac{1}{2}>0$, $0<\symstart\leq\symstart_a\stackrel{\eqref{symstart_a}}{\leq}\frac{1}{2}$ and    $\epsilon_a>0$.

We first bound the probability of $\ZnC_n$ not satisfying the lower bound in our claim (recall \eqref{events1}):
\begin{IEEEeqnarray}{lCl}
	 \IEEEnosubnumber\IEEEnonumber
     \Prb\left(\ZnC_n < 2^{-N^{\symga}}\right)&\stackrel{\eqref{recursive_lower_bound_with_Bi}}{\leq}&
    \Prb\left(\ZnC_{\nstart}^{2^{\sum_{i=\nstart}^{n-1}B_{i+1}}} < 2^{-N^{\symga}}\right)
    \\ \IEEEnonumber&\eqann[=]{a}  &
    \Prb\left(\sum_{i=\nstart}^{n-1}B_{i+1}  > n\cdot \symga -\log_2\left(-\log_2\symstart\right)   \right)  
     \\ \IEEEnonumber&\eqann[\leq]{b}  &
    \Prb\left(\sum_{i=\nstart}^{n-1}B_{i+1}  > (n-\nstart)\cdot \symga  \right)
    \\\IEEEnonumber & \eqann[=]{c} & 
    \Prb\left(\sum_{i=\nstart}^{n-1}B_{i+1}  > (n-\nstart)\cdot \left( \frac{1}{2}+\rho\right)   \right)
    \\\IEEEnonumber & \leq & 
    \Prb\left(\sum_{i=\nstart}^{n-1}B_{i+1}  \geq (n-\nstart)\cdot \left( \frac{1}{2}+\rho\right)   \right)
    \\ \IEEEyesnumber & \eqann[\leq]{d} & e^{-2(n-\nstart)\rho^2} 
    \label{hoeffding_bound_with_symga}
\end{IEEEeqnarray}
In \eqannref{a} we applied $\log_2(-\log_2(\cdot))$ to both sides of the inequality, while recalling that $\ZnC_{\nstart}\stackrel{\eqref{Zstep2new}}{=}\symstart\stackrel{\eqref{symstart_a}}{\in}\left(0,\frac{1}{2}\right]$. In \eqannref{b}, we enlarged the probability by decreasing the lower bound, since $\nstart\geq\nastartth\stackrel{\eqref{nstart_a_threshold}}{>}\frac{1}{\symga}\log_2(-\log_2\symstart)$. Inequality  \eqannref{c} holds by  our selection of $\rho=\symga-\frac{1}{2}$. Lastly, in \eqannref{d} we applied the  Hoeffding bound \cite[page 78]{MitzenmacherUpfal:17b}.

Next, we set:
\begin{equation}
    \label{nIIathnew}
    \naIIthnew \triangleq \nstart+  \Delta_a,
\end{equation}
  and address the probability of our lower bound holding for all $n\geq\naIIthnew$:
\begin{IEEEeqnarray}{lCl}
	 \IEEEnosubnumber\IEEEnonumber
    \Prb&(\ZnC_n\geq2^{-N^\symga}, &\quad \forall n\geq \naIIthnew)  \\\IEEEnonumber
    & \qquad =&
    1-\Prb\left(\exists n \geq \naIIthnew, \quad \ZnC_n<2^{-N^\symga} \right)
    \\\IEEEnonumber
     & \qquad  =  &   1-\Prb\left(\bigcup_{n=\naIIthnew}^{\infty} \ZnC_n<2^{-N^\symga}\right)
        \\\IEEEnonumber
     & \qquad  \eqann[\geq ]{a}  &  1-\sum_{n=\naIIthnew}^{\infty}\Prb\left( \ZnC_n<2^{-N^\symga}\right)
            \\\IEEEnonumber
    & \qquad  \stackrel{\eqref{hoeffding_bound_with_symga}}{\geq} &    1-\sum_{n=\naIIthnew}^{\infty}e^{-2(n-\nstart)\rho^2}
    \\\IEEEnonumber
    & \qquad  \eqann[=]{b}& 1 - \frac{1}{1-e^{-2\rho^2}} e^{-2(\naIIthnew-\nstart)\rho^2}
        \\\IEEEnonumber
    & \qquad  \stackrel{\eqref{nIIathnew}}{=}& 1 - \frac{1}{1-e^{-2\rho^2}} e^{-2\Delta_a\rho^2}
    \\ \IEEEyesnumber \label{probability_lower_bound_larger_than_nath}
    & \qquad \stackrel{\eqref{Delta_a}}{\geq}& 1-\epsilon_a
\end{IEEEeqnarray}
where \eqannref{a} follows from the union bound,  and for \eqannref{b} and we remind that $\rho>0$.

For the initial period of $n\in\left[\nstart,\naIIthnew\right)$, we lower bound $\ZnC_n$ by the extreme event of drawing only $B_{i+1}=1 \; \plustrans$ for $i=\nstart,...,n-1$. This event indeed minimizes $\ZnC_n$ for a low enough starting point $\symstart\leq\symstart_a$, as shown in Lemma \ref{lemmaSigma1} in  Appendix \ref{appendixSigma1}.  That is, for $n\in\left[\nstart,\naIIthnew\right)$:
\begin{IEEEeqnarray}{lCl}\IEEEnosubnumber\IEEEnonumber
\ZnC_n &\geq& \ZnC_n \text{ for }B_{i+1}=1 \;\plustrans,\; \forall i=\nstart,...,n-1
\\\IEEEnonumber
& \stackrel{\eqref{recursive_lower_bound_with_Bi}}{\geq}& \ZnC_{\nstart}^{2^{\sum_{i=\nstart}^{n-1}B_{i+1}}}\text{ for }B_{i+1}=1\; \plustrans,\; \forall i=\nstart,...,n-1
\\\IEEEnonumber
& =& \ZnC_{\nstart}^{2^{(n-\nstart)}}
\\\IEEEnonumber & \stackrel{\eqref{Zstep2new}}{=} &\symstart^{2^{(n-\nstart)}}
\\\IEEEnonumber & \eqann[>]{a} &\symstart^{2^{(\naIIthnew-\nstart)}}
\\\IEEEnonumber & \stackrel{\eqref{nIIathnew}}{=} &\symstart^{2^{\Delta_a}}
\\\IEEEnonumber & \stackrel{\eqref{nstart_a_threshold}}{\geq} &2^{-2^{\symga \nstart}}
\\ \label{lower_bound_smaller_than_nath} & \eqann[\geq]{b} &2^{-2^{\symga n}} = 2^{-N^{\symga}} \; .
\end{IEEEeqnarray}
For \eqannref{a} and \eqannref{b} we remind that we are considering the case of  $n\in\left[\nstart,\naIIthnew\right)$.

In \eqref{lower_bound_smaller_than_nath} we have shown that if $\nstart\geq\nastartth$, then $\ZnC_n\geq 2^{-N^{\symga}}$ for all $n \in\left[\nstart,\naIIthnew\right)$. In \eqref{probability_lower_bound_larger_than_nath} we have shown the same lower bound holds for all   $n\geq \naIIthnew$ with probability of at least $1-\epsilon_a$. By combining the two results we complete the proof of our claim.
\end{IEEEproof}
\subsection{Proof of Claim \ref{claimB}}
Claim \ref{claimB} is a specialization of \cite[Proposition 49]{Shuval_Universal}. Namely, \cite[Equation 166]{Shuval_Universal} as well as the constraint on $Z_0$ in \cite[Proposition 49]{Shuval_Universal} have inequalities which in \eqref{Zstepnew} are specialized to equalities. The fact that our indexing starts at $n_0$ as opposed to $0$ in\cite[Proposition 49]{Shuval_Universal} is of no consequence.

The above paragraph proves our claim, and the reader may elect to leave it at that. However, since the thresholds $\eta$ and $n_0$ are not given explicitly in \cite[Proposition 49]{Shuval_Universal}, we choose to do so here, for completeness. That is, we give explicit expressions for $\symstart_b$ (our analog of the $\eta$ in \cite{Shuval_Universal})  and $\nbIIthnew$ (our analog of the $n_0$ in \cite{Shuval_Universal}).

To set $\symstart_b$ and $\nbIIthnew$ explicitly, we carefully inspect the proof of \cite[Proposition 49]{Shuval_Universal} and the strong polarization proof in \cite{Tal:17.2p} on which it relies.
For the diligent reader wishing to correspond our steps to the mentioned sources,   we note that the process $\ZnC_n$ here is written without the ``bar'' in \cite[Proposition 49]{Shuval_Universal} and in \cite{Tal:17.2p}. In addition,  $\epsilonaSP$ and $\sigma_b$ which will be used here are denoted $\epsilon_a$ and $\epsilon_b$, respectively, in \cite[Proposition 49]{Shuval_Universal} and in \cite{Tal:17.2p}. This replacement was performed to avoid confusion with $\epsilon_a$ and $\epsilon_b$ from Claim~\ref{claimA} and Claim~\ref{claimB}, respectively. We also note that   \cite{Tal:17.2p} discusses a general process of the form: 
\begin{equation*}
Z_{n+1}\leq K \cdot Z_n^{d_{T_n}},
\end{equation*}
where $K\geq1$, $T_n \text{ is uniformly distributed over } 
\{1,2,\ldots,\ell\}$  and $d_1,d_2,\ldots,d_{\ell}\geq1$. In our case, and in \cite[Proposition 49]{Shuval_Universal} as well, $\ell=2$ and:
\begin{equation*}   
d_{T_n} = \begin{cases}
    1 \;\mbox{ if } T_n =1 \; , \\
    2 \;\mbox{ if } T_n =2 \; .
\end{cases}
\end{equation*}
This simplifies some of the expressions we cite from \cite{Tal:17.2p}. Specifically, we note that the parameter $E\triangleq\frac{1}{\ell}\sum_{t=1}^{\ell}\log_{\ell} d_t >\beta$, defined  in \cite{Tal:17.2p}, equals $\frac{1}{2}$ in our case. Lastly, in our case $K=2\cdot\symK$, not to be confused with the substitution of $K$ with $\symK$, used in \cite[Proposition 49]{Shuval_Universal}. We note that the multiplicative factor of 2 added to the process $\ZnC_n$ (see \eqref{Zstepnew}) essentially originates from ``getting rid'' of the additive penalty in the process $Z_n$ (see \eqref{ZstepWithTailnew}).

We now summarize in four steps how to set $\symstart_b$ and $\nbIIthnew$, in order to prove our claim.  By following these steps we satisfy two conditions. The first is that for all $n\geq\nbIIthnew$, the process $\ZnC_n$ is upper bounded by $2^{-2^{(\frac{1}{2}-\Delta)n}}$, with probability at least $1-\epsilon_b$. The expression for $\Delta$ is calculated in \cite[Equation 16]{Tal:17.2p}. We rewrite it here for our case, i.e. for $\ell=2$, $d_1 = 1$ and $d_2=2$:
\begin{align}
&\Delta = \frac{1}{2}\left(\log_2\left(\frac{1}{1-\thetaSP}\right)+\log_2\left(\frac{2}{2-\thetaSP}\right)\right) \label{eq:sPDelta1}\\
&+
  \sigma_b\cdot\left(-\log_2(1-\thetaSP)+\log_2(2-\thetaSP)\right)
   \label{eq:sPDelta2}\\& +
  \frac{m_a}{n}\left(\left(\frac{1}{2}+\sigma_b\right)\log_2(1-\thetaSP)+\left(\frac{1}{2}-\sigma_b\right)\log_2(2-\thetaSP)\right) \label{eq:sPDelta3}
\end{align}
where $\thetaSP\triangleq-\log_{\epsilonaSP}2\symK$ is defined by a small parameter $\sigma_a>0$ which we will select shortly. The parameters $\epsilonbSP>0$ and $m_a$ will be set soon as well.
 
 Just as in  \cite{Tal:17.2p}, the second condition we will satisfy is that each of the three terms in the expression of $\Delta$, i.e. \eqref{eq:sPDelta1}, \eqref{eq:sPDelta2} and \eqref{eq:sPDelta3}, is upper bounded by  $\frac{\frac{1}{2}-\beta}{3}$. Hence, $\frac{1}{2}-\Delta>\beta$. By combining these two conditions we get the desired upper bound on $\ZnC_n$.
 
Throughout these four steps,  all the introduced parameters ($\epsilonaSP,\epsilonbSP$ and $m_a$)  are functions of the parameters in our claim: $\symK,\epsilon_b,\eta, \nstart$ and $\beta$.
\begin{enumerate}
    \item 
We first set a small enough $\epsilonaSP>0$ such that, for $\thetaSP\triangleq-\log_{\epsilonaSP}2\symK\geq0$, we have 
$\thetaSP<1$  and  the term in \eqref{eq:sPDelta1} is $<\frac{\frac{1}{2}-\beta}{3}$.
For example, set:
\[\thetaSP\triangleq\frac{2}{3}\left(1-2^{-\frac{1-2\beta}{3}}\right)>0\]
and $\epsilonaSP=(2\symK)^{-\frac{1}{\thetaSP}}>0$.

\item We set a small enough $\symstart_b$  and a large enough $m_a$ such that if $\ZnC_{n_0}=\symstart\leq \symstart_b$  then
\begin{equation}\label{boazLemma_const}
\Prb\left(\ZnC_n<\epsilonaSP,\quad\forall n\geq m_a\right)>1-\frac{\epsilon_b}{2}    .
\end{equation}
The above, up to renaming of parameters, is \cite[Equation 171]{Shuval_Universal}.
A possible selection for these parameters, suggested at the very end of the proof of \cite[Proposition 49]{Shuval_Universal}, is:
\begin{equation*}  \symstart_b(\epsilon_b,\symK) \triangleq \frac{1}{2(2\symK)^2}\left(\frac{\epsilon_b}{4}\right) ^{1/r},
\end{equation*}
where $r=r(\symK)$ is the largest positive solution of $(2\symK)^r+(4\symK)^{-r}=2$, and
\begin{equation*} 
m_a= n_0+\left\lceil\max\left\{ 
\begin{array}{l}
\frac{4}{\ln 2} (\ln\symstart -\ln \epsilonaSP),\\\qquad\qquad\frac{2}{s^2}\ln\left(\frac{4}{\epsilon_b(1-e^{-s^2/2})}\right)
\end{array}
\right\}\right\rceil\;,
\end{equation*}
where $s=s(\symK)\triangleq\frac{\ln2}{2\ln(2(2\symK)^2)}$.

\item We set a small enough $\sigma_b>0$ such that the term in \eqref{eq:sPDelta2} is $<\frac{\frac{1}{2}-\beta}{3}$.
For example: 
\[\sigma_b=\frac{1}{2}\cdot\frac{\frac{\frac{1}{2}-\beta}{3}}{\log_2\left(\frac{2-\thetaSP}{1-\thetaSP}\right)}\;.\]
We remind that $\thetaSP$ was set in the first step and satisfies $\theta < 1$.
\item We set a large enough $\nbIIthnew$ such that for all $n\geq \nbIIthnew$ the term in \eqref{eq:sPDelta3} is $<\frac{\frac{1}{2}-\beta}{3}$
and 
\[
\Prb\left(\begin{array}{l}
     \left|\frac{|\{m_a\leq i<n:\;T_i=t\}|}{n-m_a}-\frac{1}{2}\right|\leq\sigma_b \\
     \quad \text{ for all } n\geq \nbIIthnew \text{ and }t\in\{1,2\}
\end{array}\right)>1-\frac{\epsilon_b}{2}.    
\]
That is, take $\nbIIthnew$ to be the maximum of
\[
1 +\left\lfloor	\frac{m_a\left(\left(\frac{1}{2}+\sigma_b\right)\log_2(1-\thetaSP)+\left(\frac{1}{2}-\sigma_b\right)\log_2(2-\thetaSP)\right)}{\frac{\frac{1}{2} - \beta}{3}} \right\rfloor
\]
and
\begin{equation}
	\label{eq:hoeffding}
	m_a+\left\lceil \frac{1}{2\sigma_b^2} \ln \left(\frac{4}{\epsilon_b(1-e^{-2\sigma_b^2})}\right) \right\rceil	\;,
\end{equation}
where \eqref{eq:hoeffding} follows from Hoeffding's inequality \cite[Theorem 4.12]{MitzenmacherUpfal:17b} and the union bound (see \eqref{hoeffding_bound_with_symga} and \eqref{probability_lower_bound_larger_than_nath} in which we applied similar arguments).
\end{enumerate}

\subsection{Proof of Claim \ref{claimC}}

\begin{IEEEproof}
Under event $\eventSigmaanew$ \eqref{events1}: $\ZnC_n\geq2^{-N^{\symga}}$ for all $ n\geq \nstart$. Since $\symga\stackrel{\eqref{symga}}{<}\symg$, for $n\geq\nstart\geq\ncIthnew(\symg,\symga)\triangleq\frac{1}{\symg-\symga}$ we get \[\symg n \geq \symga n +1 \;.\] 
Thus, if  $\eventSigmaanew$ occurs then $\ZnC_n$ satisfies:
\begin{equation} \label{ZnCtailrelation}
\begin{array}{ccccc}
      \symK\cdot\ZnC_n &\geq \ZnC_n &\geq&2^{-2^{\symga n} }&\geq2^{-2^{\symg n} }\;,\\
      \symK\cdot\ZnC_n^2&\geq\ZnC_n^2&\geq&2^{-2^{\symga n+1}} &\geq2^{-2^{\symg n} }\;.
\end{array}
\end{equation} 
From the above we get that under $\eventSigmaanew$,  if $\ZnC_n \geq Z_n$, then $\ZnC_{n+1} \geq Z_{n+1}$ as well (regardless of $B_{n+1}$). That is:
\begin{align*}
	\ZnC_{n+1}  &\stackrel{\eqref{Zstep1new}}{=}     \begin{cases}
   \symK\cdot \ZnC_n +\symK \cdot \ZnC_n
    &\mbox{if }B_{n+1}=0 \;\mintrans 
    \\
   \symK\cdot \ZnC_n^2 + \symK\cdot \ZnC_n^2 &\mbox{if }B_{n+1}=1 \;\plustrans 
    \end{cases}, \quad n\geq \nstart \\
    &\stackrel{\eqref{ZnCtailrelation}}{\geq}    \begin{cases}
   \symK\cdot \ZnC_n +2^{-2^{\symg n}}
    &\mbox{if }B_{n+1}=0 \;\mintrans 
    \\
   \symK\cdot \ZnC_n^2 +2^{-2^{\symg n}} &\mbox{if }B_{n+1}=1 \;\plustrans 
    \end{cases}, \quad n\geq \nstart 
    \\
        &\eqann[\geq]{a}    \begin{cases}
   \symK\cdot Z_n +2^{-2^{\symg n}}
    &\mbox{if }B_{n+1}=0 \;\mintrans 
    \\
   \symK\cdot {Z_n}^2 +2^{-2^{\symg n}} &\mbox{if }B_{n+1}=1 \;\plustrans 
    \end{cases}, \quad n\geq \nstart 
    \\
    &\stackrel{\eqref{ZstepWithTailnew}}{\geq} Z_{n+1}\;,
\end{align*}
where \eqannref{a} holds under the hypothesis that $\ZnC_n \geq Z_n$.

We remind that: $\ZnC_{\nstart}\stackrel{\eqref{Zstep2new}}{=}\symstart\stackrel{\eqref{Zn0}}{\geq}\Zstart$.
 Thus, we may show by induction that under $\eventSigmaanew$, $\ZnC_n \geq Z_n$ for all $n\geq \nstart$.
We have shown that under $\eventSigmaanew$: 
\[Z_n\leq  \ZnC_n, \quad \forall n\geq \nstart\;.\]   
\end{IEEEproof}

\section{Proof of our main theorem}\label{sec:deletion-proof} 
\begin{IEEEproof}
    The proof of Theorem~\ref{theoremZstrong} follows by combining %
Theorem~\ref{theoremWeakPolarization}, %
Lemma~\ref{lemmaRateWithGBs},  Lemma~\ref{lemmaStep},  Lemma~\ref{lemmaZproccess}, and Lemma~\ref{lemma:nthbetabeta'}. Let $\delta\in(0,1)$ be a fixed deletion probability of the deletion channel. Fix $\epsilon\in (0,1)$, $\xi \in (0,\frac{1}{6})$, and  $0 < \beta' < \beta < \frac{1}{2}$.
We take $\epsold$ in both %
Theorem~\ref{theoremWeakPolarization} and Lemma~\ref{lemmaZproccess} to be $\frac{\epsilon}{3}$, and fix $\symK$ and $\symg$ from Lemma~\ref{lemmaZproccess} to $2$ and $\frac{2}{3}$, respectively. We set $\symstart$ in Theorem~\ref{theoremWeakPolarization} to be  $\symstart(\epsilon',\symK)$ from   Lemma~\ref{lemmaZproccess}. Next, we take
 $\nzerothtwo(\epsilon,\delta,\xi)$ to be the maximum of 
 \onetwo
 {
	 \newcommand{\mytemp}{60pt}
 \begin{multline*}
	 \hspace{\mytemp} \nzerothtwo(\epsilon',\symstart,\delta)\text{  from 
Theorem~\ref{theoremWeakPolarization},}
\\
\nzerothtwo(\epsilon,\xi)\text{ from Lemma~\ref{lemmaRateWithGBs}, \hspace{\mytemp}}
\\
 \hspace{\mytemp} \nzerothtwo(\xi,\delta)\text{ from Lemma~\ref{lemmaStep},}
\\
\text{and }\nzerothtwo(\epsilon',\symK,\symg)\text{ from Lemma~\ref{lemmaZproccess}.} \hspace{\mytemp}
 \end{multline*}
 }
 {
 \begin{multline*}
     \nzerothtwo(\epsilon',\symstart,\delta)\text{  from 
Theorem~\ref{theoremWeakPolarization},}
\\
\nzerothtwo(\epsilon,\xi)\text{ from Lemma~\ref{lemmaRateWithGBs},}
\\
\nzerothtwo(\xi,\delta)\text{ from Lemma~\ref{lemmaStep},}
\\
\text{and }\nzerothtwo(\epsilon',\symK,\symg)\text{ from Lemma~\ref{lemmaZproccess}.}
 \end{multline*}
}
 By this selection we are promised for $n_0\geq\nzerothtwo$ that
\[\frac{N}{\Lambda} = \frac{|\bfX|}{|\bfG|} > 1-\epsilon\;,\] by Lemma~\ref{lemmaRateWithGBs}. 
For a fixed $n_0\geq\nzerothtwo$ we take $\nIIthnew(\epsilon,\beta,\beta',n_0)$ to be the maximum of $\nIIthnew(\epsilon',\beta,\symK,\symg,n_0)$ from Lemma~\ref{lemmaZproccess} and $\nIIthnew(\beta,\beta',\epsilon)$ from Lemma~\ref{lemma:nthbetabeta'}.  

We now prove that if we take $n\geq\nIIthnew$, then
\begin{equation}
	\label{eq:ZYstarAndKSmallEnough}
Z(U_i|U_1^{i-1},\bfY^*) \leq {2^{-N^{\beta}}}
\mbox{ and } K(U_i|U_1^{i-1}) \leq {2^{-N^{\beta}}}
\end{equation}
 for  at least $\mathcal{I}-\epsilon$ of the indices. For this, we define the processes  $Z_{\nstartstart},Z_{\nstartstart+1},Z_{\nstartstart+2}\ldots$ and
$K_{\nstartstart},K_{\nstartstart+1},K_{\nstartstart+2}\ldots$ which start from the  $\nstart\geq\nzerothtwo$ we fixed above. We couple these processes to the Bernoulli($\frac{1}{2}$) i.i.d process  $B_{1},B_{2},\ldots$ by defining:
 \[
    Z_n = Z(U_i|U_1^{i-1},\bfY^*)
   \mbox{ and }  K_n  = K(U_i|U_1^{i-1})
\]
 for $i=i(B_{1},B_{2},\ldots,B_{n})$, see \eqref{eq:iBinaryRepresentation}. That is, the drawn $B_1,\ldots,B_{n}$ are the index bits $b_1,\ldots,b_{n}$ of $i$.

Let $\eta$ and $n_0$ be as fixed above. We denote by $\mathcal{W}$ the set of ``weak polarization'' indices. That is, for $|\bfU|=|\bfX|=N_0$, $\mathcal{W}$ is the set of indices $1\leq j\leq N_0$ which satisfy:
\[Z(U_j|U_1^{j-1},\bfY^*)\leq \eta\mbox{ and }K(U_j|U_1^{j-1})\leq \eta\;.\]
In terms of our random processes we get
\begin{equation}
	\label{eq:setW}
i(B_1,\ldots,B_{\nstartstart})\in\mathcal{W} \;\Leftrightarrow\; Z_{n_0}\leq \eta\mbox{ and }K_{n_0}\leq \eta\;.
\end{equation}

We now show that for all $n\geq\nIIthnew$ our processes are upper bounded by $2^{-N^\beta}$ with probability $\geq\mathcal{I}-\epsilon$.
\begin{IEEEeqnarray*}{rCl}
\IEEEeqnarraymulticol{3}{l}{\Prb(Z_n <2^{-N^\beta}, \; K_n <2^{-N^\beta},\; \forall n\geq\nIIthnew)}
    \\
    \quad &\geq& \Prb(Z_n <2^{-N^\beta}, K_n <2^{-N^\beta}, \forall n\geq\nIIthnew,  Z_{n_0}\leq \eta, K_{n_0}\leq \eta)
    \\
	    &=& \!\!\! \sum_{j\in\mathcal{W}}\Prb(Z_n {<}2^{-N^\beta}, K_n {<}2^{-N^\beta}, \forall n{\geq}\nIIthnew, i(B_1,\ldots,B_{\nstartstart}){=}j)
    \\
	  &=& \!\!\! \sum_{j\in\mathcal{W}}
	  \Prb(Z_n {<} 2^{-N^\beta}, K_n {<}2^{-N^\beta}, \forall n{\geq}\nIIthnew| \; i(B_1,\ldots,B_{\nstartstart}){=}j)\\ 
\IEEEeqnarraymulticol{3}{r}{\times\; \Prb(i(B_1,\ldots,B_{\nstartstart})=j)}
   \\
	  &\eqann[\geq]{a}&  \sum_{j\in\mathcal{W}}
	   \Big[
	    \Prb(Z_n <2^{-N^\beta},\; \forall n\geq\nIIthnew|\;i(B_1,\ldots,B_{\nstartstart})=j)\\
	    \IEEEeqnarraymulticol{3}{r}{+\;\Prb(K_n < 2^{-N^\beta},\; \forall n\geq\nIIthnew|\;i(B_1,\ldots,B_{\nstartstart})=j)-1\Big]}
	    
    \\ 
    \IEEEeqnarraymulticol{3}{r}{\times\; \Prb(i(B_1,\ldots,B_{\nstartstart})=j)}
      \\  &\eqann[\geq]{b}& \sum_{j\in\mathcal{W}}
      \left(
    (1-\epsilon')+(1-\epsilon')-1
    \right)\times\; \Prb(i(B_1,\ldots,B_{\nstartstart})=j)
	\\&=&(1-2\epsilon')\cdot \sum_{j\in\mathcal{W}}\Prb(i(B_1,\ldots,B_{\nstartstart})=j)
   \\
	  &=& (1-2\epsilon')\cdot \frac{|\mathcal{W}|}{N_0} 
	\\&\eqann[\geq]{c}& (1-2\epsilon')\cdot (\mathcal{I}-\epsilon') 
       \\ &\geq& \mathcal{I}-3\epsilon'
   \\
	  &=&1-\epsilon 
\end{IEEEeqnarray*}
Inequality \eqannref{a} follows by noting that $P(A \cap B) \geq P(A) + P(B) - 1$, since $P(A) + P(B) - P(A \cap B) = P(A \cup B) \leq 1$.
Inequality \eqannref{c} is by Theorem~\ref{theoremWeakPolarization}. Inequality \eqannref{b} follows by applying Lemma~\ref{lemmaZproccess} to each\footnote{We note that the use of Lemma~\ref{lemmaZproccess} for upper bounding the process $K$ is a bit of an overkill. The evolution of $K$ in each polarization step is better than the evolution of the Bhattacharyya parameter $Z(U_I|U_1^{i-1},\bfY^*)$. 
 That is,  there is no small additive penalty of $2^{-N^{2/3}}$, contrast \eqref{ZstepBothOrig} with its analog for $K$ in \eqref{eq:Kstep}.  Thus, instead of using Lemma~\ref{lemmaZproccess} we could have used \cite[Proposition 49]{Shuval_Universal}.
} of the processes $Z_{\nstartstart},Z_{\nstartstart+1},Z_{\nstartstart+2}\ldots$ and $K_{\nstartstart},K_{\nstartstart+1},K_{\nstartstart+2}\ldots$.
Indeed, first note that our choices of $n_0$, $n$, and $\eta$ satisfy the conditions on them in Lemma~\ref{lemmaZproccess}. We now show that each of the two processes satisfy the two conditions, \eqref{ZstepWithTailnew} and \eqref{Zn0}, in Lemma~\ref{lemmaZproccess}. To see that each process satisfies \eqref{Zn0} recall that we are under the event $i(B_1,\ldots,B_{n_0})=j$ where $j\in\mathcal{W}$, and thus \eqref{Zn0} is satisfied by \eqref{eq:setW}. For the Bhattacharyya process $Z_{n_0},Z_{n_0+1},\ldots$ condition \eqref{ZstepWithTailnew} holds by Lemma~\ref{lemmaStep} ($\symK=2$ and $\symg=\frac{2}{3}$) and our choice of $n_0$. For the total variation process $K_{n_0},K_{n_0+1},\ldots$ condition \eqref{ZstepWithTailnew} holds by \cite[Proposition 4]{ShuvalTal:19.2p}.  That is, since the blocks of $\bfX$ are i.i.d, for $n>n_0$ we have  \cite[Proposition 4]{ShuvalTal:19.2p}
\begin{equation}\label{eq:Kstep}
K(U_i|U_1^{i-1})\leq\begin{cases}
    K(V_j|V_1^{j-1})^2 &\mbox{ if }b_n = 0 \;\mintrans \; ,
    \\
        2\cdot K(V_j|V_1^{j-1}) &\mbox{ if }b_n = 1 \;\plustrans \; ,
\end{cases}    
\end{equation}
where $\bfU,\bfV, i,j$ and $b_n$ are as in Lemma~\ref{lemmaStep}.  To summarize, we have proved that \eqref{eq:ZYstarAndKSmallEnough} holds for a fraction of at least $\mathcal{I} - \epsilon$ of the indices.

We now show that for  $n\geq\nIIthnew$, 
\begin{IEEEeqnarray*}{rCrCcCl}
	Z(U_i|U_1^{i-1},\bfY) &\eqann[\leq]{a} & Z(U_i|U_1^{i-1},\bfY^*) & \eqann[\leq]{b} & {2^{-N^{\beta}}} & \eqann[\leq]{c} & \frac{1}{2N} {2^{-\Lambda^{\beta'}}} \\
	\IEEEeqnarraymulticol{3}{r}{\mbox{and} \quad K(U_i|U_1^{i-1})} & \eqann[\leq]{b} & {2^{-N^{\beta}}} & \eqann[\leq]{c} & \frac{1}{2N} {2^{-\Lambda^{\beta'}}} \; ,
\end{IEEEeqnarray*}

 for  at least $\mathcal{I}-\epsilon$ of the indices $1 \leq i\leq N$. Inequality \eqannref{a} results from the TDC being a degradation of the deletion channel. That is, $\bfX-\bfY-\bfY^*$ form a Markov chain in that order. Inequalities \eqannref{b}, that is \eqref{eq:ZYstarAndKSmallEnough}, indeed hold for the above fraction of indices, as we have just proved. Lastly, \eqannref{c} is the result of combining Lemma~\ref{lemmaRateWithGBs} and Lemma~\ref{lemma:nthbetabeta'}, which may be applied by our selection of $n_0$ and $n$, respectively.
\end{IEEEproof}

\appendices

\section{High probability of $\GBM$} \label{sec:highprobGBMfullproof}
We now prove Lemma~\ref{lemma_PrnegGBM}. This lemma states that the $\GBM$ event, i.e.\ the event where the middle bit in the TDC output is a bit from the middle GB, occurs with high probability.

\begin{IEEEproof}
As mentioned previously in Subsection~\ref{subsec:boundingZusingGBM}, parts of the proof resemble the steps taken in the proof of \cite[Lemma 23]{TPFV:22p}.
We define the following length differences due to channel deletion (recall illustration of $\bfG_{\RN{1}},\bfY_{\RN{1}},\bfZ_{\RN{1}}$, etc. in Figure~\ref{fig:XGYZ}):
\begin{IEEEeqnarray}{ll}\label{alphabetagammaALL}
\IEEEyesnumber\IEEEyessubnumber\label{alphabetagamma}
\begin{array}{ll}
    & \alpha = |\bfG_{\RN{1}}|-|\bfY_{\RN{1}}| \\
    & \beta = |\bfG_{\Delta}|-|\bfY_{\Delta}| \\
        & \gamma = |\bfG_{\RN{2}}|-|\bfY_{\RN{2}}| \;,
        \end{array}
   \end{IEEEeqnarray}     
and the following length differences due to trimming:
\begin{IEEEeqnarray}{ll}
\IEEEyessubnumber\label{alpha'beta'gamma'}
\begin{array}{ll}
        & \alpha ' = |\bfY_{\RN{1}}|-|\bfZ_{\RN{1}}| \\
            & \beta ' = |\bfY_{\Delta}|-|\bfZ_{\Delta}| \\
        & \gamma ' = |\bfY_{\RN{2}}|-|\bfZ_{\RN{2}}| \;.
                \end{array}
\end{IEEEeqnarray}
We will be interested in the event: $ A\cap A' \cap B \cap B' \cap C\cap C'$,
where $A,B,C$ will be events constraining the number of deletions in $\bfG_{\RN{1}},\bfG_{\Delta},\bfG_{\RN{2}}$, respectively, and $A',B',C'$ are events constraining the number of bits trimmed in $\bfY_{\RN{1}},\bfY_{\Delta},\bfY_{\RN{2}}$, respectively. These events will be defined explicitly in a moment, but first, we note that the main property of these events is:
\[A\cap A' \cap B' \cap B \cap C\cap C' \Rightarrow \GBM\;.\]
That is, under all of the events $A,A',B,B',C,C'$ we are under the $\GBM$ event. The justification of this property will soon be given in \eqref{GBMabca'b'c'}. 

We define:
\begin{IEEEeqnarray}{lCl}
\IEEEyesnumber\label{eventsABCDA'B'C'}\IEEEyessubnumber*
A &=& \{\delta|\bfG_{\RN{1}}|-\hat{\ell}< \alpha<\delta|\bfG_{\RN{1}}|+\hat{\ell}\} \label{eventAdef}
    \\
A' &=& \{0\leq\alpha'<\hat{\ell} \}\label{eventA'def}
       \\
B &=& \{\beta < \delta|\bfG_{\Delta}|+\hat{\ell}\}\label{eventBdef}
       \\
   B' &=& \{\beta ' = 0\}\label{eventB'def}
    \\
    C &=& \{\delta|\bfG_{\RN{2}}|-\hat{\ell}<\gamma<\delta|\bfG_{\RN{2}}|+\hat{\ell}\}\label{eventCdef}
    \\
    C' &=& \{0\leq\gamma'<\hat{\ell}\}\label{eventC'def} \; .
   \end{IEEEeqnarray}
Intuitively, a small $\hat{\ell}$ implies a tight tolerance on the events defined in \eqref{eventsABCDA'B'C'} --- and a small enough tolerance will imply GBM. On the other hand, we must not take $\hat{\ell}$ too small, since we want each of the probabilities of these events to be close to $1$. As we will see, both aims are achieved by taking 
\begin{equation} \label{l_based_on_ln}
    \hat{\ell} = \frac{1-\delta}{4}\ell_n\;.
\end{equation}
For this selection, and since \[\ell_n\stackrel{\eqref{ln}}{<}2^{(n-1)}=|\bfX_\RN{1}|\leq|\bfG_\RN{1}|\;,\]  we notice that under the event $A$, the number of bits deleted from $\bfG_{\RN{1}}$ is less than $\frac{1+\delta}{2}|\bfG_\RN{1}|$.
Also, under  $A'$   we will trim less than $ \frac{\strut 1-\delta}{4}|\bfG_\RN{1}|$ bits from $\bfY_{\RN{1}}$. Thus, in total, under $A\cap A'$, the total number of bits deleted or trimmed from $\bfG_\RN{1}$ is less than 
\[
	\left( \left( \frac{1+\delta}{2} \right) + \left( \frac{1-\delta}{4} \right) \right)|\bfG_\RN{1}| =  \frac{3+\delta}{4}|\bfG_\RN{1}| \leq |\bfG_\RN{1}| \; .
\]
That is, under $A\cap A'$, the trimming of $\bfY_{\RN{1}}$ has stopped prior to the received GB bits in $\bfY_\Delta$. By symmetry, under $A\cap A' \cap C \cap C'$, we do not trim $\bfY_\Delta$ from either side. Hence,
\begin{equation} \label{eq:eventsAA'CC'giveB'}
	A\cap A' \cap  C\cap C' \Rightarrow B' \; .    
\end{equation}
We use this now to show the following:
\begin{IEEEeqnarray*}{rCl} 
\IEEEeqnarraymulticol{3}{l}{\{A  \cap A' \cap B  \cap C   \cap C'\}}
\\&\stackrel{\eqref{eq:eventsAA'CC'giveB'}}{\Rightarrow}&
\{A  \cap A' \cap B \cap B' \cap C   \cap C'\}
\\
     &\eqann[\Rightarrow]{a} &    
     \left\{\begin{array}{l}
\alpha +\alpha' <\gamma+\gamma'+\ell_n -\beta - \beta'
\\        \quad  \text{and } \gamma+\gamma' <\alpha +\alpha'+\ell_n -\beta - \beta'
     \end{array}\right\}
    \\
    &\stackrel{\eqref{alphabetagammaALL}}{\Rightarrow} &
    \{
 |\bfZ_{\RN{2}}|<|\bfZ_\Delta|+|\bfZ_{\RN{1}}|   \quad \text{and} \quad |\bfZ_{\RN{1}}|<|\bfZ_\Delta|+|\bfZ_{\RN{2}}|\}
    \\
    &\eqann[\Rightarrow]{b} &\GBM    \;.\IEEEyesnumber\label{GBMabca'b'c'}  
\end{IEEEeqnarray*}
\eqannref{a} holds since under the event $A  \cap A' \cap B \cap B' \cap C   \cap C'$:
\begin{IEEEeqnarray*}{lCl}
	\gamma+\gamma'+\ell_n -\beta - \beta' & \stackrel{\eqref{eventB'def}}{=} & \gamma+\gamma'+\ell_n -\beta \\
&\stackrel{\eqref{eventBdef},\eqref{eventCdef},\eqref{eventC'def}}{>}& \delta|\bfG_\RN{2}|-\hat{\ell}+0+\ell_n-\delta |\bfG_\Delta| -\hat{\ell} \\
			     &\stackrel{\eqref{GBrule}}{=}  & \delta|\bfG_\RN{2}|-\hat{\ell}+\ell_n-\delta \ell_n -\hat{\ell} \\
&\stackrel{\eqref{l_based_on_ln}}{=}&\delta|\bfG_\RN{2}|-\hat{\ell}+4\hat{\ell}-\hat{\ell}
\\
&=&\delta|\bfG_\RN{2}|+2\hat{\ell}
\\
& \stackrel{\eqref{eventAdef},\eqref{eventA'def}}{>} &\alpha+\alpha'\;,
\end{IEEEeqnarray*}
and $\gamma+\gamma' <\alpha +\alpha'+\ell_n -\beta - \beta' $ by the same steps. For
\eqannref{b}, the middle index $i_{\mathrm{mid}}$ does not fall in $\bfZ_{\RN{1}}$, by
\begin{IEEEeqnarray*}{rCl}
    i_{\mathrm{mid}} &\stackrel{\eqref{eq:midIndex}}{=} &\left\lfloor\frac{|\bfZ|+1}{2}\right\rfloor 
    \\
    & = &\left\lfloor\frac{|\bfZ_{\RN{1}}|+|\bfZ_{\Delta}|+|\bfZ_{\RN{2}}|+1}{2}\right\rfloor
    \\
    &\geq &\left\lfloor\frac{|\bfZ_{\RN{1}}|+|\bfZ_{\RN{1}}|+1+1}{2}\right\rfloor
    \\
&=&|\bfZ_{\RN{1}}|+1 \;,
\end{IEEEeqnarray*}
and the middle index $i_{\mathrm{mid}}$ does not fall in $\bfZ_{\RN{2}}$, by
\begin{IEEEeqnarray*}{rCl}
    i_{\mathrm{mid}} &\stackrel{\eqref{eq:midIndex}}{=} &\left\lfloor\frac{|\bfZ|+1}{2}\right\rfloor 
    \\
    & = &\left\lfloor\frac{|\bfZ_{\RN{1}}|+|\bfZ_{\Delta}|+|\bfZ_{\RN{2}}|+1}{2}\right\rfloor
    \\
    &\leq &\frac{|\bfZ_{\RN{1}}|+|\bfZ_{\Delta}|+|\bfZ_{\RN{2}}|+1}{2}
    \\
&< &\frac{|\bfZ_{\RN{1}}|+|\bfZ_{\Delta}|+|\bfZ_{\RN{1}}|+|\bfZ_{\Delta}|+1}{2}
\\
&<&|\bfZ_{\RN{1}}|+|\bfZ_{\Delta}| +1 \;.
\end{IEEEeqnarray*}
Thus, $i_{\mathrm{mid}}$ must fall within the middle guard-band $\bfZ_{\Delta}$. We note that  the outputs middle guard-band $\bfZ_{\Delta}$ is not an empty string under the events above. If it was an empty string, i.e. $|\bfZ_{\Delta}|=0$, we get from \eqref{GBMabca'b'c'}: $|\bfZ_{\RN{1}}|<|\bfZ_{\RN{2}}|$
 and $|\bfZ_{\RN{1}}|>|\bfZ_{\RN{2}}|$ which is not possible.
 
From \eqref{GBMabca'b'c'} we get: $\Prb(\GBM)\geq \Prb(A  \cap A' \cap B \cap C  \cap C')$. We are interested in the complementary event, which satisfies: 
\begin{align*}
    \Prb(\GBMc)&\leq \Prb(\neg\{A  \cap A' \cap B \cap C  \cap C'\})
    \\&\eqann[\leq]{a} \Prb(\Ac)+\Prb(\Ac')+\Prb(\Bc)+\Prb(\Cc)+\Prb(\Cc')
    \\ & \eqann[=]{b} 2\cdot\Prb(\Ac)+2\cdot\Prb(\Ac')+\Prb(\Bc)
\end{align*}
\eqannref{a} is by the union bound and \eqannref{b} results from the symmetry between events $A,A'$ and $C,C'$ respectively, by \eqref{eventsABCDA'B'C'}.

We now set:
\begin{equation}
    \label{eq:n0th4notGBMeventBound}
  \nzerothtwo(\xi,\delta) \triangleq \max{\left\{\begin{matrix*}[l]\frac{1}{\xi}\log_{2}\left(1+\frac{1}{1-2^{-\xi}}\right),\;\frac{1}{1-\xi}\log_2 \left(\frac{64\tau}{1-\delta}\right)+1,
  \\\frac{1}{\xi}\log_2{\left( \frac{(1-\delta)^2}{128\cdot D} \right)},  \frac{1}{1-2\xi-\frac{2}{3}}\log_2{\left(\frac{128\cdot2}{(1-\delta)^2}\right)}
    \\ \frac{1}{1-2\xi}\log_2{\left(\frac{128\log_2(7)}{(1-\delta)^2(\log_2(e)-1)}\right)}
\end{matrix*}\right\}}\;,\end{equation} 
and proceed to upper bound the probabilities of each of the events $\Ac$,$\Ac'$, and $\Bc$, for $n>n_0\geq\nzerothtwo$. We note that $\tau>0$ is a constant dependent on the input distribution  that will be defined later on. We will also set the constant $D>0$ later on, which is dependent on the input distribution and on the deletion rate $\delta$. In Lemma \ref{lemmaUpperBoundEventA'} we  will bound  $\Prb(\Ac')$.
$\Prb(\Ac)$ and $\Prb(\Bc)$ may be bounded using Hoeffding \cite[Theorem 4.12]{MitzenmacherUpfal:05b}, as in  \cite[equations (89),(90)]{TPFV:22p}. That is,
\begin{IEEEeqnarray}{rCl}\IEEEnonumber
    \Prb(\Ac)&\leq &2\cdot e^{-\frac{2{\hat{\ell}}^2}{|\bfG_{\RN{1}}|}}
     \\  &\eqann[\leq]{a}& 2\cdot e^{-\frac{2{\hat{\ell}}^2}{2^n}} \IEEEnonumber
    \\  &\stackrel{\eqref{l_based_on_ln}}{\leq}& 2\cdot e^{-\frac{2{((1-\delta)\ell_n/4)}^2}{2^n}} \IEEEnonumber
     \\  &=& 2\cdot e^{-\frac{(1-\delta)^2}{8}\ell_n^2 2^{-n}} \IEEEnonumber
        \\  &\stackrel{\eqref{ln}}{<}& 2\cdot e^{-\frac{(1-\delta)^2}{8}2^{2((1-\xi)(n-1)-1)}2^{-n}} \IEEEnonumber
                \\  &=& 2\cdot e^{-\frac{(1-\delta)^2}{32}2^{2(1-\xi)(n-1)-n}}\IEEEnonumber
                \\  &\leq& 2\cdot e^{-\frac{(1-\delta)^2}{128}2^{(1-2\xi)n}}\;, \IEEEyesnumber \label{eq:boundAc}
\end{IEEEeqnarray}
where in \eqannref{a} we bounded $|\bfG_{\RN{1}}|\leq2\cdot2^{n-1}$, which results from  \eqref{Glessthan1.5Na} and $n_0\geq\nzerothtwo\stackrel{\eqref{eq:n0th4notGBMeventBound}}{\geq}\frac{1}{\xi}\log_2\left(1+\frac{1}{1-2^{-\xi}}\right)>\frac{1}{\xi}\log_2\left(\frac{1}{1-2^{-\xi}}\right)$. Similarly for the  event $\Bc$ we have (using the penultimate displayed equation in the proof of \cite[Theorem 4.12]{MitzenmacherUpfal:05b})
\begin{IEEEeqnarray}{rCl}\IEEEnonumber
    \Prb(\Bc)&\leq &e^{-\frac{2{\hat{\ell}}^2}{|\bfG_{\Delta}|}}
    \\  &\stackrel{\eqref{l_based_on_ln}}{=}& e^{-\frac{2{((1-\delta)\ell_n/4)}^2}{\ell_n}} \IEEEnonumber
        \\  &=& e^{-2\frac{(1-\delta)^2}{16}\ell_n} \IEEEnonumber
        \\  &\stackrel{\eqref{ln}}{=}& e^{-2\frac{(1-\delta)^2}{16}\left\lfloor2^{(1-\xi)(n-1)}\right\rfloor}\IEEEnonumber
         \\  &\eqann[\leq]{a}& e^{-2\frac{(1-\delta)^2}{16}2^{(1-\xi)(n-1)-1}}\IEEEnonumber
                \\  &=& e^{-\frac{(1-\delta)^2}{16}2^{(1-\xi)(n-1)}} \IEEEnonumber
                \\  &\leq& e^{-\frac{(1-\delta)^2}{32}2^{(1-\xi)n}}\;, \IEEEyesnumber \label{eq:boundBc}
\end{IEEEeqnarray}
where in  \eqannref{a} we used $\left\lfloor2^x\right\rfloor> 2^{x-1}$ which holds for any $x\geq0$ (for $x\in [0,1)$ we have $1> 2^{x-1}$ and for $x\geq1$ we have $\left\lfloor2^x\right\rfloor> 2^x-1\geq 2^{x-1}$).

In total, we reach the following upper bound for $\Prb(\GBMc)$:
\begin{IEEEeqnarray*}{rCl}
    \Prb(\GBMc)&\leq& 2\cdot \Prb(\Ac)+2\cdot \Prb(\Ac')+\Prb(\Bc)\\
    & \stackrel{\eqref{eq:boundAc},\eqref{eq:boundBc},\text{ Lemma~\ref{lemmaUpperBoundEventA'}}}{\leq}& 2\cdot2\cdot e^{-\frac{(1-\delta)^2}{128}2^{(1-2\xi)n} } \\
    && \qquad \;
    {}+2\cdot e^{-D \cdot2^{(1-\xi)n}} \\
    &&
    \qquad\qquad\quad {}+ e^{-\frac{(1-\delta)^2}{32}2^{(1-\xi)n}} \;,
\end{IEEEeqnarray*}
where $D>0$ is the same constant from \eqref{eq:n0th4notGBMeventBound}. The value of $D$ will be given explicitly in the proof of Lemma \ref{lemmaUpperBoundEventA'}. We note that when bounding $\Prb(A')$ we will use the fact that $n>n_0\geq\nzerothtwo\stackrel{\eqref{eq:n0th4notGBMeventBound}}{\geq}\max\left\{\frac{1}{\xi}\log_{2}\left(1+\frac{1}{1-2^{-\xi}}\right),\;\frac{1}{1-\xi}\log_2 \left(\frac{64\tau}{1-\delta}\right)+1\right\}$, and the qualities of the input distribution we fixed.  The definition of $\tau$ will also be given in the proof of Lemma \ref{lemmaUpperBoundEventA'}.

Finally, for $0<\xi<\frac{1}{6}$ and  $n>n_0\geq\nzerothtwo$ we have
\begin{align*}
    \Prb(\GBMc)\eqann[\leq]{a} 7e^{-\frac{(1-\delta)^2}{128}2^{(1-2\xi)n} }\eqann[\leq]{b} 2^{-\frac{(1-\delta)^2}{128}2^{(1-2\xi)n} } \eqann[\leq]{c}  
    2^{-2\cdot N^{\frac{2}{3}}}\;.
\end{align*}
Specifically, %
inequality \eqannref{a} holds for
$\nzerothtwo\stackrel{\eqref{eq:n0th4notGBMeventBound}}{\geq}
\frac{1}{\xi}\log_2{\left( \frac{(1-\delta)^2}{128\cdot D} \right)}$,
 \eqannref{b} holds for 
$\nzerothtwo\stackrel{\eqref{eq:n0th4notGBMeventBound}}{\geq} \frac{1}{1-2\xi}\log_2\left(\frac{128}{(1-\delta)^2}\cdot\frac{\log_2(7)}{\log_2(e)-1}\right)$,
 and \eqannref{c} holds for
$\nzerothtwo\stackrel{\eqref{eq:n0th4notGBMeventBound}}{\geq} \frac{1}{1-2\xi-\frac{2}{3}}\log_2{\left(\frac{128\cdot2}{(1-\delta)^2}\right)}$.
\end{IEEEproof}

\subsection{Bounding the probability of event  $\Ac'$}
The following lemma was used in the proof of Lemma \ref{lemma_PrnegGBM}. In this lemma we develop a bound on $\Prb(\Ac')$, the probability that ``too many'' bits were trimmed in $\bfY_{\RN{1}}$.  The bound we reach  decays with $n$ (in contrast to the weaker bound in \cite[equation (94)]{TPFV:22p} which decays with $n_0$). 
\begin{lemma}[Upper bounding $\Prb(\Ac')$] \label{lemmaUpperBoundEventA'}
Let $A'$ be as in \eqref{eventA'def}. There exists  $\nzerothtwo(\xi)$, which is also dependent on the input distribution, such that 
for $n>n_0\geq\nzerothtwo(\xi)$
\[\Prb(\Ac')\leq e^{-D \cdot 2^{(1-\xi)n}}\;,\]
where $D>0$ is a constant dependent on the input distribution and on the deletion rate $\delta$.
\end{lemma} 
\begin{IEEEproof}
We consider the event $A''$, defined as follows.
Under the event $A''$, some index $j< \strut\hat{\ell}$ in $\bfG_\RN{1}$ is a `$1$' bit and was not deleted by the channel (where $\hat{\ell}$ was set in \eqref{l_based_on_ln}). Clearly
$A''\Rightarrow{}A'$, hence,
\[\Prb(\Ac')\leq \Prb(\Ac'')\;.\]
$\Ac''$ is the complementary event where no index $j<\hat{\ell}$ in $\bfG_\RN{1}$ is a `$1$' bit that was not deleted.

We denote by $\#_{\bfX}^{j}$ the number of bits to the left of index $j$ in $g(\bfX,n_0,\xi)$ that originate from $\bfX$, and denote by $\#_{\mathrm{GB}}^{j}$ the number of GB bits to the left of index $j$. For $n_0\geq\strut\nzerothtwo\geq \frac{1}{\xi}\log_2\left(1+\frac{1}{1-2^{-\xi}}\right)$
\begin{equation}\label{numXgeqnumGB}
    \#_{\bfX}^{j} \geq \#_{\mathrm{GB}}^{j},\quad \forall j\in\{1,2,\ldots ,\Lambda\}\;.
\end{equation}
That is, there are more bits from $\bfX$ than GB bits, for any prefix of  $\bfG$. The proof of \eqref{numXgeqnumGB} is given in Lemma \ref{Ldata>Lgb}. %

By \eqref{numXgeqnumGB}, there are at least $\frac{j}{2}$ bits from $\bfX$ prior to index $j$ in $\bfG$, i.e.
\begin{equation} \label{num_bitx_from_X_geq_half}
    \#_{\bfX}^{j} \geq \frac{j}{2}\;.
\end{equation}

For the case of $\bfX$ distributed according to a regular Markov input distribution with states $\mathcal{S}$, which we assumed is not degenerate, there exists an integer $\tau>0$ and a probability $0<p_0<1$ s.t.\ for any state $s\in\mathcal{S}$
\begin{equation} \label{notDegenerate}
    \Prb\left((X_1,X_2,\ldots,X_\tau)=(0,0,\ldots,0)|S_0=s\right)<p_0\;.
\end{equation} 
That is, the probability of a `$1$' bit in a series of $\tau$ bits in $\bfX$ is greater than $1-p_0$. For each $\tau$ bits in $\bfX$, the probability of at least one of them being a `$1$' bit that was not deleted in the channel is greater than
\[(1-p_0)(1-\delta)\;.\] 

In our setting, $\bfX$ consists of blocks of length $N_0$, each drawn independently from the regular hidden Markov input distribution. We consider two cases:  $\#_{\bfX}^{\lfloor \hat{\ell} \rfloor}\leq N_0$  and $\#_{\bfX}^{\lfloor \hat{\ell} \rfloor}> N_0$. For the first case, all the bits up to index $\lfloor \hat{\ell} \rfloor$  that originate from $\bfX$  were drawn according to the underlying regular Markov input distribution (that is, originate from the same block in $\bfX$). 
There are $\left\lfloor{\#_{\bfX}^{\lfloor \hat{\ell} \rfloor}/\tau}\right\rfloor$ segments of $\bfX$ bits of length $\tau$ up to index $\lfloor \hat{\ell} \rfloor$. 

For the case where $\#_{\bfX}^{\lfloor \hat{\ell} \rfloor}> N_0$, we count the number of segments of $\tau$ bits up to index $\lfloor \hat{\ell} \rfloor$ in $\bfG$ that originate from $\bfX$ and require that each segment was taken from the same block in $\bfX$. There are at least \[\left\lfloor{\#_{\bfX}^{\lfloor \hat{\ell} \rfloor}/N_0}\right\rfloor\cdot\left\lfloor{N_0/\tau}\right\rfloor\] such segments of length $\tau$.  We lower bound by 
\[\left\lfloor{\#_{\bfX}^{\lfloor \hat{\ell} \rfloor}/N_0}\right\rfloor\cdot\left\lfloor{N_0/\tau}\right\rfloor\eqann[\geq]{a} 
\frac{\#_{\bfX}^{\lfloor \hat{\ell} \rfloor}}{2N_0}\cdot\frac{N_0}{2\tau}=\frac{\#_{\bfX}^{\lfloor \hat{\ell} \rfloor}}{4\tau}\geq\left\lfloor\frac{\#_{\bfX}^{\lfloor \hat{\ell} \rfloor}}{4\tau}\right\rfloor\;.\]
In \eqannref{a} we applied $\lfloor x\rfloor\geq\frac{x}{2}$ which holds for $x\geq1$, which is satisfied for $\#_{\bfX}^{\lfloor \hat{\ell} \rfloor}>N_0$ and $n_0\geq\nzerothtwo\geq\log_2\tau$. 

For both cases we get that there are at least $\left\lfloor{\frac{\#_{\bfX}^{\lfloor \hat{\ell} \rfloor}}{4\tau}}\right\rfloor$ segments of $\bfX$ bits of length $\tau$ up to index $\lfloor \hat{\ell} \rfloor$. Thus, by the Markov property:
\begin{IEEEeqnarray}{lcl}\IEEEnonumber*
\Prb(\Ac^{''})&\leq &\left(1-(1-p_0)(1-\delta)\right)^{
\Bigl\lfloor
\frac{\#_{\bfX}^{\lfloor \hat{\ell} \rfloor}}{4\tau}
\Bigr\rfloor
}
\\
&\stackrel{\eqref{num_bitx_from_X_geq_half}}{\leq} &(p_0(1-\delta)+\delta)^{
\Bigl
\lfloor
\frac{\lfloor \hat{\ell} \rfloor}{8\tau}
\Bigr\rfloor
} \\ 
&\eqann{a} &(p_0(1-\delta)+\delta)^{\left\lfloor{\frac{\hat{\ell} }{8\tau}}\right\rfloor} \;,
\end{IEEEeqnarray}
where \eqannref{a} follows from \cite[equation (3.11)]{GKP:94b}.
We continue to upper bound the RHS from above:
\begin{IEEEeqnarray}{lcl}\IEEEnonumber*
    &\stackrel{\eqref{l_based_on_ln}}{=}&(p_0(1-\delta)+\delta)^{\left\lfloor{\frac{(1-\delta) \ell_n}{32\tau}}\right\rfloor}
\\
    &\stackrel{\eqref{ln} }{=}&
(p_0(1-\delta)+\delta)^{\left\lfloor{\frac{1-\delta }{32\tau}\left\lfloor2^{(1-\xi)(n-1)}\right\rfloor}\right\rfloor}
    \\
    &
\eqann[<]{a}& 
(p_0(1-\delta)+\delta)^{\left\lfloor{\frac{1-\delta}{64\tau} 2^{(1-\xi)(n-1)}}\right\rfloor}
        \\
    &
\eqann[\leq]{b}&(p_0(1-\delta)+\delta)^{{\frac{1-\delta}{128\tau} 2^{(1-\xi)(n-1)}}}
     \\
    &=&(p_0(1-\delta)+\delta)^{\frac{1-\delta }{128\tau\cdot2^{(1-\xi)}}2^{(1-\xi)n}}
         \\
    &\leq&(p_0(1-\delta)+\delta)^{\frac{1-\delta }{128\cdot 2\tau}2^{(1-\xi)n}}
    \\
    &=&e^{-\ln\left(\frac{1}{p_0(1-\delta)+\delta}\right)\frac{1-\delta}{256\tau} 2^{(1-\xi)n}}        
\end{IEEEeqnarray}
where \eqannref{a} is by $\left\lfloor2^x\right\rfloor> 2^{x-1}$ which holds for any $x\geq0$. For \eqannref{b} we apply  $\lfloor x\rfloor\geq\frac{x}{2}$ which holds for $x\geq1$, thus inequality \eqannref{b} holds for $n>n_0\geq \nzerothtwo\geq\frac{1}{1-\xi}\log_2 \left(\frac{64\tau}{1-\delta}\right)+1$.

We denote $D\triangleq
\frac{1-\delta}{256\tau}\ln\left(\frac{1}{p_0(1-\delta)+\delta}\right) $, 
where $\tau$ and $p_0$ satisfy \eqref{notDegenerate}. We note that $D>0$, since $0<p_0(1-\delta)+\delta<1$. 
Finally,
\[\Prb(\Ac')\leq\Prb(\Ac'')\leq e^{-D\cdot 2^{(1-\xi)n}} \;  \; ,\]
for 
\onetwo{
	\[
n>n_0\geq \nzerothtwo \triangleq\max\left\{\frac{1}{\xi}\log_2\left(1+\frac{1}{1-2^{-\xi}}\right),\frac{1}{1-\xi}\log_2 \left(\frac{64\tau}{1-\delta}\right)+1\right\}\;.    
\]
}{
\begin{multline*}
n>n_0\geq \nzerothtwo \\\triangleq\max\left\{\frac{1}{\xi}\log_2\left(1+\frac{1}{1-2^{-\xi}}\right),\frac{1}{1-\xi}\log_2 \left(\frac{64\tau}{1-\delta}\right)+1\right\}\;.    
\end{multline*}
}
\end{IEEEproof}

\subsection{Guard-band presence in $g(\bfX)$}
To show \eqref{numXgeqnumGB}, we state and prove the following lemma.
\begin{lemma}[Majority of $\bfX$ bits in any prefix of $\bfG$]  \label{Ldata>Lgb}
If $n_0\geq\frac{1}{\xi}\log_2\left(1+\frac{1}{1-2^{-\xi}}\right)$, then for $n> n_0$ and for any given index $j$ in $g(\bfX,n_0,\xi)$,
\[\#_{\bfX}^{j} \geq \#_{\mathrm{GB}}^{j} \;,\]
 where $\#_{\bfX}^{j}$ is the number of $\bfX$ bits in the prefix up to $j$ in $g(\bfX,n_0,\xi)$, and $\#_{\mathrm{GB}}^{j}$ is the number of GB bits up to $j$. That is, in any prefix of $\bfG$ the number of bits from $\bfX$ is greater or equal to the number of GB bits.

\end{lemma}

\begin{IEEEproof}
We divide our proof to three claims.

\begin{claim} \label{claim:GB1} Assume to the contrary that there exists an index $j_0$ for which our lemma does not hold, i.e.\ $\#_{\bfX}^{j_0} < \#_{\mathrm{GB}}^{j_0}$. Then, there must exist some index $j_1$ which is located at the \textbf{right edge of some guard-band} that also does not satisfy the lemma, i.e.\ $\#_{\bfX}^{j_1} < \#_{\mathrm{GB}}^{j_1}$.
\end{claim}
\begin{IEEEproof}
If $j_0$ is an index of a GB bit, we may continue to the right edge of the GB containing $j_0$, making the rightmost index of this GB the desired $j_1$.  This $j_1$ satisfies:
\[\#_{\bfX}^{j_1}=\#_{\bfX}^{j_0} < \#_{\mathrm{GB}}^{j_0}\leq \#_{\mathrm{GB}}^{j_1}\;.\]
If $j_0$ is an index of an $\bfX$ bit, we may continue to the left edge of the block of $\bfX$ containing $j_0$, making the rightmost index of the GB to the left of this block the desired $j_1$. This $j_1$ satisfies:
\[\#_{\bfX}^{j_1}< \#_{\bfX}^{j_0} < \#_{\mathrm{GB}}^{j_0}=\#_{\mathrm{GB}}^{j_1}\;.\]
\end{IEEEproof}

\begin{claim}\label{claim:GB2} We define index $j_{\mathrm{mid}}$ as the rightmost index of the \textbf{middle} GB of $g(\bfX)$. We remind that $g(\bfX)$ is created from $N_1=2^{n_1}$ blocks of $\bfX$, each block of length $N_0=2^{n_0}$. 
If $\#_{\bfX}^{j_{\mathrm{mid}}} \geq \#_{\mathrm{GB}}^{j_{\mathrm{mid}}} $ for all $n_1$,
then, 
 \[\#_{\bfX}^{j} \geq \#_{\mathrm{GB}}^{j} \;\]
for any index $j$ in $g(\bfX)$.
\end{claim}
\begin{IEEEproof} Recall that $g(\bfX)$ was generated according to a given $n_0$ and $n_1$. We denote  by
\[(n_1,n_0)_{\mathrm{series}}\]
the prefix of $g(\bfX)$ of length $j_{\mathrm{mid}}$. 
For a general $n_1$, the full  $g(\bfX)$ will be the concatenation:
\[
 \scalebox{0.8}{$(n_1,n_0)_{\mathrm{series}}\odot(n_1-1,n_0)_{\mathrm{series}}\odot\ldots\odot(2,n_0)_{\mathrm{series}}\odot(1,n_0)_{\ \mathrm{series}}\odot\bfX(N_1)$}\;.\]
 See Figure~\ref{fig:jmid}, for \scalebox{0.9}{$g(\bfX)=(2,n_0)_{\mathrm{series}}\odot(1,n_0)_{\mathrm{series}}\odot\bfX(4)$}. 
\begin{figure}[H]
    \centering
\input{./tikz_schemes/scheme_GB_presence_small}
\caption[The $(n_1,n_0)_{\mathrm{series}}$ portions of $g(\bfX)$]{The $(n_1,n_0)_{\mathrm{series}}$ portions of $g(\bfX)$. In this example $n=n_0+2$ and the vector $g(\bfX)$ is a concatenation of $(2,n_0)_{\mathrm{series}}$, $(1,n_0)_{\mathrm{series}}$, and $\bfX(4)$.} \label{fig:jmid}
\end{figure}

We notice the following property:
for each index $j_1$ located at the right edge of some guard-band in $g(\bfX)$, the series of bits to the left of $j_1$ are concatenations of the building-blocks
\[\{(i,n_0)_{\mathrm{series}}\}_{i\in\mathcal{J}} \; , \]
where $\mathcal{J}$ is some subset of $\{1,2,\ldots,n_1\}$. 
Therefore, if $\#_{\bfX}^{j_{\mathrm{mid}}} \geq \#_{\mathrm{GB}}^{j_{\mathrm{mid}}}$
is satisfied for any $n_1$, then each building-block $(n_1,n_0)_{\mathrm{series}}$ contains at least as many bits from $\bfX$ as guard-band bits. Thus,
\[\#_{\bfX}^{j_1} \geq \#_{\mathrm{GB}}^{j_1}\;,\]
for any rightmost index $j_1$ of a GB in $g(\bfX)$. By Claim \ref{claim:GB1}, this leads to:
$\#_{\bfX}^{j} \geq \#_{\mathrm{GB}}^{j} \quad  \forall j\in\{1,\ldots,\Lambda\}$.
\end{IEEEproof}
\begin{claim}\label{claim:GB3}
For any $n_1$,
$\#_{\bfX}^{j_{\mathrm{mid}}} \geq \#_{\mathrm{GB}}^{j_{\mathrm{mid}}}$.
\end{claim}
\begin{IEEEproof} 
In the series of bits up to $j_{\mathrm{mid}}$, there are half of the bits of $\bfX$. That is,
\begin{equation}\label{numDATAjmid}
\#_{\bfX}^{j_{\mathrm{mid}}} = \frac{1}{2}|\bfX|=2^{n-1}\;.
\end{equation}
Also,  up to $j_{\mathrm{mid}}$, there are half of the GB bits of $g(\bfX)$, plus the additional bits from the middle GB,
\begin{equation}\label{numGBjmid}
  \#_{\mathrm{GB}}^{j_{\mathrm{mid}}} = \frac{1}{2}(\Lambda-N)+\frac{1}{2}\ell_n \;.
\end{equation}
The total number of GB bits satisfies
\begin{align} \label{numGBbits}
\begin{array}{rcl}
    |g(\bfX)|-|\bfX|&=&\Lambda-N
    \\
    & \stackrel{\eqref{Glessthan1.5Na}}{\leq}&  2^{n}\cdot \left(\frac{2^{-\xi n_0}}{1-2^{-\xi}}\right) \; .
\end{array}
\end{align}
The length of the middle GB satisfies: 
\begin{equation} \label{midGBlength}
\begin{array}{rl}
\ell_n\stackrel{\eqref{ln}}{ 
\leq} 2^{(1-\xi)(n-1)}
\leq  2^{n-1}\cdot2^{-\xi n_0} \;,
\end{array}
\end{equation}
where the last inequality is by $n\geq n_0+1$. Thus,
\begin{align}
\begin{array}{lcl}
\#_{\bfX}^{j_{\mathrm{mid}}}&\stackrel{\eqref{numDATAjmid}}{=}&2^{n-1} \\
 &\eqann[\geq]{a} &\frac{1}{2}\cdot2^{n}\cdot \left(\frac{2^{-\xi n_0}}{1-2^{-\xi}}\right)+2^{n-1}\cdot 2^{-\xi n_0}
 \\
&\stackrel{\eqref{numGBbits},\eqref{midGBlength}}{\geq}& \frac{1}{2}(\Lambda-N)+\ell_n
\\&\stackrel{\eqref{numGBjmid}}{\geq}& \#_{\mathrm{GB}}^{j_{\mathrm{mid}}}   \;,
\end{array}
\end{align}
where \eqannref{a} is satisfied for $n_0\geq \frac{1}{\xi}\log_{2}\left(1+\frac{1}{1-2^{-\xi}}\right)$ and any $\xi>0$.
\end{IEEEproof}

By combining the results from Claims \ref{claim:GB2} and \ref{claim:GB3}, we have proven  Lemma.~\ref{Ldata>Lgb}.
\end{IEEEproof}

\section{Plus transforms decrease the process $\ZnC_n$}\label{appendixSigma1}
Recall that Claim~\ref{claimA} was stated and proved as part of the proof of Lemma~\ref{lemmaZproccess}. That proof relied on Lemma~\ref{lemmaSigma1}, which we prove here.
We remind the reader that $B_1,B_2,\ldots$ is a random process with i.i.d.\  entries, each with distribution Bernoulli$(\frac{1}{2})$. %
That is,
\[ B_n = 
\begin{cases}
  0 \mbox{ w.p. }\frac{1}{2} \; ,\\
  1 \mbox{ w.p. }\frac{1}{2} \; .
\end{cases}
\]
 We also remind that the process $B_n$ is coupled with the process $\ZnC_n$  in \eqref{Zstepnew}. In  Claim~\ref{claimA}  we lower bound $\ZnC_n$. A pivotal part in lower bounding $\ZnC_n$ is showing that the event of drawing only $1$'s in the process $B_n$ minimizes the process $\ZnC_n$. This will be proven herein.

For this, we denote by $\Sigma_1(n)$ the event of drawing only $1$'s up to $n$, starting from $n_0$. The parameter $n_0 $ is the starting index of the process $\ZnC_n$ \eqref{Zstepnew}. That is,
\begin{IEEEeqnarray}{rCl}
\IEEEyesnumber\IEEEyessubnumber*  \label{Sigma1}
    \Sigma_1(n) &=&\left\{B_{i+1}=1 , \quad \forall \nstart\leq i \leq n-1\right\}\;. 
\end{IEEEeqnarray}
The following lemma shows that under this event, $\ZnC_n$ is minimized.
\begin{lemma}\label{lemmaSigma1}
	Let $\ZnC_{n_0},\ZnC_{n_0+1},\ldots$ be the process defined in \eqref{Zstepnew}.  We denote by $\ZnC_{n}^{\Sigma_1}$  the value of $\ZnC_{n}$ under event $\Sigma_1(n)$ in \eqref{Sigma1}. Note that $\ZnC_{n}^{\Sigma_1}$ is a constant. Let  $\symstart_a=\frac{1}{2\cdot\symK}$ as in \eqref{symstart_a}. For  $\symstart\leq\symstart_a$ and for any $ n\geq \nstart$:
\begin{equation*}
	\ZnC_{n}\geq \ZnC_{n}^{\Sigma_1} \quad \mbox{with probability $1$}\;. 
\end{equation*}
\end{lemma}

\begin{IEEEproof}
We denote the value of $\ZnC_{n+1}$ when $B_{n+1}=1 \;\plustrans \; $ as $\ZnC_{n+1}^{+}$, and the value of $\ZnC_{n+1}$ when $B_{n+1}=0 \;\mintrans \; $ as $\ZnC_{n+1}^{-}$.
\begin{subclaim} \label{subclaim0inSigma1Lemma}
\begin{align} \label{plus_remains_low}
\text{For }\symstart\leq \symstart_a :  \quad  \ZnC_n\leq\symstart\Rightarrow \ZnC_{n+1}^{+} \leq \symstart\;. 
\end{align}
\end{subclaim}
Proof:
 \begin{IEEEeqnarray*}{lCl}
     \ZnC_{n+1}^{+} &\stackrel{\eqref{Zstep1new}}{=}& 2\cdot \symK \cdot \ZnC_n^2 \\
     &\leq &2\cdot \symK \cdot \symstart^2  \\
      &\leq &2\cdot \symK \cdot \symstart_a\cdot\symstart \\
     &\stackrel{\eqref{symstart_a}}{=}&
      \symstart\;.
 \end{IEEEeqnarray*}

\begin{subclaim} \label{subclaim1inSigma1Lemma} For  $\symstart\leq \symstart_a $:
\[\ZnC_n^{\Sigma_1}\leq\symstart, \quad \forall n\geq \nstart\;. \]
\end{subclaim}
Sub-claim \ref{subclaim1inSigma1Lemma} may be easily proven by induction, where the induction basis is given by our starting point at $\ZnC_{\nstart}\stackrel{\eqref{Zstep2new}}{=}\symstart$. The induction step is given by \eqref{plus_remains_low}. 
\begin{subclaim}\label{subclaim1_5inSigma1Lemma}
Under $\Sigma_1(n)$, for $\symstart\leq \symstart_a $:
\begin{equation}\label{plus_decreases}
    \ZnC_{n+1}^{+}\leq\ZnC_{n}<\ZnC_{n+1}^{-}\;. 
\end{equation}
\end{subclaim}
The `$-$' case in Sub-claim \ref{subclaim1_5inSigma1Lemma} is true since $\symK\geq1$ and since the process $\ZnC_n$ is positive for all $n$:
\[ \ZnC_n< 2\cdot \symK\cdot  \ZnC_n \stackrel{\eqref{Zstep1new}}{=}   \ZnC_{n+1}^{-}\;. \]
For the  `$+$' case, we must show that for $\symstart\leq \symstart_a $ and under $\Sigma_1(n)$,
\[\ZnC_{n+1}^{+}\stackrel{\eqref{Zstep1new}}{=} 2\cdot \symK \ZnC_n^2 \leq \ZnC_n\;. \]
Hence, it suffices to prove that under $\Sigma_1(n)$:
\[2\cdot \symK \cdot \ZnC_n\leq 1\;.\]
We use Sub-claim \ref{subclaim1inSigma1Lemma} to recall that $\ZnC_n^{\Sigma_1}\leq\symstart$. 
Recalling again that $\symstart\stackrel{ \eqref{symstart_a}}{\leq}\frac{1}{2\cdot\symK}$, we deduce the above.

Now to the proof of our lemma. We will prove by contradiction. Assume that $\ZnC_n$ is minimized for some series of draws $\mathcal{B}=\{B_{i+1}\}_{i=\nstart}^{n-1}$ which is not all $1$'s. For this series, for some $i$: $B_{i+1}=0\;\mintrans$. 
We denote the first index in $\mathcal{B}$ where $0$ was drawn as $i_0$, i.e,
\[B_{i_0+1}=0\;\mintrans \;  \text{ and } B_{i+1}=1\;\plustrans \;  \ \forall \nstart\leq i<{i_0}\;.\]
Notice that up to $i_0$ we have performed only `$+$' operations. Thus, by Sub-claim \ref{subclaim1inSigma1Lemma}, $\ZnC_{i_0}\leq\symstart$. We now replace $B_{i_0+1}=0$  with $B_{i_0+1}=1$. 
By \eqref{plus_decreases}, $\ZnC_{i_0+1}$ is now smaller. We perform all other transforms dictated from $\mathcal{B}$ without change.
The key thing to note is that both operations in  \eqref{Zstep1new} are monotonically increasing in $\ZnC_n$. Thus, by a simple induction argument, $\ZnC_n$ will also be decreased by our change at step $i_0$. We arrive at a contradiction to $\ZnC_n$ being minimized by the original series $\mathcal{B}$. Hence, the series minimizing $\ZnC_n$ is $B_{i+1}=1$ for all $\nstart\leq i\leq n-1$ (the draw of event $\Sigma_1(n)$).

\end{IEEEproof}
\section{Auxiliary threshold calculation for Theorem~\ref{theoremZstrong} }
The following lemma, combined with Lemma~\ref{lemmaRateWithGBs}, proves  the rightmost inequalities in  \eqref{BhatCond} and \eqref{TVCond} for a large enough $n$. This lemma is used in Section~\ref{sec:deletion-proof}, i.e. in the proof of our main theorem, Theorem~\ref{theoremZstrong}. 

\begin{lemma}\label{lemma:nthbetabeta'}
Fix $\beta>\beta'>0$ and $\epsilon\in(0,1)$. Let $N\triangleq2^n$. There exists a threshold $\nIIthnew(\beta,\beta',\epsilon)$ such that if $n\geq\nIIthnew$ then
\[2^{-N^\beta}<\frac{1}{2N}\cdot2^{-\left(\frac{N}{1-\epsilon}\right)^{\beta'}}\;.\]
\end{lemma}
\begin{IEEEproof}
    We set
    \onetwo{
    \begin{equation}\label{eq:nthbeta}
\nIIthnew(\beta,\beta',\epsilon) \triangleq  \max\left\{\frac{1}{(\ln(2)\cdot\beta')^2} \; , \; 1 + \frac{1}{\beta-\beta'}\log_2\left(1+\frac{1}{(1-\epsilon)^{\beta'}}\right)\right\}\;.  
    \end{equation}

    }{
    \begin{multline}\label{eq:nthbeta}
\nIIthnew(\beta,\beta',\epsilon) \triangleq\\  \max\left\{\frac{1}{(\ln(2)\cdot\beta')^2} \; , \; 1 + \frac{1}{\beta-\beta'}\log_2\left(1+\frac{1}{(1-\epsilon)^{\beta'}}\right)\right\}\;.  
    \end{multline}
}
    For $n\geq\nIIthnew$:
    \begin{IEEEeqnarray*}{CCCl}
       & 2^{-N^\beta}&<&\frac{1}{2N}\cdot2^{-\left(\frac{N}{1-\epsilon}\right)^{\beta'}} \\
        \stackrel{-\log_2(\cdot)}{\Leftrightarrow}& 2^{\beta n} &> &\frac{1}{(1-\epsilon)^{\beta'}}\cdot 2^{\beta'n}+1+n
        \\
        \eqann[\Leftarrow]{a} & 2^{\beta n} &>& \frac{1}{(1-\epsilon)^{\beta'}}\cdot 2^{\beta'n}+1+(\ln(2)\beta'n)^2
        \\
        \eqann[\Leftarrow]{b}&  2^{\beta n} &>& \frac{1}{(1-\epsilon)^{\beta'}}\cdot 2^{\beta'n}+e^{\ln(2)\beta'n}
         \\
       \Leftrightarrow &2^{\beta n}& > & 2^{\beta'n}\cdot\left(1+ \frac{1}{(1-\epsilon)^{\beta'}}\right)
        \\
        \Leftrightarrow& 2^{(\beta -\beta') n} &> &\left(1+ \frac{1}{(1-\epsilon)^{\beta'}}\right)
        \\
        \Leftarrow& n &\geq &\nIIthnew\stackrel{\eqref{eq:nthbeta}}{>} \frac{1}{\beta-\beta'}\log_2\left(1+\frac{1}{(1-\epsilon)^{\beta'}}\right)
        \;.
    \end{IEEEeqnarray*}
Transition \eqannref{a} is satisfied for $n\geq \nIIthnew \stackrel{\eqref{eq:nthbeta}}{\geq} \frac{1}{(\ln(2)\cdot\beta')^2}$. For transition \eqannref{b}, recall that by Taylor's theorem applied to $f(x)=e^x$, for $x \geq 0$ we have:
\begin{equation}\label{taylor_f(x)}
    e^x =1+x+\frac{x^2}{2}+E_2(x),
\end{equation}
where $E_2(x)\triangleq\frac{f^{(3)}(c)}{3!}x^{3}=\frac{e^c}{3!}x^{3}$ for $c\in[0,x]$\cite[p. 283]{Apostol:67b}. 
Since all the terms on the RHS of \eqref{taylor_f(x)} are nonnegative, we have:
\begin{equation}
    \label{exp_geq_xplus1}
        e^x\geq 1+x^2, \quad \forall x\geq0\;.
\end{equation}
That is,    for $x=\ln(2)\beta'n\geq0$  inequality \eqref{exp_geq_xplus1} yields transition \eqannref{b}.

\end{IEEEproof}

\bibliographystyle{IEEEtran}
\bibliography{main}

\end{document}

%% file: tikz_schemes/scheme_TDC.tex
\centerline{
\begin{tikzpicture}[every node/.style={scale=1},
	start chain=going right,
	box/.style={
		on chain,join,draw,
		minimum height=1cm,
		text centered,
		minimum width=1.5cm,
	}
	, every join/.style=->,
	node distance=4mm
]

\node [on chain] {$\bfG$};%
\node [box,xshift=3mm, align=center,red!70!black](dc){DC \\
\textit{\small deletion}
};

\node [on chain,join] (y){$\bfY$};

\node [box, align=center,green!50!black](tc){TC \\
\textit{\small trimming}
};
\node [on chain,join,xshift=3mm]{$\bfY^*$};
\node [rectangle,below=4.5mm of y] (midBottom){};

\node [
	rectangle,draw,dashed,
	above=0mm of midBottom,
	minimum width=5.25cm,
	minimum height=1.5cm, label= TDC
] (tdc) {};
\end{tikzpicture}
}

%% file: tikz_schemes/scheme_gX_small.tex
 \centerline{
\hspace*{3em}\begin{tikzpicture}
[scale=0.85, every node/.style={scale=0.85}]
\draw[white] (-1,-1)--(11,-1)--(11,1)--(-1,1)--(-1,-1);

\node[rectangle,draw, minimum width = 1.5cm, 
    minimum height = 0.75cm] (D1) at (0,0) {$\bfX(1)$};
\node[rectangle,draw,   text = white, fill = blue!70!black,
    minimum width = 1.25cm,
    minimum height = 0.75cm, right = -\the\pgflinewidth of D1.north east, anchor = north west] (GB1) {$00...0$};
\node[rectangle,draw, minimum width = 1.5cm, 
    minimum height = 0.75cm, right = -\the\pgflinewidth of GB1.north east, anchor = north west] (D2) {$\bfX(2)$};
\node[rectangle,draw,   text = white, fill = blue!70!black,
    minimum width = 2cm,
    minimum height = 0.75cm, right = -\the\pgflinewidth of D2.north east, anchor = north west] (GB2) {$00........0$};
    \node[rectangle,draw, minimum width = 1.5cm, 
    minimum height = 0.75cm, right = -\the\pgflinewidth of GB2.north east, anchor = north west] (D3) {$\bfX(3)$};
\node[rectangle,draw,   text = white, fill = blue!70!black,
    minimum width = 1.25cm,
    minimum height = 0.75cm, right = -\the\pgflinewidth of D3.north east, anchor = north west] (GB3) {$00...0$};
\node[rectangle,draw, minimum width = 1.5cm, 
    minimum height = 0.75cm, right = -\the\pgflinewidth of GB3.north east, anchor = north west] (D4) {$\bfX(4)$};

\draw [decorate,
    decoration = {calligraphic brace,
        amplitude=4pt},thick] (-0.75,0.5)--(9.75,0.5)  node[pos=0.5,above  =4pt]{$g(\bfX)$} ;
\draw [{Bar[scale=0.5pt]}-{Bar[scale=0.5]}] (-0.75,-0.5)--(0.75,-0.5)  node[pos=0.5,below  =1pt]{$N_0$} ;
  \draw[{Bar[scale=0.5pt]}-{Bar[scale=0.5]}](0.75,-0.7)--(2,-0.7)  node[pos=0.5,below  =1pt]{$\ell_{n_0+1}$} ;
\draw [{Bar[scale=0.5pt]}-{Bar[scale=0.5]}] (2,-0.5)--(3.5,-0.5)  node[pos=0.5,below  =1pt]{$N_0$} ;
        \draw [{Bar[scale=0.5pt]}-{Bar[scale=0.5]}] (3.5,-0.7)--(5.5,-0.7)  node[pos=0.5,below  =1pt]{$\ell_{n_0+2}$} ;
\draw [{Bar[scale=0.5pt]}-{Bar[scale=0.5]}] (5.5,-0.5)--(7,-0.5)  node[pos=0.5,below  =1pt]{$N_0$} ;
         \draw [{Bar[scale=0.5pt]}-{Bar[scale=0.5]}] (7,-0.7)--(8.25,-0.7)  node[pos=0.5,below  =1pt]{$\ell_{n_0+1}$} ;       
 \draw [{Bar[scale=0.5pt]}-{Bar[scale=0.5]}] (8.25,-0.5)--(9.75,-0.5)  node[pos=0.5,below  =1pt]{$N_0$} ;

\commentBlock{
\draw [decorate,
    decoration = {calligraphic brace,mirror, 
        amplitude=3pt},thick] (-0.75,-0.5)--(0.75,-0.5)  node[pos=0.5,below  =3pt]{$N_0$} ;
        \draw [decorate,
    decoration = {calligraphic brace,mirror, 
        amplitude=3pt},thick] (0.75,-0.5)--(2,-0.5)  node[pos=0.5,below  =3pt]{$\ell_{n_0+1}$} ;
\draw [decorate,
    decoration = {calligraphic brace,mirror, 
        amplitude=3pt},thick] (2,-0.5)--(3.5,-0.5)  node[pos=0.5,below  =3pt]{$N_0$} ;
        \draw [decorate,
    decoration = {calligraphic brace,mirror, 
        amplitude=3pt},thick] (3.5,-0.5)--(5.5,-0.5)  node[pos=0.5,below  =3pt]{$\ell_{n_0+2}$} ;
\draw [decorate,
    decoration = {calligraphic brace,mirror, 
        amplitude=3pt},thick] (5.5,-0.5)--(7,-0.5)  node[pos=0.5,below  =3pt]{$N_0$} ;
         \draw [decorate,
    decoration = {calligraphic brace,mirror, 
        amplitude=3pt},thick] (7,-0.5)--(8.25,-0.5)  node[pos=0.5,below  =3pt]{$\ell_{n_0+1}$} ;       
 \draw [decorate,
    decoration = {calligraphic brace,mirror, 
        amplitude=3pt},thick] (8.25,-0.5)--(9.75,-0.5)  node[pos=0.5,below  =3pt]{$N_0$} ; }      
        \end{tikzpicture}
}

%% file: tikz_schemes/scheme_XGYZ.tex
        \begin{tikzpicture}[xscale=1.2,yscale=1]
		\node[anchor=west] at (-4.1,5.25) {$\bfX$};
        \draw [very thick,draw=black] (-1.3,5.5) rectangle (0,5) node[pos=.5] {$\bfX_{\RN{1}}$};
        \draw [very thick,draw=black] (0,5.5) rectangle (1.3,5) node[pos=.5] {$\bfX_{\RN{2}}$};
        
\node[anchor=west] at (-4.1,3.75) {$\bfG = g(\bfX)$};
        \draw [very thick,draw=black] (-2.5,4) rectangle (-0.5,3.5) node[pos=.5] {$\bfG_{\RN{1}}$};
        \draw [very thick,draw=black,fill=blue!70!black] (-0.5,4) rectangle (0.5,3.5) node[pos=.5,white] {$\bfG_{\Delta}$};
        \draw [very thick,draw=black] (0.5,4) rectangle (2.5,3.5) node[pos=.5] {$\bfG_{\RN{2}}$};
        
\draw[dashed] (-1.3,5) -- (-2.5,4);
        \draw[dashed] (1.3,5) -- (2.5,4);
        \draw[dashed] (0,5) -- (-0.5,4);
        \draw[dashed] (0,5) -- (0.5,4);
        
\node[anchor=west] at (-4.1,2.25) {$\bfY$};
        \draw [very thick,draw=black] (-2.2,2.5) rectangle (-0.60,2) node[pos=.5] {$\bfY_{\RN{1}}$};
        \draw [very thick,draw=black,fill=blue!70!black] (-0.60,2.5) rectangle (0.30,2) node[pos=.5,white] {$\bfY_{\Delta}$};
        \draw [very thick,draw=black] (0.30,2.5) rectangle (2.2,2) node[pos=.5] {$\bfY_{\RN{2}}$};
        
\draw[dashed] (-2.5,3.5) -- (-2.2,2.4);
        \draw[dashed] (2.5,3.5) -- (2.2,2.5);
        \draw[dashed] (-0.5,3.5) -- (-0.60,2.5);
        \draw[dashed] (0.5,3.5) -- (0.30,2.5);
        
\node[anchor=west] at (-4.1,1) {$\bfZ = \bfY^*$};
        \draw [very thick,draw=black] (-2.0,1.25) rectangle (-0.60,0.75) node[pos=.5] {$\bfZ_{\RN{1}}$};
        \draw [very thick,draw=black,fill=blue!70!black] (-0.60,1.25) rectangle (0.30,0.75) node[pos=.5,white] {$\bfZ_{\Delta}$};
        \draw [very thick,draw=black] (0.30,1.25) rectangle (2.0,0.75) node[pos=.5] {$\bfZ_{\RN{2}}$};
        
\draw[dashed] (-2.2,2) -- (-2.0,1.25);
        \draw[dashed] (2.2,2) -- (2.0,1.25);
        \draw[dashed] (-0.60,2) -- (-0.6,1.25);
        \draw[dashed] (0.30,2) -- (0.3,1.25);

\draw[-{latex'[scale=1pt]},blue!70!black, very thick] (-3.88,5.05)--(-3.88,4) node[left,pos=.4] {$g$};
\draw[-{latex'[scale=1pt]},thick,blue!70!black,opacity =0.5] (-0.8,4.95)--(-1.4,4.05)node[above=0.25,scale=1pt] {$g$};
\draw[-{latex'[scale=1pt]},thick,blue!70!black,opacity =0.5] (0.8,4.95)--(1.4,4.05)node[above=0.25,scale=1pt] {$g$};
\draw[-{latex'[scale=1pt]},red!70!black, very thick] (-3.88,3.55)--(-3.88,2.5) node[left,pos=.4] {DC};
\draw[-{latex'[scale=1pt]},thick,red!70!black,opacity =0.5] (-1.5,3.45)--(-1.4,2.55)node[left,pos=0.4,scale=0.9pt] {DC};
\draw[-{latex'[scale=1pt]},thick,red!70!black,opacity =0.5] (1.5,3.45)--(1.3,2.55)node[right,pos=0.4,scale=0.9pt] {DC};
\draw[-{latex'[scale=1pt]},thick,red!70!black,opacity =0.5] (0,3.45)--(-0.1,2.55)node[left,pos=0.4,scale=0.9pt] {DC};
\draw[-{latex'[scale=1pt]},green!50!black, very thick] (-3.88,2)--(-3.88,1.25) node[left,pos=.4] {TC};
        \end{tikzpicture}

%% file: tikz_schemes/scheme_trellis_polarization.tex
\centerline{
\begin{tikzpicture}[scale=0.55%
, every node/.style={scale=0.47}]
\draw[{Circle[scale=0.5]}-{Stealth[scale=0.5]}] (1.3,0) node[left,scale=2]{$\mathcal{T}$} -- (2.3,1) node[right,scale=2]{$\mathcal{T}^{[0]}$} node[above,midway,scale=1.5,rotate=45]{$-$} ;

\draw[dashed,-{Stealth[scale=0.5]}] (1.3,0) node[left,scale=2]{$\mathcal{T}$} -- (2.3,-1) node[right= 0.5pt,scale=2]{$\mathcal{T}^{[1]}$} node[below,midway,scale=1.5,rotate=-45]{$+$} ;

\draw[{Circle[scale=0.5]}-{Stealth[scale=0.5]}] (3.8,1)  -- (5.3,1.5) node[ right,scale=2]{$\mathcal{T}^{[00]}$} ;

\draw[dashed,-{Stealth[scale=0.5]}] (3.8,1) -- (5.3,0.5) node[right = 0.5pt,scale=2]{$\mathcal{T}^{[01]}$} ;

\draw[{Circle[scale=0.5]}-{Stealth[scale=0.5]}] (3.8,-1)  -- (5.3,-0.5) node[ right,scale=2]{$\mathcal{T}^{[10]}$} ;

\draw[dashed,-{Stealth[scale=0.5]}] (3.8,-1) -- (5.3,-1.5) node[right = 0.5pt,scale=2]{$\mathcal{T}^{[11]}$} ;

\draw (7.6,0) node[scale=3]{$...$};

\draw[{Circle[scale=0.5]}-{Stealth[scale=0.5]}] (8.5,2)  -- (10,3) ;

\draw[{Circle[scale=0.5]}-{Stealth[scale=0.5]}] (10,3)  -- (11.5,3.5) node[ right,scale=2]{$\mathcal{T}^{[000...00]}$} ;
\draw[-{Stealth[scale=0.5]},green!50!black] (14.4,3.5)  -- (15.1,3.5) node[ right,scale=1.5]{decide $\hat{u}_1$} ;

\draw[-{Stealth[scale=0.5]},dashed] (10,3)  -- (11.5,2.5) node[ right,scale=2]{$\mathcal{T}^{[000...01]}$} ;
\draw[-{Stealth[scale=0.5]},green!50!black] (14.4,2.5)  -- (15.1,2.5) node[ right,scale=1.5]{decide $\hat{u}_2$} ;

\draw[-{Stealth[scale=0.5]},dashed] (8.5,2)  -- (10,1) ;
\draw[{Circle[scale=0.5]}-{Stealth[scale=0.5]}] (10,1)  -- (11.5,1.5) ;
\draw[-{Stealth[scale=0.5]},dashed] (10,1)  -- (11.5,0.5)  ;

\draw[{Circle[scale=0.5]}-{Stealth[scale=0.5]}] (8.5,-2)  -- (10,-1) ;

\draw[{Circle[scale=0.5]}-{Stealth[scale=0.5]}] (10,-1)  -- (11.5,-0.5)  ;
\draw[-{Stealth[scale=0.5]},dashed] (10,-1)  -- (11.5,-1.5) ;

\draw[-{Stealth[scale=0.5]},dashed] (8.5,-2)  -- (10,-3) ;
\draw[{Circle[scale=0.5]}-{Stealth[scale=0.5]}] (10,-3)  -- (11.5,-2.5) node[ right,scale=2]{$\mathcal{T}^{[111...10]}$} ;
\draw[-{Stealth[scale=0.5]},green!50!black] (14.4,-2.5)  -- (15.1,-2.5) node[ right,scale=1.5]{decide $\hat{u}_{N-1}$} ;
\draw[-{Stealth[scale=0.5]},dashed] (10,-3)  -- (11.5,-3.5) node[ right,scale=2]{$\mathcal{T}^{[111...11]}$} ;
\draw[-{Stealth[scale=0.5]},green!50!black] (14.4,-3.5)  -- (15.1,-3.5) node[ right,scale=1.5]{decide $\hat{u}_{N}$} ;
\draw (8.5,0) node[scale=3,rotate=90]{$.....$};

\draw (14,0) node[scale=3,rotate=90]{$.......$};

\draw[-{Stealth[scale=1.5]},blue!70!black]  (1.3,-4.5) -- (14,-4.5) node[below=9.5pt,midway,scale=1.5]{\textit{trellis polarization depth}} node[right,scale=1.5]{$\lambda$};

\draw[|-|,blue!70!black]  (1.3,-4.5)node[below=1.5pt,scale=1.5]{$0$} -- (12.5,-4.5)  node[below=1.5pt,scale=1.5]{$n$};
\draw[|-|,blue!70!black]  (3.8,-4.5)node[below=1.5pt,scale=1.5]{$1$} -- (10,-4.5)  node[below=1.5pt,scale=1.5]{$n-1$};
\draw[|-|,blue!70!black]  (5.6,-4.5)node[below=1.5pt,scale=1.5]{$2$} -- (8.5,-4.5)  node[below=1.5pt,scale=1.5]{$n-2$};

\draw[dotted,white,line width=0.5mm] (6,-4.5)--(8,-4.5);
\end{tikzpicture}
}

%% file: tikz_schemes/scheme_trellis_evolution_spaced.tex
\begin{tikzpicture}
[scale=1, every node/.style={scale=1}]

\node[rectangle,draw, minimum width = 1.5cm, 
    minimum height = 2cm] (D1) at (0,0) {$\mathcal{T}^{\mathrm{B-1}}$};
\node[rectangle,draw,   text = white, fill = blue!70!black,
    minimum width = 1.25cm,
    minimum height = 2cm, right = -\the\pgflinewidth of D1.north east, anchor = north west] (GB1){$\mathcal{T}^{\mathrm{GB-1}}$};
\node[rectangle,draw, minimum width = 1.5cm, 
    minimum height = 2cm, right = -\the\pgflinewidth of GB1.north east, anchor = north west] (D2){$\mathcal{T}^{\mathrm{B-2}}$} ;
\node[rectangle,draw,   text = white, fill = blue!70!black,
    minimum width = 2cm,
    minimum height = 2cm, right = -\the\pgflinewidth of D2.north east, anchor = north west] (GB2){$\mathcal{T}^{\mathrm{GB-2}}$};
    \node[rectangle,draw, minimum width = 1.5cm, 
    minimum height = 2cm, right = -\the\pgflinewidth of GB2.north east, anchor = north west] (D3) {$\mathcal{T}^{\mathrm{B-3}}$};
\node[rectangle,draw,   text = white, fill = blue!70!black,
    minimum width = 1.25cm,
    minimum height = 2cm, right = -\the\pgflinewidth of D3.north east, anchor = north west] (GB3) {$\mathcal{T}^{\mathrm{GB-3}}$};
\node[rectangle,draw, minimum width = 1.5cm, 
    minimum height = 2cm, right = -\the\pgflinewidth of GB3.north east, anchor = north west] (D4) {$\mathcal{T}^{\mathrm{B-4}}$};

\node[anchor=west] at ($(D1.north)+(-0.45,0.25)$) {$\bfx(1)$};
\node[anchor=west] at ($(D2.north)+(-0.45,0.25)$) {$\bfx(2)$};
\node[anchor=west] at ($(D3.north)+(-0.45,0.25)$) {$\bfx(3)$};
\node[anchor=west] at ($(D4.north)+(-0.45,0.25)$) {$\bfx(4)$};

\node[anchor=west] at ($(GB1.north)+(-0.45,0.25)$) {GB};
\node[anchor=west] at ($(GB2.north)+(-0.45,0.25)$) {GB};
\node[anchor=west] at ($(GB3.north)+(-0.45,0.25)$) {GB};

\node(Y)[anchor=west] at (-1.35,0) {$\bfy$};

\draw [decorate,decoration = {calligraphic brace,        amplitude=4pt},thick] ($(D1.north west)+(0,0.65)$)--($(D4.north east)+(0,0.65)$)  node[pos=0.5,above  =4pt]{$\mathcal{T}$} ;
\draw [{Bar[scale=0.5pt]}-{Bar[scale=0.5]}] ($(D1.south east)+(0,-0.15)$)--($(D1.south west)+(0,-0.15)$)  node[pos=0.5,below  =-1pt]{$N_0$} ;
  \draw[{Bar[scale=0.5pt]}-{Bar[scale=0.5]}]($(GB1.south east)+(0,-0.25)$)--($(GB1.south west)+(0,-0.25)$) node[pos=0.5,below  =-1pt]{$\ell_{n_0+1}$} ;
\draw [{Bar[scale=0.5pt]}-{Bar[scale=0.5]}] ($(D2.south east)+(0,-0.15)$)--($(D2.south west)+(0,-0.15)$)  node[pos=0.5,below  =-1pt]{$N_0$} ;
        \draw [{Bar[scale=0.5pt]}-{Bar[scale=0.5]}] ($(GB2.south east)+(0,-0.25)$)--($(GB2.south west)+(0,-0.25)$)  node[pos=0.5,below  =-1pt]{$\ell_{n_0+2}$} ;
\draw [{Bar[scale=0.5pt]}-{Bar[scale=0.5]}] ($(D3.south east)+(0,-0.15)$)--($(D3.south west)+(0,-0.15)$)  node[pos=0.5,below  =-1pt]{$N_0$} ;
         \draw [{Bar[scale=0.5pt]}-{Bar[scale=0.5]}] ($(GB3.south east)+(0,-0.25)$)--($(GB3.south west)+(0,-0.25)$)  node[pos=0.5,below  =-1pt]{$\ell_{n_0+1}$} ;       
 \draw [{Bar[scale=0.5pt]}-{Bar[scale=0.5]}] ($(D4.south east)+(0,-0.15)$)--($(D4.south west)+(0,-0.15)$)  node[pos=0.5,below  =-1pt]{$N_0$} ;

 \node(firstPolarTrans)[anchor=north west] at (-1.8,-3.5) {\parbox{200pt}{Perform $n_0$ ``without GB'' transforms, and merge sections in GB locations:}};

\node[rectangle,draw, minimum width = 0.5cm, 
    minimum height = 2cm] (D1_n0_1) at (6.5,-4.55) {a};
\node[rectangle,draw,   text = white, fill = blue!70!black,
    minimum width = 0.5cm,
    minimum height = 2cm, right = -\the\pgflinewidth of D1_n0_1.north east, anchor = north west] (GB1_n0_1) {b};
\node[rectangle,draw, minimum width = 0.5cm, 
    minimum height = 2cm, right = -\the\pgflinewidth of GB1_n0_1.north east, anchor = north west] (D2_n0_1) {c};
\node[rectangle,draw,   text = white, fill = blue!70!black,
    minimum width = 0.5cm,
    minimum height = 2cm, right = -\the\pgflinewidth of D2_n0_1.north east, anchor = north west] (GB2_n0_1) {d};
    \node[rectangle,draw, minimum width = 0.5cm, 
    minimum height = 2cm, right = -\the\pgflinewidth of GB2_n0_1.north east, anchor = north west] (D3_n0_1) {e};
\node[rectangle,draw,   text = white, fill = blue!70!black,
    minimum width = 0.5cm,
    minimum height = 2cm, right = -\the\pgflinewidth of D3_n0_1.north east, anchor = north west] (GB3_n0_1) {f};
\node[rectangle,draw, minimum width = 0.5cm, 
    minimum height = 2cm, right = -\the\pgflinewidth of GB3_n0_1.north east, anchor = north west] (D4_n0_1) {g};

\draw [decorate,decoration = {calligraphic brace,        amplitude=3pt},thick] (6.25,-3)--(9.75,-3)  node[pos=0.5,above  =2pt]{$\mathcal{T}^{[b_1\dots b_{n_0}]}$} ;
\draw [decorate,decoration = {calligraphic brace,        amplitude=2pt},thick] (6.25,-3.45)--(7.75,-3.45)  node[pos=0.5,above  =1pt]{$s\mathcal{T}$} ;
\draw [decorate,decoration = {calligraphic brace,        amplitude=2pt},thick] (8.25,-3.45)--(9.75,-3.45)  node[pos=0.5,above  =1pt]{$s\mathcal{T}$} ;

\draw [{Bar[scale=0.5pt]}-{Bar[scale=0.5]}]  ($(D1_n0_1.south east)+(0,-0.15)$)--($(D1_n0_1.south west)+(0,-0.15)$)   node[pos=0.5,below  =-0.5pt]{$1$} ;
  \draw[{Bar[scale=0.5pt]}-{Bar[scale=0.5]}] ($(GB1_n0_1.south east)+(0,-0.25)$)--($(GB1_n0_1.south west)+(0,-0.25)$)  node[pos=0.5,below  =-0.5pt]{$1$} ;
\draw [{Bar[scale=0.5pt]}-{Bar[scale=0.5]}] ($(D2_n0_1.south east)+(0,-0.15)$)--($(D2_n0_1.south west)+(0,-0.15)$)    node[pos=0.5,below  =-0.5pt]{$1$} ;
        \draw [{Bar[scale=0.5pt]}-{Bar[scale=0.5]}] ($(GB2_n0_1.south east)+(0,-0.25)$)--($(GB2_n0_1.south west)+(0,-0.25)$)  node[pos=0.5,below  =-0.5pt]{$1$} ;
\draw [{Bar[scale=0.5pt]}-{Bar[scale=0.5]}] ($(D3_n0_1.south east)+(0,-0.15)$)--($(D3_n0_1.south west)+(0,-0.15)$)    node[pos=0.5,below  =-0.5pt]{$1$} ;
         \draw [{Bar[scale=0.5pt]}-{Bar[scale=0.5]}] ($(GB3_n0_1.south east)+(0,-0.25)$)--($(GB3_n0_1.south west)+(0,-0.25)$)   node[pos=0.5,below  =-0.5pt]{$1$} ;       
 \draw [{Bar[scale=0.5pt]}-{Bar[scale=0.5]}]($(D4_n0_1.south east)+(0,-0.15)$)--($(D4_n0_1.south west)+(0,-0.15)$)    node[pos=0.5,below  =-0.5pt]{$1$} ;  

\node(nextPolarTrans) [anchor=north west] at (-1.8,-7.5) {\parbox{230pt}{Perform ``with GB'' transform:

Step 1: in each sub trellis $s\mathcal{T}$, merge the left block trellis and the middle GB trellis:\\
a${}\triangleq \mathcal{T}^{\mathrm{B-1}}$ merge b${}\triangleq \mathcal{T}^{\mathrm{GB-1}}$, \\ e${}\triangleq \mathcal{T}^{\mathrm{B-3}}$ merge f${}\triangleq \mathcal{T}^{\mathrm{GB-3}}$}};

\node[rectangle,draw, minimum width = 0.5cm, 
    minimum height = 2cm] (D1_n0) at (7.5,-8.55) {\parbox{7pt}{a \\  b}};

\node[rectangle,draw, minimum width = 0.5cm, 
    minimum height = 2cm, right = -\the\pgflinewidth of D1_n0.north east, anchor = north west] (D2_n0) {c};
\node[rectangle,draw,   text = white, fill = blue!70!black,
    minimum width = 0.5cm,
    minimum height = 2cm, right = -\the\pgflinewidth of D2_n0.north east, anchor = north west] (GB2_n0) {d};
    \node[rectangle,draw, minimum width = 0.5cm, 
    minimum height = 2cm, right = -\the\pgflinewidth of GB2_n0.north east, anchor = north west] (D3_n0) {\parbox{7pt}{e \\  f}};
\node[rectangle,draw, minimum width = 0.5cm, 
    minimum height = 2cm, right = -\the\pgflinewidth of D3_n0.north east, anchor = north west] (D4_n0) {g};

\draw [decorate,decoration = {calligraphic brace,        amplitude=2pt},thick] (7.25,-7.45)--(8.25,-7.45)  node[pos=0.5,above  =1pt]{$s\mathcal{T}$} ;
\draw [decorate,decoration = {calligraphic brace,        amplitude=2pt},thick] (8.75,-7.45)--(9.75,-7.45)  node[pos=0.5,above  =1pt]{$s\mathcal{T}$} ;

\draw [{Bar[scale=0.5pt]}-{Bar[scale=0.5]}] ($(D1_n0.south east)+(0,-0.15)$)--($(D1_n0.south west)+(0,-0.15)$)   node[pos=0.5,below  =-0.5pt]{$1$} ;
  \draw[{Bar[scale=0.5pt]}-{Bar[scale=0.5]}]($(D2_n0.south east)+(0,-0.25)$)--($(D2_n0.south west)+(0,-0.25)$)  node[pos=0.5,below  =-0.5pt]{$1$} ;
\draw [{Bar[scale=0.5pt]}-{Bar[scale=0.5]}] ($(GB2_n0.south east)+(0,-0.15)$)--($(GB2_n0.south west)+(0,-0.15)$)     node[pos=0.5,below  =-0.5pt]{$1$} ;
        \draw [{Bar[scale=0.5pt]}-{Bar[scale=0.5]}] ($(D3_n0.south east)+(0,-0.25)$)--($(D3_n0.south west)+(0,-0.25)$)     node[pos=0.5,below  =-0.5pt]{$1$} ;
\draw [{Bar[scale=0.5pt]}-{Bar[scale=0.5]}] ($(D4_n0.south east)+(0,-0.15)$)--($(D4_n0.south west)+(0,-0.15)$)   node[pos=0.5,below  =-0.5pt]{$1$} ;

\node(nextPolarTrans2) [anchor=north west] at (-1.8,-11) {\parbox{230pt}{Step 2: in each sub trellis $s\mathcal{T}$, sum over two-edge paths as in the ``without GB'' transform:\\
sum over paths in (a b) and c,\\ sum over  paths in (e  f) and g}};

\node[rectangle,draw, minimum width = 0.5cm, 
    minimum height = 2cm] (D1_n0) at (8.5,-12.05) {\parbox{7pt}{a \\  b \\  c}};

\node[rectangle,draw,   text = white, fill = blue!70!black,
    minimum width = 0.5cm,
    minimum height = 2cm, right = -\the\pgflinewidth of D1_n0.north east, anchor = north west] (GB2_n0) {d};
    \node[rectangle,draw, minimum width = 0.5cm, 
    minimum height = 2cm, right = -\the\pgflinewidth of GB2_n0.north east, anchor = north west] (D3_n0) {\parbox{7pt}{ e\\  f \\ g}};

\draw [decorate,decoration = {calligraphic brace,        amplitude=3pt},thick] (8.25,-10.95)--(9.75,-10.95)  node[pos=0.5,above  =2pt]{$\mathcal{T}^{[b_1\dots b_{n_0+1}]}$} ;

\draw [{Bar[scale=0.5pt]}-{Bar[scale=0.5]}] ($(D1_n0.south east)+(0,-0.15)$)--($(D1_n0.south west)+(0,-0.15)$)   node[pos=0.5,below  =-0.5pt]{$1$} ;
\draw [{Bar[scale=0.5pt]}-{Bar[scale=0.5]}] ($(GB2_n0.south east)+(0,-0.25)$)--($(GB2_n0.south west)+(0,-0.25)$)     node[pos=0.5,below  =-0.5pt]{$1$} ;
        \draw [{Bar[scale=0.5pt]}-{Bar[scale=0.5]}] ($(D3_n0.south east)+(0,-0.15)$)--($(D3_n0.south west)+(0,-0.15)$)     node[pos=0.5,below  =-0.5pt]{$1$} ;

\draw[rounded corners] (-2.5, -1.8) rectangle (11, 2.5) {};
\draw[rounded corners] (-2.5, -6.3) rectangle (11, -2.3) {};
\draw[rounded corners] (-2.5, -13.8) rectangle (11, -6.8) {};

\node at (-2, 2) {(a)};
\node at (-2, -2.8) {(b)};
\node at (-2, -7.3) {(c)};
        \end{tikzpicture}

%% file: tikz_schemes/scheme_trellis_evolution.tex
\begin{tikzpicture}
[scale=0.8, every node/.style={scale=0.8}]

\node[rectangle,draw, minimum width = 1.5cm, 
    minimum height = 2cm] (D1) at (0,0) {$\mathcal{T}^{\mathrm{B-1}}$};
\node[rectangle,draw,   text = white, fill = blue!70!black,
    minimum width = 1.25cm,
    minimum height = 2cm, right = -\the\pgflinewidth of D1.north east, anchor = north west] (GB1){$\mathcal{T}^{\mathrm{GB-1}}$};
\node[rectangle,draw, minimum width = 1.5cm, 
    minimum height = 2cm, right = -\the\pgflinewidth of GB1.north east, anchor = north west] (D2){$\mathcal{T}^{\mathrm{B-2}}$} ;
\node[rectangle,draw,   text = white, fill = blue!70!black,
    minimum width = 2cm,
    minimum height = 2cm, right = -\the\pgflinewidth of D2.north east, anchor = north west] (GB2){$\mathcal{T}^{\mathrm{GB-2}}$};
    \node[rectangle,draw, minimum width = 1.5cm, 
    minimum height = 2cm, right = -\the\pgflinewidth of GB2.north east, anchor = north west] (D3) {$\mathcal{T}^{\mathrm{B-3}}$};
\node[rectangle,draw,   text = white, fill = blue!70!black,
    minimum width = 1.25cm,
    minimum height = 2cm, right = -\the\pgflinewidth of D3.north east, anchor = north west] (GB3) {$\mathcal{T}^{\mathrm{GB-3}}$};
\node[rectangle,draw, minimum width = 1.5cm, 
    minimum height = 2cm, right = -\the\pgflinewidth of GB3.north east, anchor = north west] (D4) {$\mathcal{T}^{\mathrm{B-4}}$};

\node[anchor=west] at (-0.5,1.25) {$\bfx(1)$};
\node[anchor=west] at (2.25,1.25) {$\bfx(2)$};
\node[anchor=west] at (5.75,1.25) {$\bfx(3)$};
\node[anchor=west] at (8.5,1.25) {$\bfx(4)$};

\node[anchor=west] at (1,1.25) {GB};
\node[anchor=west] at (4,1.25) {GB};
\node[anchor=west] at (7.25,1.25) {GB};

\node(Y)[anchor=west] at (-1.35,0) {$\bfy$};

\draw [decorate,decoration = {calligraphic brace,        amplitude=4pt},thick] (-1.15,1.6)--(9.75,1.6)  node[pos=0.5,above  =4pt]{$\mathcal{T}$} ;
\draw [{Bar[scale=0.5pt]}-{Bar[scale=0.5]}] (-0.75,-1.15)--(0.75,-1.15)  node[pos=0.5,below  =-1pt]{$N_0$} ;
  \draw[{Bar[scale=0.5pt]}-{Bar[scale=0.5]}](0.75,-1.3)--(2,-1.3)  node[pos=0.5,below  =-1pt]{$\ell_{n_0+1}$} ;
\draw [{Bar[scale=0.5pt]}-{Bar[scale=0.5]}] (2,-1.15)--(3.5,-1.15)  node[pos=0.5,below  =-1pt]{$N_0$} ;
        \draw [{Bar[scale=0.5pt]}-{Bar[scale=0.5]}] (3.5,-1.3)--(5.5,-1.3)  node[pos=0.5,below  =-1pt]{$\ell_{n_0+2}$} ;
\draw [{Bar[scale=0.5pt]}-{Bar[scale=0.5]}] (5.5,-1.15)--(7,-1.15)  node[pos=0.5,below  =-1pt]{$N_0$} ;
         \draw [{Bar[scale=0.5pt]}-{Bar[scale=0.5]}] (7,-1.3)--(8.25,-1.3)  node[pos=0.5,below  =-1pt]{$\ell_{n_0+1}$} ;       
 \draw [{Bar[scale=0.5pt]}-{Bar[scale=0.5]}] (8.25,-1.15)--(9.75,-1.15)  node[pos=0.5,below  =-1pt]{$N_0$} ;

 \node(firstPolarTrans)[anchor=north west] at (-1.15,-3) {\parbox{190pt}{Perform $n_0$ ``without GB'' transforms, and merge sections in GB locations:}};

\node[rectangle,draw, minimum width = 0.5cm, 
    minimum height = 2cm] (D1_n0_1) at (6.5,-4.05) {a};
\node[rectangle,draw,   text = white, fill = blue!70!black,
    minimum width = 0.5cm,
    minimum height = 2cm, right = -\the\pgflinewidth of D1_n0_1.north east, anchor = north west] (GB1_n0_1) {b};
\node[rectangle,draw, minimum width = 0.5cm, 
    minimum height = 2cm, right = -\the\pgflinewidth of GB1_n0_1.north east, anchor = north west] (D2_n0_1) {c};
\node[rectangle,draw,   text = white, fill = blue!70!black,
    minimum width = 0.5cm,
    minimum height = 2cm, right = -\the\pgflinewidth of D2_n0_1.north east, anchor = north west] (GB2_n0_1) {d};
    \node[rectangle,draw, minimum width = 0.5cm, 
    minimum height = 2cm, right = -\the\pgflinewidth of GB2_n0_1.north east, anchor = north west] (D3_n0_1) {e};
\node[rectangle,draw,   text = white, fill = blue!70!black,
    minimum width = 0.5cm,
    minimum height = 2cm, right = -\the\pgflinewidth of D3_n0_1.north east, anchor = north west] (GB3_n0_1) {f};
\node[rectangle,draw, minimum width = 0.5cm, 
    minimum height = 2cm, right = -\the\pgflinewidth of GB3_n0_1.north east, anchor = north west] (D4_n0_1) {g};

\draw [decorate,decoration = {calligraphic brace,        amplitude=3pt},thick] (6.25,-2.5)--(9.75,-2.5)  node[pos=0.5,above  =2pt]{$\mathcal{T}^{[b_1\dots b_{n_0}]}$} ;
\draw [decorate,decoration = {calligraphic brace,        amplitude=2pt},thick] (6.25,-2.95)--(7.75,-2.95)  node[pos=0.5,above  =1pt]{$s\mathcal{T}$} ;
\draw [decorate,decoration = {calligraphic brace,        amplitude=2pt},thick] (8.25,-2.95)--(9.75,-2.95)  node[pos=0.5,above  =1pt]{$s\mathcal{T}$} ;

\draw [{Bar[scale=0.5pt]}-{Bar[scale=0.5]}] (6.25,-5.2)--(6.75,-5.2)  node[pos=0.5,below  =-0.5pt]{$1$} ;
  \draw[{Bar[scale=0.5pt]}-{Bar[scale=0.5]}](6.75,-5.35)--(7.25,-5.35)  node[pos=0.5,below  =-0.5pt]{$1$} ;
\draw [{Bar[scale=0.5pt]}-{Bar[scale=0.5]}] (7.25,-5.2)--(7.75,-5.2)  node[pos=0.5,below  =-0.5pt]{$1$} ;
        \draw [{Bar[scale=0.5pt]}-{Bar[scale=0.5]}] (7.75,-5.35)--(8.25,-5.35)  node[pos=0.5,below  =-0.5pt]{$1$} ;
\draw [{Bar[scale=0.5pt]}-{Bar[scale=0.5]}] (8.25,-5.2)--(8.75,-5.2)  node[pos=0.5,below  =-0.5pt]{$1$} ;
         \draw [{Bar[scale=0.5pt]}-{Bar[scale=0.5]}] (8.75,-5.35)--(9.25,-5.35)  node[pos=0.5,below  =-0.5pt]{$1$} ;       
 \draw [{Bar[scale=0.5pt]}-{Bar[scale=0.5]}] (9.25,-5.2)--(9.75,-5.2)  node[pos=0.5,below  =-0.5pt]{$1$} ;  

\node(nextPolarTrans) [anchor=north west] at (-1.15,-6.5) {\parbox{190pt}{Perform ``with GB'' transform:

Step 1: in each sub trellis $s\mathcal{T}$, merge the left block trellis and the middle GB trellis:\\
a${}\triangleq \mathcal{T}^{\mathrm{B-1}}$ merge b${}\triangleq \mathcal{T}^{\mathrm{GB-1}}$, \\ e${}\triangleq \mathcal{T}^{\mathrm{B-3}}$ merge f${}\triangleq \mathcal{T}^{\mathrm{GB-3}}$}};

\node[rectangle,draw, minimum width = 0.5cm, 
    minimum height = 2cm] (D1_n0) at (7.5,-7.55) {\parbox{7pt}{a \\  b}};

\node[rectangle,draw, minimum width = 0.5cm, 
    minimum height = 2cm, right = -\the\pgflinewidth of D1_n0.north east, anchor = north west] (D2_n0) {c};
\node[rectangle,draw,   text = white, fill = blue!70!black,
    minimum width = 0.5cm,
    minimum height = 2cm, right = -\the\pgflinewidth of D2_n0.north east, anchor = north west] (GB2_n0) {d};
    \node[rectangle,draw, minimum width = 0.5cm, 
    minimum height = 2cm, right = -\the\pgflinewidth of GB2_n0.north east, anchor = north west] (D3_n0) {\parbox{7pt}{e \\  f}};
\node[rectangle,draw, minimum width = 0.5cm, 
    minimum height = 2cm, right = -\the\pgflinewidth of D3_n0.north east, anchor = north west] (D4_n0) {g};

\draw [decorate,decoration = {calligraphic brace,        amplitude=2pt},thick] (7.25,-6.45)--(8.25,-6.45)  node[pos=0.5,above  =1pt]{$s\mathcal{T}$} ;
\draw [decorate,decoration = {calligraphic brace,        amplitude=2pt},thick] (8.75,-6.45)--(9.75,-6.45)  node[pos=0.5,above  =1pt]{$s\mathcal{T}$} ;

\draw [{Bar[scale=0.5pt]}-{Bar[scale=0.5]}] (7.25,-8.7)--(7.75,-8.7)  node[pos=0.5,below  =-0.5pt]{$1$} ;
  \draw[{Bar[scale=0.5pt]}-{Bar[scale=0.5]}](7.75,-8.85)--(8.25,-8.85)  node[pos=0.5,below  =-0.5pt]{$1$} ;
\draw [{Bar[scale=0.5pt]}-{Bar[scale=0.5]}] (8.25,-8.7)--(8.75,-8.7)  node[pos=0.5,below  =-0.5pt]{$1$} ;
        \draw [{Bar[scale=0.5pt]}-{Bar[scale=0.5]}] (8.75,-8.85)--(9.25,-8.85)  node[pos=0.5,below  =-0.5pt]{$1$} ;
\draw [{Bar[scale=0.5pt]}-{Bar[scale=0.5]}] (9.25,-8.7)--(9.75,-8.7)  node[pos=0.5,below  =-0.5pt]{$1$} ;

\node(nextPolarTrans2) [anchor=north west] at (-1.15,-10) {\parbox{190pt}{Step 2: in each sub trellis $s\mathcal{T}$, sum over two-edge paths as in the ``without GB'' transform:\\
sum over paths in (a b) and c,\\ sum over  paths in (e  f) and g}};

\node[rectangle,draw, minimum width = 0.5cm, 
    minimum height = 2cm] (D1_n0) at (8.5,-11.05) {\parbox{7pt}{a \\  b \\  c}};

\node[rectangle,draw,   text = white, fill = blue!70!black,
    minimum width = 0.5cm,
    minimum height = 2cm, right = -\the\pgflinewidth of D1_n0.north east, anchor = north west] (GB2_n0) {d};
    \node[rectangle,draw, minimum width = 0.5cm, 
    minimum height = 2cm, right = -\the\pgflinewidth of GB2_n0.north east, anchor = north west] (D3_n0) {\parbox{7pt}{ e\\  f \\ g}};

\draw [decorate,decoration = {calligraphic brace,        amplitude=3pt},thick] (8.25,-9.95)--(9.75,-9.95)  node[pos=0.5,above  =2pt]{$\mathcal{T}^{[b_1\dots b_{n_0+1}]}$} ;
\draw [{Bar[scale=0.5pt]}-{Bar[scale=0.5]}] (8.25,-12.2)--(8.75,-12.2)  node[pos=0.5,below  =-0.5pt]{$1$} ;
  \draw[{Bar[scale=0.5pt]}-{Bar[scale=0.5]}](8.75,-12.35)--(9.25,-12.35)  node[pos=0.5,below  =-0.5pt]{$1$} ;
\draw [{Bar[scale=0.5pt]}-{Bar[scale=0.5]}] (9.25,-12.2)--(9.75,-12.2)  node[pos=0.5,below  =-0.5pt]{$1$} ;

\draw[rounded corners] (-1.3, -1.8) rectangle (10, 2.5) {};
\draw[rounded corners] (-1.3, -5.8) rectangle (10, -1.8) {};
\draw[rounded corners] (-1.3, -12.8) rectangle (10, -5.8) {};

\node at (-1, 2.2) {(a)};
\node at (-1, -2.1) {(b)};
\node at (-1, -6.1) {(c)};
        \end{tikzpicture}

%% file: tikz_schemes/smallTrellises-noBackground.tex
\begin{tikzpicture}
[scale=0.7, every node/.style={scale=0.7}]

\node[rectangle,draw,fill=white, minimum width = 1.25cm, 
    minimum height = 0.5cm] (D1) at (0,0) {};
\node[rectangle, draw, minimum width=1.25cm, fill=white, minimum height=1.1cm, line width=\pgflinewidth, anchor=north west] (D2) at ($(D1.south east)+(1,-0.5)$) {};
\node[rectangle, draw,fill=white, minimum width=1.25cm, minimum height=0.2cm, line width=\pgflinewidth, anchor=north west] (D3) at ($(D2.south east)+(1.4,-0.5)$) {};
\node[rectangle, draw,fill=white, minimum width=1.25cm, minimum height=0.5cm, line width=\pgflinewidth, anchor=north west] (D4) at ($(D3.south east)+(1,-0.5)$) {};

\node[anchor=west](X1) at ($(D1.north)+(-0.5,0.25)$) {$\bfx(1)$};
\node[anchor=west](X2) at ($(D2.north)+(-0.5,0.25)$) {$\bfx(2)$};
\node[anchor=west](X3) at ($(D3.north)+(-0.5,0.25)$) {$\bfx(3)$};
\node[anchor=west](X4) at ($(D4.north)+(-0.5,0.25)$) {$\bfx(4)$};

\node[anchor=west](Y1) at ($(D1.west)+(-1.5,0)$) {$\bfy^*(1)$};
\node[anchor=west](Y2) at ($(Y1.west)+(0,-1.2)$) {$\bfy^*(2)$};
\node[anchor=west](Y3) at ($(Y2.west)+(0,-1.2)$) {$\bfy^*(3)$};
\node[anchor=west](Y4) at ($(Y3.west)+(0,-1)$)  {$\bfy^*(4)$};

\node[anchor=west](Y) at ($(Y2.west)+(-3,-0.5)$) {$\bfy$};

\node[single arrow, draw=lightgray, fill=lightgray, 
      minimum width = 2cm, single arrow head extend=5pt,      single arrow tip angle=120, 
      minimum height=1.5cm, align=center] at ($(Y.east)+(1,0)$) {partition \\ \& \\ trimming}; %

\draw [{Bar[scale=0.5pt]}-{Bar[scale=0.5]}] ($(D1.south east)+(0,-0.15)$)--($(D1.south west)+(0,-0.15)$)
node[pos=0.5,below  =-1pt]{$N_0$} ;
\draw [{Bar[scale=0.5pt]}-{Bar[scale=0.5]}] ($(D2.south east)+(0,-0.15)$)--($(D2.south west)+(0,-0.15)$)  node[pos=0.5,below  =-1pt]{$N_0$} ;
\draw [{Bar[scale=0.5pt]}-{Bar[scale=0.5]}] ($(D3.south east)+(0,-0.15)$)--($(D3.south west)+(0,-0.15)$) node[pos=0.5,below  =-1pt]{$N_0$} ;   
 \draw [{Bar[scale=0.5pt]}-{Bar[scale=0.5]}] ($(D4.south east)+(0,-0.15)$)--($(D4.south west)+(0,-0.15)$)  node[pos=0.5,below  =-1pt]{$N_0$} ;

\end{tikzpicture}

%% file: tikz_schemes/tikz_blockTrellis.tex
 \begin{tikzpicture}[scale=1,dot/.style={draw,circle,minimum size=1mm,inner sep=0pt,outer sep=0pt,fill=black}, >=latex]
        
          \pgfmathsetmacro{\xlim}{5}
          \pgfmathsetmacro{\ylim}{9}
          \pgfmathsetmacro{\xxlim}{\xlim-1}
          \pgfmathsetmacro{\yylim}{\ylim-1}
          \newcommand{\zz}{0}
          \newcommand{\recv}{01100010}
          
          \foreach \i in {0,...,\yylim}
            \node (i\i) at (0.65,\ylim-\i) {\scriptsize$i\!=\!\i$};
          \foreach \i in {1,...,\yylim}
            \node (y\i) at (0.25,\ylim-\i+0.5) {$y_{\i} \!=\! \StrChar{\recv}{\i}$};
        
          \foreach \j in {0,...,\xxlim}
            \node (j\j) at (\j+1,\ylim+0.30) {\scriptsize$j\!=\!\j$};    
          \foreach \j in {1,...,\xxlim}
            \node (x\j) at (\j+0.5,\ylim+.7) {$g_\j$};
            \foreach \j in {1,...,\xxlim}
            \node (x\j) at (\j+0.7,\ylim+1.3) {$x_\j$};
            \foreach \j in {1,...,\xxlim}
            \node[ rotate=60] (x\j) at (\j+0.6,\ylim+1) {$=$};
        
          \node (cg) at (0.5,\ylim+0.7) {$g_j$};
          \node[ rotate=60] (cequal) at (0.6,\ylim+1) {$=$};
           \node (cx) at (0.7,\ylim+1.3) {$x_{\indj}$};
          \node (cy) at (0,\ylim+0.25) {$y_i$};
          \draw (0,\ylim+0.7) -- (0.5, \ylim+0.25); 
        
          \foreach \i in {0,...,\yylim}
            \foreach \j in {0,...,\xxlim}
        	  \node [dot] (\j-\i) at (\j+1,\ylim-\i) {};
        
          \pgfmathsetmacro{\xxxlim}{\xlim-2}
          \pgfmathsetmacro{\yyylim}{\ylim-2}
          \foreach \i in {0,...,\yylim}
            \foreach \j in {0,...,\xxxlim}
        	{
              \pgfmathtruncatemacro{\ii}{\i+1}
              \pgfmathtruncatemacro{\jj}{\j+1}
        	      \draw[->,blue,thick] (\j-\i) to [in=150,out=30] (\jj-\i);
        	      \draw[->,red,thick] (\j-\i) to [in=210,out=330] (\jj-\i);
        	}
        	
          \pgfmathsetmacro{\xxxlim}{\xlim-2}
          \pgfmathsetmacro{\yyylim}{\ylim-2}
          \foreach \i in {0,...,\yyylim}
            \foreach \j in {0,...,\xxxlim}
        	{
              \pgfmathtruncatemacro{\ii}{\i+1}
              \pgfmathtruncatemacro{\jj}{\j+1}
              \pgfmathtruncatemacro{\iit}{\i+\xlim-\ylim}
                  \StrChar{\recv}{\ii}[\temp]
        		  \ifthenelse{\equal{\temp}{\zz}}
        		  {\draw[->,thick,blue] (\j-\i) to (\jj-\ii);}
        		  {\draw[->,thick,red] (\j-\i) to (\jj-\ii);}
        	}

        \end{tikzpicture}
                \begin{tikzpicture}[scale=1,dot/.style={draw,circle,minimum size=1mm,inner sep=0pt,outer sep=0pt,fill=black}, >=latex]
        
          \pgfmathsetmacro{\xlim}{2}
          \pgfmathsetmacro{\columnstart}{4}
          \pgfmathsetmacro{\ylim}{9}
          \pgfmathsetmacro{\xxlim}{\xlim-1}
          \pgfmathsetmacro{\yylim}{\ylim-1}
          \newcommand{\zz}{0}
          \newcommand{\recv}{01100010}
          
          \foreach \i in {0,...,\yylim}
            \node (i\i) at (0.65,\ylim-\i) {\scriptsize$i\!=\!\i$};
          \foreach \i in {1,...,\yylim}
            \node (y\i) at (0.25,\ylim-\i+0.5) {$y_{\i} \!=\! \StrChar{\recv}{\i}$};

          \foreach \i in {0,...,\yylim}
            \foreach \j in {0,...,\xxlim}
            \pgfmathsetmacro{\jj}{\j*1}
        	\node [dot] (\j-\i) at (\jj+1,\ylim-\i) {};

          \pgfmathsetmacro{\xxxlim}{\xlim-2}
          \pgfmathsetmacro{\yyylim}{\ylim-2}
          \foreach \i in {0,...,\yylim}
            \foreach \j in {0,...,\xxxlim}
        	{
              \pgfmathtruncatemacro{\ii}{\i+1}
              \pgfmathtruncatemacro{\jj}{\j+1}

          \draw[->,gray,thick] (\j-\i) to (\jj-\i);
        	}

          \pgfmathsetmacro{\xxxlim}{\xlim-2}
          \pgfmathsetmacro{\yyylim}{\ylim-2}
          \foreach \i in {0,...,\yyylim}
            \foreach \j in {0,...,\xxxlim}
        	{
              \pgfmathtruncatemacro{\ii}{\i+1}
              \pgfmathtruncatemacro{\jj}{\j+1}
        \draw[->,gray,thick] (\j-\i) to (\jj-\ii);	
         }

          \pgfmathsetmacro{\xxxlim}{\xlim-2}
          \pgfmathsetmacro{\yyylim}{\ylim-2}
          \foreach \i in {0,...,\yyylim}{
            \foreach \j in {0,...,\xxxlim}
        	{
              \pgfmathtruncatemacro{\iii}{\i+2}
              \pgfmathtruncatemacro{\jj}{\j+1}
              \ifthenelse{\iii>\yylim}{}{
            \draw[->,gray,thick] (\j-\i) to (\jj-\iii);	
         }} }  
               
          \pgfmathsetmacro{\xxxlim}{\xlim-2}
          \pgfmathsetmacro{\yyylim}{\ylim-2}
          \foreach \i in {0,...,\yyylim}{
            \foreach \j in {0,...,\xxxlim}
        	{
              \pgfmathtruncatemacro{\iiii}{\i+3}
              \pgfmathtruncatemacro{\jj}{\j+1}
             
                  \ifthenelse{\iiii>\yylim}{}{
               \draw[->,gray,thick] (\j-\i) to (\jj-\iiii);
        	}
        	}  }  
            \newcommand{\diagfourlabels}{00111}
          \pgfmathsetmacro{\xxxlim}{\xlim-2}
          \pgfmathsetmacro{\yyylim}{\ylim-2}
          \foreach \i in {0,...,\yyylim}{
            \foreach \j in {0,...,\xxxlim}
        	{
              \pgfmathtruncatemacro{\ii}{\i+1}
              \pgfmathtruncatemacro{\iii}{\i+2}
              \pgfmathtruncatemacro{\iiii}{\i+3}
              \pgfmathtruncatemacro{\iiii}{\i+4}
              \pgfmathtruncatemacro{\jj}{\j+1}
               \ifthenelse{\iiii>\yylim}{}{
              \StrChar{\diagfourlabels}{\ii}[\curlabel]
        		  \ifthenelse{\equal{\curlabel}{\zz}}{
            	\draw[->,thick,blue] (\j-\i) to (\jj-\iiii);}
        		  {\draw[->,thick,red] (\j-\i) to (\jj-\iiii);} }
            }}
        \end{tikzpicture}
       

%% file: tikz_schemes/tikz_GBtrellis.tex
   \begin{tikzpicture}[scale=1,dot/.style={draw,circle,minimum size=1mm,inner sep=0pt,outer sep=0pt,fill=black}, >=latex]
        
          \pgfmathsetmacro{\xlim}{4}
          \pgfmathsetmacro{\columnstart}{4}
          \pgfmathsetmacro{\ylim}{9}
          \pgfmathsetmacro{\xxlim}{\xlim-1}
          \pgfmathsetmacro{\yylim}{\ylim-1}
          \newcommand{\zz}{0}
          \newcommand{\recv}{01100010}
          
          \foreach \i in {0,...,\yylim}
            \node (i\i) at (0.65,\ylim-\i) {\scriptsize$i\!=\!\i$};
          \foreach \i in {1,...,\yylim}
            \node (y\i) at (0.25,\ylim-\i+0.5) {$y_{\i} \!=\! \StrChar{\recv}{\i}$};
        
          \foreach \j in {0,...,\xxlim}
            \node (j\j) at (\j+1,\ylim+0.30) {\scriptsize$j\!=\!\the\numexpr\j+\columnstart\relax$};    
          \foreach \j in {1,...,\xxlim}
            \node (x\j) at (\j+0.5,\ylim+.7) {$g_{\the\numexpr\j+\columnstart\relax}$};
        
          \node (cx) at (0.5,\ylim+0.7) {$g_j$};
          \node (cy) at (0,\ylim+0.25) {$y_i$};
          \draw (0,\ylim+0.7) -- (0.5, \ylim+0.25); 
        
          \foreach \i in {0,...,\yylim}
            \foreach \j in {0,...,\xxlim}
        	  \node [dot] (\j-\i) at (\j+1,\ylim-\i) {};
        
          \pgfmathsetmacro{\xxxlim}{\xlim-2}
          \pgfmathsetmacro{\yyylim}{\ylim-2}
          \foreach \i in {0,...,\yylim}
            \foreach \j in {0,...,\xxxlim}
        	{
              \pgfmathtruncatemacro{\ii}{\i+1}
              \pgfmathtruncatemacro{\jj}{\j+1}
        	      \draw[->,blue,thick] (\j-\i) to [in=150,out=30] (\jj-\i);
        	
        	}
        	
          \pgfmathsetmacro{\xxxlim}{\xlim-2}
          \pgfmathsetmacro{\yyylim}{\ylim-2}
          \foreach \i in {0,...,\yyylim}
            \foreach \j in {0,...,\xxxlim}
        	{
              \pgfmathtruncatemacro{\ii}{\i+1}
              \pgfmathtruncatemacro{\jj}{\j+1}
              \pgfmathtruncatemacro{\iit}{\i+\xlim-\ylim}
                  \StrChar{\recv}{\ii}[\temp]
        		  \ifthenelse{\equal{\temp}{\zz}}
        		  {\draw[->,thick,blue] (\j-\i) to (\jj-\ii);}
        		  {}
        	}

        \end{tikzpicture}
        \begin{tikzpicture}[scale=1,dot/.style={draw,circle,minimum size=1mm,inner sep=0pt,outer sep=0pt,fill=black}, >=latex]
        
          \pgfmathsetmacro{\xlim}{2}
          \pgfmathsetmacro{\columnstart}{4}
          \pgfmathsetmacro{\ylim}{9}
          \pgfmathsetmacro{\xxlim}{\xlim-1}
          \pgfmathsetmacro{\yylim}{\ylim-1}
          \newcommand{\zz}{0}
          \newcommand{\recv}{01100010}
          
          \foreach \i in {0,...,\yylim}
            \node (i\i) at (0.65,\ylim-\i) {\scriptsize$i\!=\!\i$};
          \foreach \i in {1,...,\yylim}
            \node (y\i) at (0.25,\ylim-\i+0.5) {$y_{\i} \!=\! \StrChar{\recv}{\i}$};

          \foreach \i in {0,...,\yylim}
            \foreach \j in {0,...,\xxlim}
        	\node [dot] (\j-\i) at (\j+1,\ylim-\i) {};

          \pgfmathsetmacro{\xxxlim}{\xlim-2}
          \pgfmathsetmacro{\yyylim}{\ylim-2}
          \foreach \i in {0,...,\yylim}
            \foreach \j in {0,...,\xxxlim}
        	{
              \pgfmathtruncatemacro{\ii}{\i+1}
              \pgfmathtruncatemacro{\jj}{\j+1}
        	      \draw[->,blue,thick] (\j-\i) to [in=150,out=30] (\jj-\i);
        	
        	}

          \pgfmathsetmacro{\xxxlim}{\xlim-2}
          \pgfmathsetmacro{\yyylim}{\ylim-2}
          \foreach \i in {0,...,\yyylim}
            \foreach \j in {0,...,\xxxlim}
        	{
              \pgfmathtruncatemacro{\ii}{\i+1}
              \pgfmathtruncatemacro{\jj}{\j+1}
              \pgfmathtruncatemacro{\iit}{\i+\xlim-\ylim}
                  \StrChar{\recv}{\ii}[\temp]
        		  \ifthenelse{\equal{\temp}{\zz}}
        		  {\draw[->,thick,blue] (\j-\i) to (\jj-\ii);}
        		  {}
        	}

          \pgfmathsetmacro{\xxxlim}{\xlim-2}
          \pgfmathsetmacro{\yyylim}{\ylim-2}
          \foreach \i in {0,...,\yyylim}
            \foreach \j in {0,...,\xxxlim}
        	{
              \pgfmathtruncatemacro{\ii}{\i+1}
              \pgfmathtruncatemacro{\iii}{\i+2}
              \pgfmathtruncatemacro{\jj}{\j+1}
             
                  \StrChar{\recv}{\ii}[\temp]
        		  \ifthenelse{\equal{\temp}{\zz}}{
                     \StrChar{\recv}{\iii}[\temp]
        		   \ifthenelse{\equal{\temp}{\zz}}
            		  {\draw[->,thick,blue] (\j-\i) to (\jj-\iii);}{}}
        		  {}
        	}    
               
          \pgfmathsetmacro{\xxxlim}{\xlim-2}
          \pgfmathsetmacro{\yyylim}{\ylim-2}
          \foreach \i in {0,...,\yyylim}
            \foreach \j in {0,...,\xxxlim}
        	{
              \pgfmathtruncatemacro{\ii}{\i+1}
              \pgfmathtruncatemacro{\iii}{\i+2}
              \pgfmathtruncatemacro{\iiii}{\i+3}
              \pgfmathtruncatemacro{\jj}{\j+1}
             
                  \StrChar{\recv}{\ii}[\temp]
        		  \ifthenelse{\equal{\temp}{\zz}}{
                     \StrChar{\recv}{\iii}[\temp]
        		   \ifthenelse{\equal{\temp}{\zz}}
            		  {\StrChar{\recv}{\iiii}[\temp]
        		   \ifthenelse{\equal{\temp}{\zz}}
            		  {\draw[->,thick,blue] (\j-\i) to (\jj-\iiii);}{}}{}}
        		  {}
        	}    
               
        \end{tikzpicture}

%% file: tikz_schemes/scheme_trellis_T00.tex
\centering 
           \begin{tikzpicture}[scale=0.85,dot/.style={draw,circle,minimum size=1mm,inner sep=0pt,outer sep=0pt,fill=black}, >=latex]
        
          \pgfmathsetmacro{\xlim}{2}
          \pgfmathsetmacro{\columnstart}{4}
          \pgfmathsetmacro{\ylim}{9}
          \pgfmathsetmacro{\xxlim}{\xlim-1}
          \pgfmathsetmacro{\yylim}{\ylim-1}
          \newcommand{\zz}{0}
          \newcommand{\recv}{01100010}
          
          \foreach \i in {0,...,\yylim}
            \node (i\i) at (0.65,\ylim-\i) {\scriptsize$i\!=\!\i$};
          \foreach \i in {1,...,\yylim}
            \node (y\i) at (0.25,\ylim-\i+0.5) {$y_{\i} \!=\! \StrChar{\recv}{\i}$};

          \foreach \i in {0,...,\yylim}
            \foreach \j in {0,...,3}
            \pgfmathsetmacro{\jj}{\j}
        	\node [dot] (\j-\i) at (\jj*1.5+1,\ylim-\i) {};

          \pgfmathsetmacro{\xxxlim}{\xlim-2}
          \pgfmathsetmacro{\yyylim}{\ylim-2}
          \foreach \i in {0,...,\yylim}
            \foreach \j in {0,...,\xxxlim}
        	{
              \pgfmathtruncatemacro{\ii}{\i+1}
              \pgfmathtruncatemacro{\jj}{\j+1}

          \draw[->,gray,thick] (\j-\i) to (\jj-\i);
        	}

          \pgfmathsetmacro{\xxxlim}{\xlim-2}
          \pgfmathsetmacro{\yyylim}{\ylim-2}
          \foreach \i in {0,...,\yyylim}
            \foreach \j in {0,...,\xxxlim}
        	{
              \pgfmathtruncatemacro{\ii}{\i+1}
              \pgfmathtruncatemacro{\jj}{\j+1}
        \draw[->,gray,thick] (\j-\i) to (\jj-\ii);	
         }

          \pgfmathsetmacro{\xxxlim}{\xlim-2}
          \pgfmathsetmacro{\yyylim}{\ylim-2}
          \foreach \i in {0,...,\yyylim}{
            \foreach \j in {0,...,\xxxlim}
        	{
              \pgfmathtruncatemacro{\iii}{\i+2}
              \pgfmathtruncatemacro{\jj}{\j+1}
              \ifthenelse{\iii>\yylim}{}{
            \draw[->,gray,thick] (\j-\i) to (\jj-\iii);	
         }} }  
               
          \pgfmathsetmacro{\xxxlim}{\xlim-2}
          \pgfmathsetmacro{\yyylim}{\ylim-2}
          \foreach \i in {0,...,\yyylim}{
            \foreach \j in {0,...,\xxxlim}
        	{
              \pgfmathtruncatemacro{\iiii}{\i+3}
              \pgfmathtruncatemacro{\jj}{\j+1}
             
                  \ifthenelse{\iiii>\yylim}{}{
               \ifthenelse{\equal{\i}{4}}{}{
               \draw[->,gray,thick] (\j-\i) to (\jj-\iiii);
        	}}
        	}  } 
         \draw[->,gray,thick, dashed] (0-4) to (1-7);
            \newcommand{\diagfourlabels}{00111}
          \pgfmathsetmacro{\xxxlim}{\xlim-2}
          \pgfmathsetmacro{\yyylim}{\ylim-2}
          \foreach \i in {0,...,\yyylim}{
            \foreach \j in {0,...,\xxxlim}
        	{
              \pgfmathtruncatemacro{\ii}{\i+1}
              \pgfmathtruncatemacro{\iii}{\i+2}
              \pgfmathtruncatemacro{\iiii}{\i+3}
              \pgfmathtruncatemacro{\iiii}{\i+4}
              \pgfmathtruncatemacro{\jj}{\j+1}
               \ifthenelse{\iiii>\yylim}{}{
               \ifthenelse{\i>2}{}{
              \StrChar{\diagfourlabels}{\ii}[\curlabel]
        		  \ifthenelse{\equal{\curlabel}{\zz}}{
            	\draw[->,thick,blue] (\j-\i) to (\jj-\iiii);}
        		  {\draw[->,thick,red] (\j-\i) to (\jj-\iiii);} }}
            }}

\draw[->,ultra thick,red] (0-3) to (1-7);
\draw[->,thick,red,dashed] (0-4) to (1-8);

          \pgfmathsetmacro{\xxxlim}{\xlim-2}
          \pgfmathsetmacro{\yyylim}{\ylim-2}
          \foreach \i in {0,...,\yylim}
            \foreach \j in {1}
        	{
              \pgfmathtruncatemacro{\ii}{\i+1}
              \pgfmathtruncatemacro{\jj}{\j+1}
        	 \ifthenelse{\i>7}{}{     
               \draw[->,blue,thick] (\j-\i) to [in=150,out=30] (\jj-\i);
        	}
        	}
        	\draw[->,blue,thick,dashed] (1-8) to [in=150,out=30] (2-8);

          \pgfmathsetmacro{\xxxlim}{\xlim-2}
          \pgfmathsetmacro{\yyylim}{\ylim-2}
          \foreach \i in {0,...,\yyylim}
            \foreach \j in {1}
        	{
              \pgfmathtruncatemacro{\ii}{\i+1}
              \pgfmathtruncatemacro{\jj}{\j+1}
              \pgfmathtruncatemacro{\iit}{\i+\xlim-\ylim}
                  \StrChar{\recv}{\ii}[\temp]
        		  \ifthenelse{\equal{\temp}{\zz}}
        		  {
            \ifthenelse{\ii>7}{}{ 
            \draw[->,thick,blue] (\j-\i) to (\jj-\ii);}
        		  }{}
        	}    
 \draw[->,ultra thick,blue,dashed] (1-7) to (2-8);

          \pgfmathsetmacro{\xxxlim}{\xlim-2}
          \pgfmathsetmacro{\yyylim}{\ylim-2}
          \foreach \i in {0,...,\yyylim}
            \foreach \j in {1}
        	{
              \pgfmathtruncatemacro{\ii}{\i+1}
              \pgfmathtruncatemacro{\iii}{\i+2}
              \pgfmathtruncatemacro{\jj}{\j+1}
             
                  \StrChar{\recv}{\ii}[\temp]
        		  \ifthenelse{\equal{\temp}{\zz}}{
                     \StrChar{\recv}{\iii}[\temp]
        		   \ifthenelse{\equal{\temp}{\zz}}
            		  {\draw[->,thick,blue] (\j-\i) to (\jj-\iii);}{}}
        		  {}
        	}    
               
          \pgfmathsetmacro{\xxxlim}{\xlim-2}
          \pgfmathsetmacro{\yyylim}{\ylim-2}
          \foreach \i in {0,...,\yyylim}
            \foreach \j in {1}
        	{
              \pgfmathtruncatemacro{\ii}{\i+1}
              \pgfmathtruncatemacro{\iii}{\i+2}
              \pgfmathtruncatemacro{\iiii}{\i+3}
              \pgfmathtruncatemacro{\jj}{\j+1}
             
                  \StrChar{\recv}{\ii}[\temp]
        		  \ifthenelse{\equal{\temp}{\zz}}{
                     \StrChar{\recv}{\iii}[\temp]
        		   \ifthenelse{\equal{\temp}{\zz}}
            		  {\StrChar{\recv}{\iiii}[\temp]
        		   \ifthenelse{\equal{\temp}{\zz}}
            		  {
                                 \draw[->,thick,blue] (\j-\i) to (\jj-\iiii);}{}}{}}
        		  {}
        	}

          \pgfmathsetmacro{\xstart}{\xlim-1}
          \pgfmathsetmacro{\xlim}{\xstart+2}
          \pgfmathsetmacro{\xxlim}{\xlim-1}

          \pgfmathsetmacro{\xxxlim}{\xlim-2}
          \pgfmathsetmacro{\yyylim}{\ylim-2}
          \foreach \i in {0,...,\yylim}
            \foreach \j in {2}
        	{
              \pgfmathtruncatemacro{\ii}{\i+1}
              \pgfmathtruncatemacro{\jj}{\j+1}

          \draw[->,gray,thick] (\j-\i) to (\jj-\i);
        	}

          \pgfmathsetmacro{\xxxlim}{\xlim-2}
          \pgfmathsetmacro{\yyylim}{\ylim-2}
          \foreach \i in {0,...,\yyylim}
            \foreach \j in {2}
        	{
              \pgfmathtruncatemacro{\ii}{\i+1}
              \pgfmathtruncatemacro{\jj}{\j+1}
        \draw[->,gray,thick] (\j-\i) to (\jj-\ii);	
         }

          \pgfmathsetmacro{\xxxlim}{\xlim-2}
          \pgfmathsetmacro{\yyylim}{\ylim-2}
          \foreach \i in {0,...,\yyylim}{
            \foreach \j in {2}
        	{
              \pgfmathtruncatemacro{\iii}{\i+2}
              \pgfmathtruncatemacro{\jj}{\j+1}
              \ifthenelse{\iii>\yylim}{}{
            \draw[->,gray,thick] (\j-\i) to (\jj-\iii);	
         }} }  
               
          \pgfmathsetmacro{\xxxlim}{\xlim-2}
          \pgfmathsetmacro{\yyylim}{\ylim-2}
          \foreach \i in {0,...,\yyylim}{
            \foreach \j in {2}
        	{
              \pgfmathtruncatemacro{\iiii}{\i+3}
              \pgfmathtruncatemacro{\jj}{\j+1}
             
                  \ifthenelse{\iiii>\yylim}{}{
               \draw[->,gray,thick] (\j-\i) to (\jj-\iiii);
        	}
        	}  }  
          \pgfmathsetmacro{\xxxlim}{\xlim-2}
          \pgfmathsetmacro{\yyylim}{\ylim-2}
          \foreach \i in {0,...,\yyylim}{
            \foreach \j in {2}
        	{
              \pgfmathtruncatemacro{\ii}{\i+1}
              \pgfmathtruncatemacro{\iii}{\i+2}
              \pgfmathtruncatemacro{\iiii}{\i+3}
              \pgfmathtruncatemacro{\iiii}{\i+4}
              \pgfmathtruncatemacro{\jj}{\j+1}
               \ifthenelse{\iiii>\yylim}{}{
              \StrChar{\diagfourlabels}{\ii}[\curlabel]
        		  \ifthenelse{\equal{\curlabel}{\zz}}{
            	\draw[->,thick,blue] (\j-\i) to (\jj-\iiii);}
        		  {\draw[->,thick,red] (\j-\i) to (\jj-\iiii);} }
            }}
            
          \pgfmathsetmacro{\xlim}{2}
          \pgfmathsetmacro{\ylim}{9}
          \pgfmathsetmacro{\xxlim}{\xlim-1}
          \pgfmathsetmacro{\yylim}{\ylim-1}
          
          \foreach \i in {0,...,\yylim}
            \node (i\i) at (0.65+6.75,\ylim-\i) {\scriptsize$i\!=\!\i$};

          \foreach \i in {0,...,\yylim}
            \foreach \j in {0,...,2}
            \pgfmathsetmacro{\jj}{\j}
        	\node [dot] (\j-\i) at          (\jj*2+7.75,\ylim-\i) {};

          \pgfmathsetmacro{\xxxlim}{\xlim-2}
          \pgfmathsetmacro{\yyylim}{\ylim-2}
          \foreach \i in {0,...,\yylim}
            \foreach \j in {0,...,\xxxlim}
        	{
              \pgfmathtruncatemacro{\ii}{\i+1}
              \pgfmathtruncatemacro{\jj}{\j+1}

          \draw[->,gray,thick] (\j-\i) to (\jj-\i);
        	}

          \pgfmathsetmacro{\xxxlim}{\xlim-2}
          \pgfmathsetmacro{\yyylim}{\ylim-2}
          \foreach \i in {0,...,\yyylim}
            \foreach \j in {0,...,\xxxlim}
        	{
              \pgfmathtruncatemacro{\ii}{\i+1}
              \pgfmathtruncatemacro{\jj}{\j+1}
        \draw[->,gray,thick] (\j-\i) to (\jj-\ii);	
         }

          \pgfmathsetmacro{\xxxlim}{\xlim-2}
          \pgfmathsetmacro{\yyylim}{\ylim-2}
          \foreach \i in {0,...,\yyylim}{
            \foreach \j in {0,...,\xxxlim}
        	{
              \pgfmathtruncatemacro{\iii}{\i+2}
              \pgfmathtruncatemacro{\jj}{\j+1}
              \ifthenelse{\iii>\yylim}{}{
            \draw[->,gray,thick] (\j-\i) to (\jj-\iii);	
         }} }  
               
          \pgfmathsetmacro{\xxxlim}{\xlim-2}
          \pgfmathsetmacro{\yyylim}{\ylim-2}
          \foreach \i in {0,...,\yyylim}{
            \foreach \j in {0,...,\xxxlim}
        	{
              \pgfmathtruncatemacro{\iiii}{\i+3}
              \pgfmathtruncatemacro{\jj}{\j+1}
             
                  \ifthenelse{\iiii>\yylim}{}{
               
               \draw[->,gray,thick] (\j-\i) to (\jj-\iiii);
        	}
        	}  }  

          \pgfmathsetmacro{\xxxlim}{\xlim-2}
          \pgfmathsetmacro{\yyylim}{\ylim-2}
          \foreach \i in {0,...,\yyylim}{
            \foreach \j in {0,...,\xxxlim}
        	{
              \pgfmathtruncatemacro{\iiii}{\i+4}
              \pgfmathtruncatemacro{\jj}{\j+1}
             
                  \ifthenelse{\iiii>\yylim}{}{
               \ifthenelse{\i>3}{}{
                \ifthenelse{\i<1}{}{
               \draw[->,gray,thick] (\j-\i) to (\jj-\iiii);
        	}}}
        	}  }  
         
         \draw[->,gray,thick,dotted] (0-0) to (1-5);
        \draw[->,gray,thick,dotted] (0-0) to (1-6);
        \draw[->,gray,thick] (0-1) to (1-6);
        \draw[->,red,thick] (0-3) to (1-7);
        \draw[->,red,ultra thick] (0-3) to (1-8);
        \draw[->,gray,thick, dashed] (0-4) to (1-8);

        \draw[->,gray,thick,dotted] (0-0) to (1-4);

          \pgfmathsetmacro{\xstart}{\xlim-1}
          \pgfmathsetmacro{\xlim}{\xstart+2}
          \pgfmathsetmacro{\xxlim}{\xlim-1}

          \pgfmathsetmacro{\xxxlim}{\xlim-2}
          \pgfmathsetmacro{\yyylim}{\ylim-2}
          \foreach \i in {0,...,\yylim}
            \foreach \j in {1}
        	{
              \pgfmathtruncatemacro{\ii}{\i+1}
              \pgfmathtruncatemacro{\jj}{\j+1}

          \draw[->,gray,thick] (\j-\i) to (\jj-\i);
        	}

          \pgfmathsetmacro{\xxxlim}{\xlim-2}
          \pgfmathsetmacro{\yyylim}{\ylim-2}
          \foreach \i in {0,...,\yyylim}
            \foreach \j in {1}
        	{
              \pgfmathtruncatemacro{\ii}{\i+1}
              \pgfmathtruncatemacro{\jj}{\j+1}
        \draw[->,gray,thick] (\j-\i) to (\jj-\ii);	
         }

          \pgfmathsetmacro{\xxxlim}{\xlim-2}
          \pgfmathsetmacro{\yyylim}{\ylim-2}
          \foreach \i in {0,...,\yyylim}{
            \foreach \j in {1}
        	{
              \pgfmathtruncatemacro{\iii}{\i+2}
              \pgfmathtruncatemacro{\jj}{\j+1}
              \ifthenelse{\iii>\yylim}{}{
                 \ifthenelse{\i>5}{}{
            \draw[->,gray,thick] (\j-\i) to (\jj-\iii);	}
         }} }  
      \draw[->,gray,thick,dotted] (1-6) to (2-8);            
          \pgfmathsetmacro{\xxxlim}{\xlim-2}
          \pgfmathsetmacro{\yyylim}{\ylim-2}
          \foreach \i in {0,...,\yyylim}{
            \foreach \j in {1}
        	{
              \pgfmathtruncatemacro{\iiii}{\i+3}
              \pgfmathtruncatemacro{\jj}{\j+1}
             
                  \ifthenelse{\iiii>\yylim}{}{
               \ifthenelse{\i>4}{}{
               \draw[->,gray,thick] (\j-\i) to (\jj-\iiii);
        	}}
        	}  }
      \draw[->,gray,thick,dotted] (1-5) to (2-8);
         
          \pgfmathsetmacro{\xxxlim}{\xlim-2}
          \pgfmathsetmacro{\yyylim}{\ylim-2}
          \foreach \i in {0,...,\yyylim}{
            \foreach \j in {1}
        	{
              \pgfmathtruncatemacro{\ii}{\i+1}
              \pgfmathtruncatemacro{\iii}{\i+2}
              \pgfmathtruncatemacro{\iiii}{\i+3}
              \pgfmathtruncatemacro{\iiii}{\i+4}
              \pgfmathtruncatemacro{\jj}{\j+1}
               \ifthenelse{\iiii>\yylim}{}{
              \StrChar{\diagfourlabels}{\ii}[\curlabel]
              \ifthenelse{\i>3}{}{
        		  \ifthenelse{\equal{\curlabel}{\zz}}{
            	\draw[->,thick,blue] (\j-\i) to (\jj-\iiii);}
        		  {\draw[->,thick,red] (\j-\i) to (\jj-\iiii);} }}
            }}
            \draw[->,thick,red,dotted] (1-4) to (2-8);

          \pgfmathsetmacro{\xlim}{2}
          \pgfmathsetmacro{\ylim}{9}
          \pgfmathsetmacro{\xxlim}{\xlim-1}
          \pgfmathsetmacro{\yylim}{\ylim-1}
          
          \foreach \i in {0,...,\yylim}
            \node (i\i) at (0.65+13,\ylim-\i) {\scriptsize$i\!=\!\i$};

          \foreach \i in {0,...,\yylim}
            \foreach \j in {0,...,\xxlim}
            \pgfmathsetmacro{\jj}{\j}
        	\node [dot] (\j-\i) at (\jj*3.5+14,\ylim-\i) {};
         
          \pgfmathsetmacro{\xxxlim}{\xlim-2}
          \pgfmathsetmacro{\yyylim}{\ylim-2}
          \foreach \i in {0,...,\yylim}
            \foreach \j in {0,...,\xxxlim}
        	{
              \pgfmathtruncatemacro{\ii}{\i+1}
              \pgfmathtruncatemacro{\jj}{\j+1}

          \draw[->,gray,thick] (\j-\i) to (\jj-\i);
        	}

          \pgfmathsetmacro{\xxxlim}{\xlim-2}
          \pgfmathsetmacro{\yyylim}{\ylim-2}
          \foreach \i in {0,...,\yyylim}
            \foreach \j in {0,...,\xxxlim}
        	{
              \pgfmathtruncatemacro{\ii}{\i+1}
              \pgfmathtruncatemacro{\jj}{\j+1}
        \draw[->,gray,thick] (\j-\i) to (\jj-\ii);	
         }

          \pgfmathsetmacro{\xxxlim}{\xlim-2}
          \pgfmathsetmacro{\yyylim}{\ylim-2}
          \foreach \i in {0,...,\yyylim}{
            \foreach \j in {0,...,\xxxlim}
        	{
              \pgfmathtruncatemacro{\iii}{\i+2}
              \pgfmathtruncatemacro{\jj}{\j+1}
              \ifthenelse{\iii>\yylim}{}{
            \draw[->,gray,thick] (\j-\i) to (\jj-\iii);	
         }} }  
               
          \pgfmathsetmacro{\xxxlim}{\xlim-2}
          \pgfmathsetmacro{\yyylim}{\ylim-2}
          \foreach \i in {0,...,\yyylim}{
            \foreach \j in {0,...,\xxxlim}
        	{
              \pgfmathtruncatemacro{\iiii}{\i+3}
              \pgfmathtruncatemacro{\jj}{\j+1}
             
                  \ifthenelse{\iiii>\yylim}{}{
               \draw[->,gray,thick] (\j-\i) to (\jj-\iiii);
        	}
        	}  }  

          \pgfmathsetmacro{\xxxlim}{\xlim-2}
          \pgfmathsetmacro{\yyylim}{\ylim-2}
          \foreach \i in {0,...,\yyylim}{
            \foreach \j in {0,...,\xxxlim}
        	{
              \pgfmathtruncatemacro{\iiii}{\i+3}
              \pgfmathtruncatemacro{\jj}{\j+1}
             
                  \ifthenelse{\iiii>\yylim}{}{
               \draw[->,gray,thick] (\j-\i) to (\jj-\iiii);
        	}
        	}  }  

          \pgfmathsetmacro{\xxxlim}{\xlim-2}
          \pgfmathsetmacro{\yyylim}{\ylim-2}
          \foreach \i in {0,...,\yyylim}{
            \foreach \j in {0,...,\xxxlim}
        	{
              \pgfmathtruncatemacro{\iiii}{\i+4}
              \pgfmathtruncatemacro{\jj}{\j+1}
             
                  \ifthenelse{\iiii>\yylim}{}{
               \draw[->,gray,thick] (\j-\i) to (\jj-\iiii);
        	}
        	}  }
          \pgfmathsetmacro{\xxxlim}{\xlim-2}
          \pgfmathsetmacro{\yyylim}{\ylim-2}
          \foreach \i in {0,...,\yyylim}{
            \foreach \j in {0,...,\xxxlim}
        	{
              \pgfmathtruncatemacro{\iiii}{\i+5}
              \pgfmathtruncatemacro{\jj}{\j+1}
             
                  \ifthenelse{\iiii>\yylim}{}{
               \draw[->,gray,thick] (\j-\i) to (\jj-\iiii);
        	}
        	}  }
          \pgfmathsetmacro{\xxxlim}{\xlim-2}
          \pgfmathsetmacro{\yyylim}{\ylim-2}
          \foreach \i in {0,...,\yyylim}{
            \foreach \j in {0,...,\xxxlim}
        	{
              \pgfmathtruncatemacro{\iiii}{\i+6}
              \pgfmathtruncatemacro{\jj}{\j+1}
             
                  \ifthenelse{\iiii>\yylim}{}{
               \draw[->,gray,thick] (\j-\i) to (\jj-\iiii);
        	}
        	}  }

\draw[->,gray,thick,dotted] (0-0) to (1-8);

\draw[rounded corners] (-0.5, 11) rectangle (6, 0.5) {};
\draw[rounded corners] (6.5, 11) rectangle (12.5, 0.5) {};
\draw[rounded corners] (13, 11) rectangle (18, 0.5) {};

\draw [decorate,decoration = {calligraphic brace,        amplitude=4pt},thick] (1,10)--(5.5,10)  node[pos=0.5,above  =4pt]{$\mathcal{T}^{[00]}$} ;
\draw [decorate,decoration = {calligraphic brace,        amplitude=4pt},thick] (1,9.3)--(2.5,9.3)  node[pos=0.5,above  =4pt]{\small $\mathcal{T}^{\mathrm{B-1}}$} ;
\draw [decorate,decoration = {calligraphic brace,        amplitude=4pt},thick] (2.5,9.3)--(4,9.3)  node[pos=0.5,above  =4pt]{\small $\mathcal{T}^{\mathrm{GB-1}}$} ;
\draw [decorate,decoration = {calligraphic brace,        amplitude=4pt},thick] (4,9.3)--(5.5,9.3)  node[pos=0.5,above  =4pt]{\small $\mathcal{T}^{\mathrm{B-2}}$} ;
\draw [decorate,decoration = {calligraphic brace,        amplitude=4pt},thick] (14,10)--(17.5,10)  node[pos=0.5,above  =4pt]{$\mathcal{T}^{[000]}$} ;
\node at (0,10.5){(a)};
\node at (7,10.5){(b)};
\node at (13.5,10.5){(c)};
       \end{tikzpicture}

%% file: tikz_schemes/GBMc-noBackground.tex
        \begin{tikzpicture}[xscale=1.2,yscale=1]
\node at (-3.15,1) {$\bfZ=\bfY^*$};
        \draw [very thick,draw=black,fill=white] (-2.0,1.25) rectangle (0.7,0.75) node[pos=.5] {$\bfZ_{\RN{1}}$};
        \draw [very thick,draw=black,fill=blue!70!black] (0.7,1.25) rectangle (1.30,0.75) node[pos=.5,white] {$\bfZ_{\Delta}$};
        \draw [very thick,draw=black,fill=white] (1.30,1.25) rectangle (2.0,0.75) node[pos=.5] {$\bfZ_{\RN{2}}$};

    \draw [{-latex'},draw=black] (0,1.6) --(0,1.25) node[above=8pt] {$i_\mathrm{mid}$};
    
    \draw [decorate,
        decoration = {calligraphic brace,mirror, 
            amplitude=6pt},thick] (-2,0.75)--(0,0.75)  node[pos=0.5,below  =5pt]{$\bfZL$} ;
    \draw [decorate,
        decoration = {calligraphic brace, mirror,
            amplitude=6pt},thick] (0,0.75)--(2,0.75)  node[pos=0.5,below  =5pt]{$\bfZR$} ;
    \draw[-{latex'[scale=1pt]},green!50!black, thick,opacity=1] (-1,0)--(-1.25,-0.75) node[left,scale=0.9pt,pos=.4] {TC};    
    \draw[-{latex'[scale=1pt]},green!50!black,opacity=1, thick] (1,0)--(1.25,-0.75) node[left,scale=0.9pt,pos=.4] {TC}; 

        \draw [very thick,draw=black,fill=white] (0.4,-0.8) rectangle (2,-1.3)node[pos=.5] {} ;
        \draw [very thick,draw=black,fill=white] (-2.0,-0.8) rectangle (-0.3,-1.3)node[pos=.5] {$\bfZL^*$} ; 
        \draw [very thick,draw=black,fill=blue!70!black] (0.7,-0.8) rectangle (1.30,-1.3) ;
        \node at (1.25,-1.05) {\contour{white}{\protect\color{black}$\bfZR^*$}};
    \draw[dashed] (-0.3,-0.8) -- (0,0.75); 
    \draw[dashed] (0.4,-0.8) -- (0,0.75);
        \draw[dashed] (2,-0.8) -- (2,0.75);
                \draw[dashed] (-2,-0.8) -- (-2,0.75);
        \end{tikzpicture}

%% file: tikz_schemes/scheme_GB_presence_small.tex
 \centerline{
 \hspace*{3em}\begin{tikzpicture}
[scale=0.8, every node/.style={scale=0.8}]
\draw[white] (-1,-1)--(11,-1)--(11,1)--(-1,1)--(-1,-1);

\node[rectangle,draw, minimum width = 1.5cm, 
    minimum height = 0.75cm] (D1) at (0,0) {$\bfX(1)$};
\node[rectangle,draw,   text = white, fill = blue!70!black,
    minimum width = 1.25cm,
    minimum height = 0.75cm, right = -\the\pgflinewidth of D1.north east, anchor = north west] (GB1) {$00...0$};
\node[rectangle,draw, minimum width = 1.5cm, 
    minimum height = 0.75cm, right = -\the\pgflinewidth of GB1.north east, anchor = north west] (D2) {$\bfX(2)$};
\node[rectangle,draw,   text = white, fill = blue!70!black,
    minimum width = 2cm,
    minimum height = 0.75cm, right = -\the\pgflinewidth of D2.north east, anchor = north west] (GB2) {$00........0$};
    \node[rectangle,draw, minimum width = 1.5cm, 
    minimum height = 0.75cm, right = -\the\pgflinewidth of GB2.north east, anchor = north west] (D3) {$\bfX(3)$};
\node[rectangle,draw,   text = white, fill = blue!70!black,
    minimum width = 1.25cm,
    minimum height = 0.75cm, right = -\the\pgflinewidth of D3.north east, anchor = north west] (GB3) {$00...0$};
\node[rectangle,draw, minimum width = 1.5cm, 
    minimum height = 0.75cm, right = -\the\pgflinewidth of GB3.north east, anchor = north west] (D4) {$\bfX(4)$};

\commentBlock{
\draw [{Bar[scale=0.5pt]}-{Bar[scale=0.5]}] (-0.75,-0.5)--(0.75,-0.5)  node[pos=0.5,below  =1pt]{$N_0$} ;
  \draw[{Bar[scale=0.5pt]}-{Bar[scale=0.5]}](0.75,-0.7)--(2,-0.7)  node[pos=0.5,below  =1pt]{$\ell_{n_0+1}$} ;
\draw [{Bar[scale=0.5pt]}-{Bar[scale=0.5]}] (2,-0.5)--(3.5,-0.5)  node[pos=0.5,below  =1pt]{$N_0$} ;
        \draw [{Bar[scale=0.5pt]}-{Bar[scale=0.5]}] (3.5,-0.7)--(5.5,-0.7)  node[pos=0.5,below  =1pt]{$\ell_{n_0+2}$} ;
\draw [{Bar[scale=0.5pt]}-{Bar[scale=0.5]}] (5.5,-0.5)--(7,-0.5)  node[pos=0.5,below  =1pt]{$N_0$} ;
         \draw [{Bar[scale=0.5pt]}-{Bar[scale=0.5]}] (7,-0.7)--(8.25,-0.7)  node[pos=0.5,below  =1pt]{$\ell_{n_0+1}$} ;       
 \draw [{Bar[scale=0.5pt]}-{Bar[scale=0.5]}] (8.25,-0.5)--(9.75,-0.5)  node[pos=0.5,below  =1pt]{$N_0$} ;  

}

\draw[-{Stealth}] (2,0.9) -- (2,0.4) node[above=0.4]{$j_1$};
\draw[-{Stealth}] (5.5,1.9) -- (5.5,1.4) node[above=0.4]{$j_{\mathrm{mid}}$};
\draw[-{Stealth}] (5.5,0.9) -- (5.5,0.4) node[above=0.4]{$j_1$};
\draw[-{Stealth}] (8.2,0.9) -- (8.2,0.4) node[above=0.4]{$j_1$};

 \draw [pen colour={blue!70!black}, decorate,   decoration = {calligraphic brace,mirror,
        amplitude=7pt},thick](5.55,-0.45)--(8.22,-0.45)  node[pos=0.5,below  =6pt, blue!70!black]{$(1,n_0)_{\mathrm{series}}$} ;    
 \draw [pen colour={blue!50!white}, decorate,    decoration = {calligraphic brace,mirror, 
        amplitude=10pt}, thick](-0.75,-0.5)--(2,-0.5)  node[pos=0.5,below  =8pt, blue!50!white]{$(1,n_0)_{\mathrm{series}}$} ;          

\draw [pen colour={blue!70!black},decorate,    decoration = {calligraphic brace,mirror,
        amplitude=7pt},thick](-0.75,-0.45)--(5.5,-0.45)  node[pos=0.5,below  =6pt,blue!70!black]{$(2,n_0)_{\mathrm{series}}$} ; 
                \end{tikzpicture}
        }